\documentclass[lettersize,journal]{IEEEtran}
\usepackage[dvipsnames]{xcolor}
\usepackage{enumitem}
\usepackage{amsmath,amsfonts}
\usepackage{booktabs}
\usepackage{graphicx}
\usepackage{subcaption}
\usepackage{verbatim}
\usepackage{xspace}
\usepackage{multirow}
\usepackage{threeparttable} 
\usepackage[bb=boondox]{mathalfa}
\usepackage{pifont}
\usepackage{marvosym}
\usepackage[colorlinks]{hyperref}
\hypersetup{
  colorlinks,
  citecolor=NavyBlue,
  linkcolor=Maroon}
\usepackage{xurl}
\usepackage{makecell}
\usepackage{algorithmic}
\usepackage{array}
\usepackage{textcomp}
\usepackage{stfloats}
\usepackage{verbatim}
\usepackage{lipsum}
\usepackage{doi}

\def\BibTeX{{\rm B\kern-.05em{\sc i\kern-.025em b}\kern-.08em
    T\kern-.1667em\lower.7ex\hbox{E}\kern-.125emX}}
\usepackage{balance}
\usepackage{etoolbox}
\makeatletter
\patchcmd{\hyper@makecurrent}{%
    \ifx\Hy@param\Hy@chapterstring
        \let\Hy@param\Hy@chapapp
    \fi
}{%
    \iftoggle{inappendix}{%true-branch
        \@checkappendixparam{chapter}%
        \@checkappendixparam{section}%
        \@checkappendixparam{subsection}%
        \@checkappendixparam{subsubsection}%
        \@checkappendixparam{paragraph}%
        \@checkappendixparam{subparagraph}%
    }{}%
}{}{\errmessage{failed to patch}}

\newcommand*{\@checkappendixparam}[1]{%
    \def\@checkappendixparamtmp{#1}%
    \ifx\Hy@param\@checkappendixparamtmp
        \let\Hy@param\Hy@appendixstring
    \fi
}
\makeatletter

\newtoggle{inappendix}
\togglefalse{inappendix}

\apptocmd{\appendix}{\toggletrue{inappendix}}{}{\errmessage{failed to patch}}
\newcommand{\customref}[1]{\hyperref[#1]{C\ref*{#1}}}

\setlist[itemize]{topsep=0pt, itemsep=0pt, parsep=0pt, partopsep=0pt}
\setlist[itemize]{leftmargin=8pt}
\usepackage[framemethod=TikZ]{mdframed}
\mdfsetup{skipabove=7pt,skipbelow=0pt}
\definecolor{darkcyan}{HTML}{0091A4}  
\definecolor{brightcyan}{HTML}{dcf0f2}
\newcounter{obscounter}
\mdfdefinestyle{obs}{
    linecolor=darkcyan, % Blue line color for left line
    outerlinewidth=3pt,
    leftline=true, % Only show the left line
    rightline=false, % No right line
    topline=false, % No top line
    bottomline=false, % No bottom line
    innertopmargin=10pt, % Space at the start of the box
    innerbottommargin=10pt, % Space at the end of the box
    innerrightmargin=13pt, % Space at the right of the box
    innerleftmargin=10pt, % Space at the left of the box
    backgroundcolor=brightcyan % Background color of the box
}
\newcommand{\obs}{\refstepcounter{obscounter}\textbf{Observation\text{ }\theobscounter\text{ }(OBS.\theobscounter).}\xspace}

\newcommand{\adddiscuss}[1]{\textcolor{black}{#1}}

\newcommand*\circled[1]{\tikz[baseline=(char.base)]{
            \node[shape=circle,draw,inner sep=1pt] (char) {#1};}}

\usepackage{orcidlink}

\begin{document}

\title{Identify As A Human Does: A Pathfinder of Next-Generation Anti-Cheat Framework for First-Person Shooter Games}

% \author{Jiayi Zhang\orcidlink{0009-0001-4452-6464}, Chenxin Sun\orcidlink{0009-0001-5057-6089}, Yue Gu\orcidlink{0009-0005-7827-7098}, Qingyu Zhang\orcidlink{0009-0009-4422-3971}, Jiayi Lin\orcidlink{0009-0004-6790-1302}, Xiaojiang Du\orcidlink{0000-0003-4235-9671}, Chenxiong Qian\orcidlink{0000-0002-6201-6011}}

\author{
% \IEEEauthorblockN{Jiayi Zhang\IEEEauthorrefmark{1}\IEEEauthorrefmark{2}\orcidlink{0009-0001-4452-6464}, Chenxin Sun\IEEEauthorrefmark{1}\IEEEauthorrefmark{2}\orcidlink{0009-0001-5057-6089}, Yue Gu\IEEEauthorrefmark{2}\orcidlink{0009-0005-7827-7098}, Qingyu Zhang\IEEEauthorrefmark{2}\orcidlink{0009-0009-4422-3971}, Jiayi Lin\IEEEauthorrefmark{2}\orcidlink{0009-0004-6790-1302}, Xiaojiang Du\IEEEauthorrefmark{4}\orcidlink{0000-0003-4235-9671}, Chenxiong Qian\IEEEauthorrefmark{3}\IEEEauthorrefmark{5}\orcidlink{0000-0002-6201-6011}}
Jiayi Zhang\orcidlink{0009-0001-4452-6464}, 
Chenxin Sun\orcidlink{0009-0001-5057-6089}, 
Yue Gu\orcidlink{0009-0005-7827-7098}, 
Qingyu Zhang\orcidlink{0009-0009-4422-3971},
Jiayi Lin\orcidlink{0009-0004-6790-1302}, 
\thanks{
Jiayi Zhang, Chenxin Sun, Yue Gu, Qingyu Zhang, Jiayi Lin, and Chenxiong Qian are with the School of Computing and Data Science, The University of Hong Kong, Hong Kong 999077, China (e-mail: \href{mailto:cqian@cs.hku.hk}{\color{black}cqian@cs.hku.hk}). Jiayi Zhang and Chenxin Sun contributed equally to this research.
}

Xiaojiang Du\orcidlink{0000-0003-4235-9671}, 
\IEEEmembership{Fellow, IEEE,}
\thanks{Xiaojiang Du is with the Department of Electrical and Computer Engineering, Stevens Institute of Technology, NJ 07030, USA.}and Chenxiong Qian\orcidlink{0000-0002-6201-6011},
\IEEEmembership{Member, IEEE}
\thanks{
Received 6 March 2025; revised 24 August 2025; accepted 10 November 2025. Date of publication $<$TBD HERE$>$; date of current version $<$TBD HERE$>$. This work is partially supported by the NSFC for Young Scientists of China (No.62202400) and the RGC for Early Career Scheme (No.27210024). The associate editor coordinating the review of this article and approving it for publication was Prof. Abderrahim Benslimane. \textit{(Corresponding author: Chenxiong Qian.)}

This paper has supplementary downloadable material available at \href{http://ieeexplore.ieee.org}{\color{black}http://ieeexplore.ieee.org}., provided by the authors. The material includes detailed feature designs, experimental settings, additional evaluation details. Contact \href{mailto:brucejiayi@connect.hku.hk}{\color{black}brucejiayi@connect.hku.hk} for further questions about this work.

Digital Object Identifier 10.1109/TIFS.$<$TBD HERE$>$
}
% \vspace{0.3em}
% \IEEEauthorblockA{\IEEEauthorrefmark{2}\IEEEauthorrefmark{5}\textit{The University of Hong Kong}
%                     \\\IEEEauthorrefmark{2}\textit{\{brucejiayi, roniny, guyue22, z1anqy22, linjy01\}@connect.hku.hk}, \IEEEauthorrefmark{5}\textit{cqian@cs.hku.hk} \\ \vspace{0.2em}
%                   \IEEEauthorrefmark{4}\textit{Stevens Institute of Technology}
%                   \\\IEEEauthorrefmark{4}\textit{xdu16@stevens.edu}
%                 }
}%end author

\maketitle

\begingroup
\renewcommand\thefootnote{\fnsymbol{footnote}}
% \footnotetext[1]{Equal contribution.}
% \footnotetext[3]{Corresponding author.}
\endgroup

\markboth{IEEE TRANSACTIONS ON INFORMATION FORENSICS AND SECURITY,~Vol.~X, No.~Y, November~2025}%
{Identify As A Human Does: A Pathfinder of Next-Generation Anti-Cheat Framework for First-Person Shooter Games}

\newcommand\CQ[1]{\textbf{\textcolor{purple}{CQ: #1}}}
\newcommand\Jiayi[1]{\textbf{\textcolor{orange}{Jiayi: #1}}}
\newcommand\Du[1]{\textbf{\textcolor{brown}{#1}}}

\newcommand{\syspov}{\textsc{RevPov}\xspace}
\newcommand{\sysstats}{\textsc{RevStats}\xspace}
\newcommand{\sysinteg}{\textsc{Mvin}\xspace}
\newcommand{\sysspc}{\textsc{ExSPC}\xspace}
\newcommand{\sys}{\textsc{Hawk}\xspace}

\maketitle

\begin{abstract}
% Game security has become increasingly critical, particularly in first-person shooter (FPS) games, one of the most popular genres, which are frequently targeted by cheaters, leading to substantial financial losses.
% %
% Anti-cheat can be classified as client-side and server-side.
% %
% Existing client-side anti-cheats face limitations such as hardware constraints, security risks, and system overheads. 
% %
% Server-side anti-cheats, though avoiding client-side issues, are still in their infancy, primarily relying on naive features, lacking robust frameworks, requiring modification on the game engine, introducing concurrent server overheads, and often questioning by the community.
% %
% The existing works on both sides never use large and real-world datasets and are unable to detect multiple types of cheats simultaneously.
% %
% To address these limitations, we propose \sys, the first server-side FPS anti-cheat framework equipped with a well-defined workflow, leveraging machine learning techniques to mimic human experts' identification process.
% %
% We evaluate \sys with CS:GO, a leading commercial FPS game, using large and real-world data containing multiple cheat types.
% %
% \sys demonstrates promising efficiency and overheads, shorter ban times compared to the official platform, a significant reduction in manual labor, and the ability to capture cheaters who evaded official inspections.

The gaming industry has experienced substantial growth, but cheating in online games poses a significant threat to the integrity of the gaming experience. 
Cheating, particularly in first-person shooter (FPS) games, can lead to substantial losses for the game industry. 
Existing anti-cheat solutions have limitations, such as client-side hardware constraints, security risks, server-side unreliable methods, and both-sides suffer from a lack of comprehensive real-world datasets.
To address these limitations, the paper proposes \sys, a server-side FPS anti-cheat framework for the popular game CS:GO.
\sys utilizes machine learning techniques to mimic human experts' identification process, leverages novel multi-view features, and is equipped with a well-defined workflow. 
\sys is evaluated with the first large and real-world datasets containing multiple cheat types and cheating sophistication, and it exhibits promising efficiency and acceptable overheads, shorter ban times, higher recall and similar false positive rate compared to the in-use anti-cheat, and the ability to capture cheaters who evaded official inspections. 
%
%The authors open-source HAWK and their datasets.

%Cheatings like aimbots and wallhacks, which assist players in gaining unfair advantages in First Person Shooter (FPS) games, result in substantial losses to the game industry.
%
%Although abundant efforts have been invested in anti-cheating, existing works face limitations such as client-side hardware constraints and security risks, server-side unreliable methods, and both-sides worrying detection performance, datasets scarcity, concurrent overheads and they can only detect a single type of cheat. 
%
%Thus we propose \sys, a server-side FPS anti-cheat framework for a popular FPS game CS:GO. 
%
%\sys utilizes machine learning techniques to mimic human experts' identification process, leverages novel multi-view features, and is equipped with a well-defined workflow.
%
%We evaluate \sys with the first large and real-world datasets containing multiple cheat types and sophistication.
%
%\sys exhibits promising efficiency and overheads, shorter ban times compared to the in-use anti-cheat, a significant reduction in manual labor, and the ability to capture cheaters who evaded official inspections.
%
%We open-source \sys and our datasets.
\end{abstract}

\begin{IEEEkeywords}
Game security, anti-cheat, intrusion detection, machine learning.
\end{IEEEkeywords}

\section{Introduction}\label{sec:intro}
\IEEEPARstart{T}{he} gaming industry has experienced substantial growth recently, with the global games market generating \$187.7 billion in 2024 and projected to generate \$213.3 billion by 2027 \cite{newzoo2024}. 
However, cheating in online games poses a significant threat to the integrity of gaming experiences and disrupts gameplay balance. 
In 2023 alone, cheating led to an estimated \$29 billion losses, with 78\% of players discouraged from playing due to cheating \cite{irdeto2023}. 
%
%This underscores the critical importance of game security \cite{gianvecchio2009battle}.
%
In particular, first-person shooter (FPS) games, which account for 20.9\% of total game sales \cite{wepc2023}, are particularly targeted by cheat developers because their stringent latency requirements make key computations on the client side, creating fertile ground for cheats to exploit.

%seeking monetary gains due to their competitive and multiplayer nature. 
%
The most prevalent forms of cheating involve the use of \textit{aimbots} to aid aiming and \textit{wallhacks} to reveal opponents' positions, thus compromising game integrity \cite{belyaeva2022stakeholder, liu2021lower, choi2023botscreen}.

Realizing the critical importance of game security \cite{gianvecchio2009battle}, both academia and industry have proposed many anti-cheats solutions, applied either on the client side~\cite{choi2023botscreen,invisibilitycloak,10.1145/3372297.3417890} or server side \cite{yeung2006detecting,yu2012statistical,6633617,liu2017detecting}.
%
%For instance, BotScreen \cite{choi2023botscreen} uses recurrent neural networks (RNN) to detect aiming anomalies in Counter-Strike: Global Offensive (CS:GO).
%
%Additionally, Valve, the publisher of CS:GO, has adopted deep learning methods for cheat detection \cite{gdc2018}.
%
However, current solutions have the following limitations.

Client-side solutions face hardware constraints, security and privacy concerns, or unacceptable system overheads.
Two state-of-the-art systems \cite{choi2023botscreen,10.1145/3372297.3417890} require a specific Trusted Execution Environment (TEE), Intel SGX. 
This dependence on particular hardware, which was deprecated in 2021 \cite{bleepingcomputer2024, intelcommunity2024}, limits the compatibility and practicality for industrial use.
Some industrial solutions \cite{EasyAC, BattlEye, SteamSupport} involve scanning hard drives and gaining root privileges, posing risks of unauthorized access and personal data leakage \cite{mikkelsen2017information,9566108,10.1145/3380786.3391397}. 
Vulnerabilities, e.g., RCE exploits \cite{BackEngineering2021} and CVEs \cite{CVE-2019-16098, CVE-2020-36603, CVE-2021-3437, CVE-2023-38817}, demonstrate the potential for system instability and privacy breaches.
And most client-side works rely on concurrent surveillance \cite{10.1145/3372297.3417890,invisibilitycloak} leading to increased client-side overheads, or frequent patching to counter varying cheat signatures \cite{EasyAC, BattlEye, SteamSupport,choi2023botscreen}, which requires extra engineering efforts and is considered to be less effective \cite{liu2017detecting} since it only identifies known and deprecated cheats.
\adddiscuss{
Importantly, client-side data and decision results are unreliable and easily tampered with \cite{gaffer2016nevertrust,guardian2016hackerscheats,i3d2024counteringcheating,liu2017detecting}.
Thus, the industry has shifted towards server-side anti-cheat solutions \cite{gaffer2016nevertrust,guardian2016hackerscheats}.
}

However, existing server-side solutions use basic features and have low recall and accuracy.
%
% \cite{yeung2006detecting, inproceedings, yu2012statistical, 6633617, liu2017detecting, han2015online}
For instance, previous works \cite{yeung2006detecting,yu2012statistical,liu2017detecting,orlova2024estimation,han2015online} use naive features like winning rates, playtime, or count shots on an invisible bait which are insufficient or inapplicable for detecting today's complex cheats or specific cheat types. 
These works leverage thresholds to classify, which is not robust enough for practical use, especially in noise-rich, real-world datasets.
Additionally, some commercial anti-cheats integrate such server-side approaches \cite{EasyAC}. 
Yet, they show low recall and accuracy, with extended ban cycles and increasing community grievances \cite{zleague_warzone_anticheat, pushtotalk_anticheat}. 
Incidents like the 2024 APEX Legends Tournament hack \cite{algshack2024} and the study \cite{orlova2024estimation} highlight the ineffectiveness of existing commercial anti-cheats, e.g., EAC \cite{EasyAC} and VAC \cite{SteamSupport}.

%\noindent\textbf{\textit{(L4)}} Current server-side anti-cheat works \cite{yeung2006detecting, inproceedings,yu2012statistical,6633617,liu2017detecting,han2015online} utilize machine learning techniques but primarily rely on simple and monospecific features, and their frameworks and workflows are hard to meet real-world anti-cheat requirements.
%
%For instance, basic features such as winning rates or playtime in prior works \cite{han2015online, yu2012statistical}, are not applicable for detecting \textit{aimbots} or \textit{wallhacks}, or unable to cover complex behaviors.
%
% Importantly, prior server-side works are evaluated on small-scale simulated datasets, which may lead to performance degradation or even failure on larger and more complex real-world datasets.
%
%Besides, these works \cite{yeung2006detecting, inproceedings,yu2012statistical,6633617,liu2017detecting}, they did not come up with a well-defined workflow that considered the occurrence of false positives, given that they worked well on small and non-real-world datasets.
%
%But for noise-rich, more complex, and exponentially larger real-world datasets, this has a high possibility of issuing false bans, and these works' real-world performance remains unclear.

%\noindent\textbf{\textit{(L5)}} Current server-side solutions \cite{yeung2006detecting, inproceedings,yu2012statistical,6633617,liu2017detecting} require in-game real-time data collection, introducing concurrent overheads to the server.

More broadly, prior research on anti-cheat techniques suffers from the lack of comprehensive real-world datasets, with most studies relying on limited or synthetic data \cite{choi2023botscreen, 10.1145/3372297.3417890, yeung2006detecting, yu2012statistical, 6633617, liu2017detecting}, which are hard to capture diverse cheating scenarios and varying levels of sophistication (explained in \autoref{sec:CT}).
%Moreover, solutions on both sides lack of comprehensive datasets and rely on limited or synthetic data \cite{choi2023botscreen, 10.1145/3372297.3417890, yeung2006detecting, yu2012statistical, 6633617, liu2017detecting}, which is hard to represent real-world cheating scenarios and lack varying levels of cheating sophistication (explained in \autoref{sec:CT}).
%
This could result in performance degradation or even failure in complex real-world scenarios.

To address the aforementioned limitations, we propose \sys, a server-side approach designed to analyze data transmitted to the server, thereby avoiding the inherent limitations of client-side solutions. 
\sys introduces a comprehensive set of multi-view features and a robust framework to meticulously monitor players' points of view, statistics, and behavioral consistencies. 
This approach includes a corresponding workflow that aims to shorten the ban cycle and considers the occurrence of false positives. 
\sys achieves high detection recall and acceptable accuracy in large real-world data sets with varying levels of sophistication of cheating.
\autoref{tab:work_comp} listed the comparison between \sys and other detection methods.

\definecolor{lightred}{HTML}{cb2424}
\definecolor{lightgreen}{HTML}{1e8b7a}

\begin{table*}[htbp]
\centering
\caption{
Detection works comparison. `Availability': public availability of technical details (T), source code (C), and datasets (D), \textcolor{lightred}{red} and \textcolor{lightgreen}{green} indicate unavailable and available. `Dynamic Tuning': ability to adjust different anti-cheat demands (e.g., prioritize different metrics, introduced in \autoref{sec:MVIN}) without retraining the entire system. `Data Extraction-Game Coupling Degree': difficulty of acquiring the game's raw data, `high coupling' indicates the process is highly coupled with the game logic (e.g., modify engine, hooking process), making it challenging to generalize or reuse across different game versions. 
}
% \vspace{-1em}
\label{tab:work_comp}
\begin{threeparttable}
\resizebox{1\textwidth}{!}{%
\begin{tabular}{@{}cllcccccccc@{}}
\toprule
\multirow{2}{*}{\textbf{Side}}   & \multicolumn{1}{l}{\multirow{2}{*}{\textbf{Method}}} & \multicolumn{1}{l}{\multirow{2}{*}{\textbf{Work(s)}}}                                                                      & \multirow{2}{*}{\textbf{Availability}} & \multirow{2}{*}{\makecell{\textbf{Detection Feature}\\\textbf{Comprehensiveness}}} & \multirow{2}{*}{\makecell{\textbf{Data Extraction-Game}\\\textbf{Coupling Degree}}} & \multicolumn{5}{c}{\textbf{Real-world Functionality}}                                                                                                    \\ \cmidrule(l){7-11} 
                        & \multicolumn{1}{c}{}                        & \multicolumn{1}{c}{}                                                                                             &                              &                                            &                                             & \textbf{Dataset}    & \textbf{Deployment Constraint}     & \textbf{Robustness} & \textbf{Dynamic Tuning}             & \textbf{Performance}                                                  \\ \midrule
\multirow{2}{*}{\textbf{Client}} & ML-based                                    & \cite{choi2023botscreen}                                                                        & \textbf{\textcolor{lightgreen}{T}}, \textbf{\textcolor{lightgreen}{C}}, \textbf{\textcolor{lightgreen}{D}}   & Single aspect \& data type                                     & High Coupling                              & SSL\tnote{1}        & TEE (Intel SGX) & Unknown          & \ding{55} & Unknown\tnote{2}                                                            \\
                        & Commercial signature-based               & \cite{EasyAC, BattlEye, SteamSupport,RiotGamesInc,WELLBIA}                                                           & \textbf{\textcolor{lightred}{T}}, \textbf{\textcolor{lightred}{C}}, \textbf{\textcolor{lightred}{D}}   & Unknown                                          & High Coupling                       & Real-world & OS and kernel drivers & Unknown          & Unknown                          & Low recall, acceptable FPR                             \\ \midrule
\multirow{3}{*}{\textbf{Server}} & Statistical/Threshold-based                 & \cite{yeung2006detecting,yu2012statistical,liu2017detecting,orlova2024estimation,han2015online} & \textbf{\textcolor{lightgreen}{T}}, \textbf{\textcolor{lightred}{C}}, \textbf{\textcolor{lightred}{D}}   & Single aspect \& data type                                     & High Coupling                     & SSL        & Concurrent overheads & Unknown          & \ding{55} & Unknown                                                            \\
                        & \multirow{2}{*}{ML-based}                   & \cite{6633617}                                                                                  & \textbf{\textcolor{lightgreen}{T}}, \textbf{\textcolor{lightred}{C}}, \textbf{\textcolor{lightred}{D}}   & Single aspect \& data type                                     & High Coupling                     & SSL        & Concurrent overheads & Unknown          & \ding{55} & Unknown                                                            \\ \cmidrule(l){3-11} 
                        &                                             & \textbf{\sys}                                                                                       & \textbf{\textcolor{lightgreen}{T}}, \textbf{\textcolor{lightgreen}{C}}, \textbf{\textcolor{lightgreen}{D}}   & Multiple aspects \& data types                                    & Low Coupling                        & Real-world & None & CE\tnote{3}, N\tnote{4}       & \ding{51} & High recall, similar FPR \\ \bottomrule
\end{tabular}%
}
\newline
\raggedright \footnotesize
\textsuperscript{1}\textbf{SSL}: Simulated, small-scale, low complexity;
\textsuperscript{2}\textbf{Unknown}: Never been tested or disclosed;
\textsuperscript{3}\textbf{CE}: Robust to cheat evolution (an adversarial attack in \autoref{appx:robustness});
\textsuperscript{4}\textbf{N}: Robust to noise.
\end{threeparttable}
% \vspace{-2em}
\end{table*}

\adddiscuss{
The key innovation of \sys lies in its ability to mimic how human experts identify cheaters by focusing on three dimensions: analyzing the player's point of view, statistically evaluating the player's performance, and assessing the player's gaming sense and performance consistency. 
Building on these dimensions, we design structured and temporal features that precisely describe player behaviors. 
These features are selected from hundreds of candidates to ensure that they effectively represent diverse behavioral patterns.
%
% These features are selected from hundreds of candidates to effectively represent different behaviors.
%
\sys replicates the human identification process by leveraging machine learning techniques, including Long Short-Term Memory (LSTM) with attention mechanisms, ensemble learning, and deep neural networks, each tailored to specific aspects of the analysis.
\sys addresses the challenges of integrating heterogeneous data structures and dynamically adjusting anti-cheating priorities by developing a dedicated integration subsystem, \sysinteg.
}
%The key innovation of \sys lies in its ability to mimic how human experts identify cheaters by focusing on three main aspects, analyzing the player's point-of-view, statistically evaluating the player's performance, and assessing the player's gaming sense and performance consistency.
%
%Based on the three aspects, we design structured and temporal features that correspondingly represent player behaviors. 
%
%And reproduce the human identification process with the help of Long Short-Term Memory (LSTM) and attention mechanisms, ensemble learning, and deep learning networks respectively.
% \begin{figure}[t]
% \centering
% % \vspace{-0.5em}
% \includegraphics[width=0.3\textwidth]{Figures/comparison_metrics.pdf}
% % \vspace{-1em}
% \caption{Recall and FPR performance comparison among \sys-Recall priority mode, \sys-Balance mode, and our partner in-use anti-cheat\protect\footnotemark. \sys's different modes are introduced and evaluated in \autoref{sec:MVIN} and \autoref{appx:MVINOpt}.}
% % \Description{Recall and FPR performance comparison among \sys-Recall priority mode, \sys-Balance mode, and our partner in-use anti-cheat. \sys's different modes are introduced and evaluated in \autoref{sec:MVIN} and \autoref{appx:MVINOpt}.}
% \vspace{-1.5em}
% \label{fig:comparison_metrics}
% \end{figure}
% \footnotetext{Commercial anti-cheat (detailed in \autoref{sec:AC_flaws}), not open-source, leverages reverse engineering on the known cheats to detect cheating signatures.}

We evaluate \sys on a mainstream FPS game, CS:GO.
To the best of our knowledge, this paper is the first server-side work using a substantial volume of real-world datasets with different cheat types and levels of sophistication, making \sys and the datasets more convincing for industrial use.
Specifically, our dataset includes 2,979 \textit{aimbots} and 2,971 \textit{wallhacks}, totaling 56,041 players.
The scale of the dataset has increased by two orders of magnitude compared to the state-of-the-art work \cite{choi2023botscreen}.
\sys achieves at most 84\% recall and 80\% accuracy in detecting \textit{aimbots} and \textit{wallhacks}.
\sys outperforms baselines in the separate comparisons across different subsystems, given \sys's diverse input data structures.
%
% \adddiscuss{As evident in \autoref{fig:comparison_metrics}, \sys achieves higher recall and lower false positive rate (FPR) compared to the current industry in-use anti-cheat (blue bars).}
%
It successfully identifies cheaters who had evaded previous detections and those who were never caught by official systems.
We created a demo website\footnote{Demo website: \href{https://hawk-anticheat.github.io/}{https://hawk-anticheat.github.io/}.} to showcase some escapees' illegal actions.
\sys validates its robustness against cheat evolution (introduced in \autoref{appx:robustness}) in real-world scenarios and showcases acceptable overheads.
%
% \autoref{sec:discussion} contains more discussion about the applicable scope and the generalizability of this work.
We particularly discuss \sys's generalizability in \autoref{sec:discussion}. 
The contributions of our work are summarized below.
\begin{itemize}[topsep=6pt, itemsep=2pt, parsep=0pt, partopsep=0pt]
\item Innovative anti-cheat observations and feature designs.
\item Novel framework \sys and a supporting workflow.
\item Real-world \textit{data} evaluations with strong performance.
\item Open-source \sys and dataset\footnote{Public repository and dataset: \href{https://doi.org/10.6084/m9.figshare.25940818}{10.6084/m9.figshare.25940818}. Ethics considerations are shown in Supplemental Material A.} for future research.
\end{itemize}

\section{Background}
The section introduces anti-cheat, cheating techniques, the current state of cheating, and the replay design in FPS games. 

\subsection{Server-Side and Client-Side Anti-Cheat}\label{sec:anti-cheat}
Client-side anti-cheats focus on data, process, and hardware protection \cite{10.1145/3372297.3417890, invisibilitycloak}, or detection by examining the existing cheats' signatures \cite{EasyAC, BattlEye} or classifying through aiming behavior \cite{choi2023botscreen}.
Yet, attackers can bypass it with higher privileges, by sniffing the detection report through side-channel attack \cite{10.1145/3456631} and altering them in the transport layer, or by changing the program's signature \cite{9774028}.
Server-side anti-cheats \cite{yeung2006detecting,yu2012statistical,6633617,inproceedings, liu2017detecting,han2015online} emphasize data analysis and anomaly detection.
Although the data might be altered on the client, they must secure victory by outperforming other players.
Thus, detecting cheats on the server is easier because it records all clients' final-state actions influencing the gameplay.
But the prior works have the limitations mentioned in \autoref{sec:intro}.

\subsection{Cheating in FPS Games}\label{sec:CT}
%In FPS games, the most significant issue that plagues the player experience is the use of cheats, which allow certain players to obtain an unfair advantage. 
%
Wallhack and aimbot are the two most prevalent types of cheats in FPS games \cite{afkgaming}. 
Both types of cheats are mainly done by accessing the client's memory data or image information.
\textbf{Wallhack} enables players to visually penetrate obstructions like walls, which allows players to spot opponents across the map and effortlessly track their movements, providing a significant tactical advantage. 
\textbf{Aimbot} assists a player's aims and shots so that the cheaters gain transcendent precision and reaction speed. 
\textit{Aimbot}'s functionality can be customized according to player preference, allowing one to obtain different aiming performance, ranging from brutal-force aims to smooth aims like human \cite{choi2023botscreen}.
\textit{Aimbot} can be subdivided into \textit{pure aimbot} (the primitive \textit{aimbot}), \textit{triggerbot} (auto-shoot only when the cross-hair is on the opponent), \textit{micro-settings} (counteract recoil with Hotkeys), and \textit{computer-vision-based aimbot} (aim with object detection models). 
In our paper, \textit{aimbot} refers to all aforementioned subcategories.
Cheating provides different levels of sophistication \cite{10.1145/3380786.3391397}. 
The cheater might cheat cautiously, pretending to be a normal player, switching the cheat on and off, or cheating unabashedly with different configurations, various brands of cheating software, and categories.
However, cheaters' behaviors will always differ from the normal players \cite{choi2023botscreen}.
Thus, cheaters must outperform the average to gain unfair advantages.
This is bound to create anomalies at the data level and therefore should be identifiable by models.
Our evaluations in \autoref{sec:eval} include different levels of cheating sophistication.

% \subsection{Current State of Rampant Cheating}
% \label{sec:current_stats}
% \noindent\adddiscuss{
% %
% Cheating is rampant in FPS games, with analyses showing that over 21\% of high-ranked players on Counter-Strike's official competitive servers are cheaters \cite{reddit2024cs2cheat}.
% %
% Statistically, this indicates nearly every match includes at least two cheaters, undermining fair play.
% %
% Despite the presence of Valve Anti-Cheat (VAC) \cite{SteamSupport}, its effectiveness remains limited.
% %
% A study~\cite{orlova2024estimation} revealed that VAC identified only 19 out of 63 cheaters in a sample of 74 players, while also producing 5 false positives. 
% %
% This poor performance leads to significant financial losses and player attrition \cite{Andy2022cheat,Ryan2023cheat,9772248}.
% %
% Similar issues extend beyond CS:GO, affecting all online FPS games \cite{zleague_warzone_anticheat, pushtotalk_anticheat}.
% %
% Therefore, platforms (e.g., Faceit~\cite{faceit}, 5EPlay~\cite{5EPlay}) equipped with dedicated anti-cheats have emerged. 
% %
% Unlike intrusion detection systems for most areas that demand low FPR, the gaming industry now desires high recall to ensure the removal of most cheaters \cite{Gabriele2024}.
% %
% The cost of manually verifying false positives is negligible compared to the financial and reputational damage caused by undetected rampant cheating \cite{9772248}.
% }

\subsection{Replay System and Replay File}\label{sec:replay}
% Replay technique has been extensively adopted in notable FPS games such as Tom Clancy’s Rainbow Six Siege, PlayerUnknown's Battlegrounds, Overwatch, Battlefield, Call of Duty: Modern Warfare and Warzone, Fortnite, and CS:GO.
%
\textbf{Replay system} is a dynamic toolset for player interaction with recorded content post-game and has been extensively adopted in notable FPS games. 
Unlike video recordings, this system enables users to explore various perspectives, adjust camera angles, and control playback speed for analysis by reconstructing the match with the \textit{replay file}. 
\textbf{Replay file} refers to a comprehensive record of a match, encapsulated in a unique data format that can be interpreted by the \textit{replay system}. 
\textbf{Demo} denotes the \textit{replay file} in CS:GO.
\sys innovatively utilizes \textit{demo} to extract data and avoid concurrent overheads on the server.
We believe the \textit{replay file} can be used as a novel data source for FPS server-side anti-cheat.
\section{Overview}\label{sec:overview}

This section provides the key observations on the human experts' identification process and the workflow of \sys.

\subsection{Fact Observations} 
\label{sec: Fact OBS}
Although the anti-cheat tasks are still challenging for automation systems in FPS games, the illegal activities are identifiable for seasoned human players \cite{consalvo2007cheating}. 
Therefore, by harnessing such inspiration, we discovered three observations that human experts always focus on to identify cheaters.
% \vspace{1mm}
% \begin{mdframed}[shadow=true, shadowsize=2pt, roundcorner=9pt]
% \textbf{Observation 1 (OBS.1).} Review a player's point-of-view (POV).
% \end{mdframed}
% \vspace{1mm}
\begin{mdframed}[style=obs]
\obs Review a player's point-of-view (POV).
\end{mdframed}
Akin to watching a player's game replay unfold over time, this review process involves analyzing various in-game behaviors, such as aim and shot, positioning and movement, and the in-game economy management and props utilization. 
These behaviors can provide temporal insights into the legitimacy of a player's gameplay.
% \vspace{1mm}
% \begin{mdframed}[shadow=true, shadowsize=2pt, roundcorner=9pt]
% \textbf{Observation 2 (OBS.2).} Conduct an in-depth analysis of a player's statistical data.
% \end{mdframed}
% \vspace{1mm}
\begin{mdframed}[style=obs]
\obs Conduct an in-depth analysis of a player's statistical data.
\end{mdframed}
Cheaters often exhibit statistical anomalies that set them apart from normal players. 
These anomalies include abnormally high accuracy and eliminations, a low death rate, and a high number of kills made through obstacles, etc.
%
%Even the number of kills made through obstacles can be abnormally high. 
%
% Also, cheaters often disregard the use of in-game props, resulting in a lower usage rate compared to normal players. 
%
These statistical deviations serve as strong indicators of cheating.
% %
% The observations provide a comprehensive understanding of the strategies employed to identify potential cheaters in FPS games. 
% %
% They form the basis for the development of automated cheat detection systems that can mimic the keen eye and intuition of seasoned players. \CQ{Don't understand the point of the last two sentences.}
% \vspace{1mm}
% \begin{mdframed}[shadow=true, shadowsize=2pt, roundcorner=9pt]
% \textbf{Observation 3 (OBS.3).} Examine the consistency between a player's game sense and combat performance. 
% \end{mdframed}
% \vspace{1mm}
\begin{mdframed}[style=obs]
\obs Examine the consistency between a player's game sense and combat performance. 
\end{mdframed}
A player requires substantial game experience to have better performances, which results in a positive correlation between a player's game sense and their performance. 
Thus, the inconsistency between the two aforementioned factors indicates potential cheating.
For instance, if a player moves like a rookie but shoots and aims incredibly accurately, it is suspicious.
% \subsection{Threat Model} \label{sec:threat_model}
% \Jiayi{TBA HERE.}
\subsection{Workflow} \label{sec:workflow}
\begin{figure}[htbp]
\centering
% \vspace{-1em}
\includegraphics[width=0.47\textwidth]{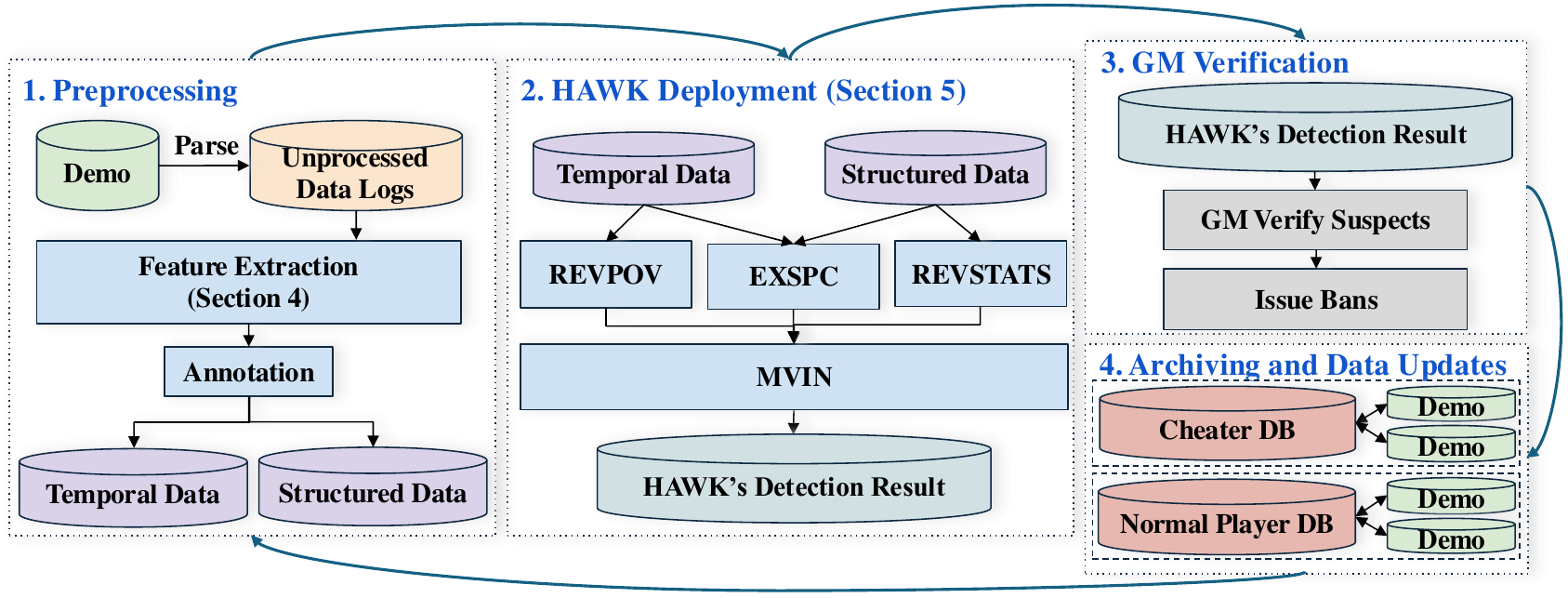}
% \vspace{-1em}
\caption{Cyclical workflow of \sys.}
%Stage 1, preprocessing and features extraction; Stage 2, \sys deployment; Stage 3, manual verification to screen out false positives; Stage 4, data update and reuse.
% \Description{Cyclical workflow of \sys.}
% \vspace{-1em}
\label{fig:workflow}
\end{figure}

\autoref{fig:workflow} illustrates \sys's interlinked workflow, which enhances the system's robustness and provides a dynamic solution to multi-type cheat detection (discussed in \autoref{appx:robustness}).

\noindent\circled{1} \textbf{Preprocessing}. 
%
% The \textit{demo} is parsed from the banned databases. 
%
The multi-view features are extracted from the parsed \textit{demo} and then annotated for training.
We discuss the feature constructions in \autoref{sec:features}.

\noindent\circled{2} \textbf{\sys deployment}. 
\sys utilizes the extracted features as inputs for its subsystems \syspov (\autoref{sec:RevPOV}), \sysstats (\autoref{sec:RevStats}), and \sysspc (\autoref{sec:ExSPC}) to mimic human identification process, respectively.
The outputs are incorporated into \sysinteg subsystem (\autoref{sec:MVIN}) that presents a final cheating report.
\sys's evaluation results in \autoref{sec:eval} are based on the results after this step.

\noindent\circled{3} \textbf{Game Master (GM) team verification}.
Given the potential personal assets tied to a player's account, a ban should not be issued solely based on an algorithmic verdict. 
It is imperative to provide substantial evidence before initiating such an action. 
Hence, the final verdict requires GMs' verification to avoid false bans.
This step is to display the entire process from data collection to the ban.

\noindent\circled{4} \textbf{Archiving and data updates}. 
Upon the affirmation of cheating, the relevant \textit{demos} are added to the banned databases and ready for retraining that may introduce novel patterns. 
%
% This act enriches and updates the reservoir of training samples which may contain novel patterns and behaviors.
%
% Thus, four stages constitute a cyclical workflow, with each stage contributing to the subsequent one.
%
\section{\sys's Feature Design}\label{sec:features}
%
% \Jiayi{add technical contributions this section TBA}
%
The feature design is one of our contributions.
This section introduces the construction of structured features, temporal features, and the classification of sense and performance features.
%
% These features constitute the backbone of the dataset as well as the inputs of \sys. 
%
These features are selected among hundreds of our designed naive features and intended to be comprehensive representations of the three aspects of human identification of cheating in \autoref{sec: Fact OBS} to overcome the challenge of the potentially biased classifications brought by the monospecific features in the prior works mentioned in \autoref{sec:intro}.
\subsection{Structured Features}
Structured features are single-value data calculated with well-designed algorithms to represent integrated in-game behavioral information per match and ensure time complexity.
%
% For example, the mean value and variance of reaction time and angle reflect the aiming performance and distribution, the first blood percentage represents the eliminations, etc.
%
This section describes the components of structured features and visualizes the structured feature differences between normal players and cheaters.
\subsubsection{Structured Features Construction}\label{sec:Statistical Features Construction} 
The structured features for \sys are categorized into four main categories that comprise 28 features. 
For the sake of brevity and coherence, only the introduction of categories is outlined below. 
A meticulous breakdown is shown in Supplemental Material C.
\begin{itemize}[topsep=6pt, itemsep=2pt, parsep=0pt, partopsep=0pt]
    \item \textbf{Aiming features} represents the efficiency and efficacy of aims.
    We break down the aiming process into three moments (initial spot, first-time fire, and first-time hit) and two stages (reaction and adjustment).
    We extract different elapsed durations and angle variations with statistical representations (e.g., average or variance) to indicate the player's different performances within a match.
    For example, the average of \textit{reaction duration} reflects a player's level of reaction speed, and the variance of that denotes the distribution of the reaction duration.
    Typically, in terms of the reaction stage, normal players obtain a higher mean value and more diverse variance compared with cheaters.
    \item \textbf{Firing features} reflects shot proficiency.
    For example, the \textit{hit group distribution variance} describes the dispersion of body part that suffers damage for the current player in a match.
    Most cheaters with \textit{aimbot} prefer to lock on specific \textit{hit group} (i.e., \textit{head}) to achieve faster elimination.
    \item \textbf{Elimination features} emphasizes metrics related to in-game kill events.
    For example, the \textit{occluder-penetration index} is a weighted index that reflects the extent of the average occluder-penetration eliminations (e.g., eliminate the opponent behind walls or smoke) conducted by the player.
    In terms of the cheaters exploiting \textit{wallhacks}, this index is more significant than the normal players'.
    \item \textbf{Props utilization features} captures a player's adeptness at using in-game auxiliary and offensive props like flashbang, grenades, etc.
    For instance, the \textit{props utilization index} denotes how frequently a player utilizes auxiliary and offensive props such as incendiaries and smoke grenades.
    This can help distinguish a player's familiarity with the game and their skillfulness.
    %
    % Seasoned players should use props frequently and skilfully, and be able to time their use correctly.
\end{itemize}
\begin{figure}[htbp]
\centering
% \vspace{-1em}
\includegraphics[width=0.47\textwidth]{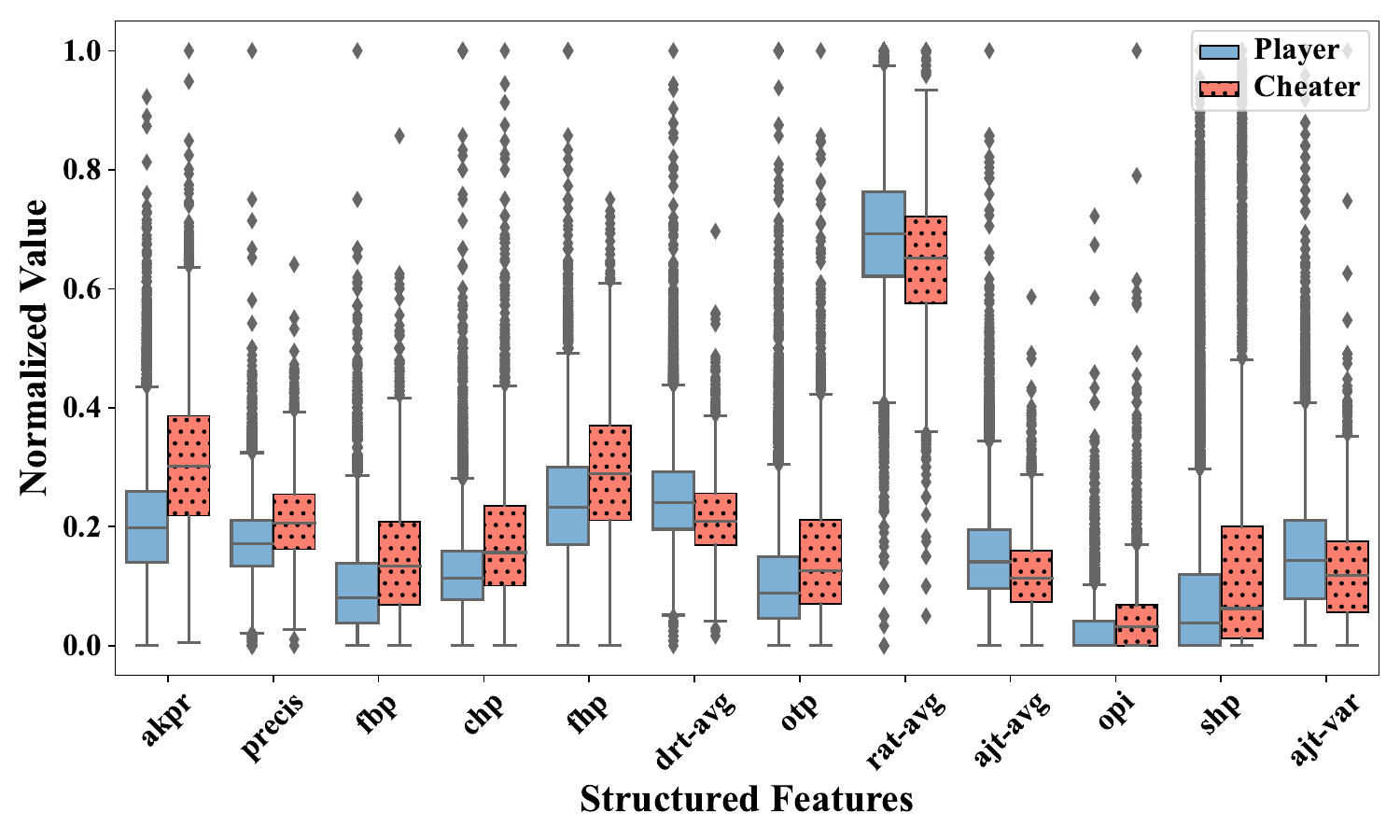}
\caption{Box plots of comparative visualization between honest players and cheaters on top-12 structured features under the Mann-Whitney U test with distinction descending order.}
% \Description{Box plots of comparative visualization between honest players and cheaters on top-12 structured features under Mann-Whitney U (Wilcoxon Rank-Sum) test with P-Value ascending (distinction descending) order.}
% \vspace{-0.5em}
\label{fig:structured_boxplot}
\end{figure}
\subsubsection{Structured Data Comparative Visualization}
\autoref{fig:structured_boxplot} offers a box-plot on structured features to analyze between honest players and cheaters comparatively. 
Each feature on the x-axis (we select top-12 features for illustration purposes) is ordered based on the outcomes of the Mann-Whitney U test, with the leftmost features, such as \textit{akpr} (introduced in Supplemental Material C), marking pronounced disparities between the two groups. 
This implies that features towards the left end, showcased by their lower p-values, are more distinguishable, whereas those on the right are less so. 
The y-axis is normalized values between 0 and 1. 
The box plots, distinguished by light blue for honest players and hatched red for cheaters, provide insights into the median, interquartile range, and outlier distributions. 
In which, diamond shapes represent outliers, emphasizing the spread and distribution of data across the features. 
Despite considering dimensionality reduction techniques, such as PCA (Supplemental Material Figure 4), empirical tests show an optimal performance when all 28 features were employed. 
%
% Therefore, our final analyses rely on this comprehensive feature set, emphasizing its integral role in distinguishing player behaviors.
%
Therefore, cheaters often exhibit deviations in their static data compared to normal players. 
Even when cheaters deliberately attempt to conceal their behavior, the subtle differences tend to accumulate over time and reveal disparities statistically. 

\subsection{Temporal Features}
Temporal features are derived in the time domain.
Each temporal data tuple is the state information (e.g., coordinates, view directions, events, etc.) associated with the present \textit{tick}.
Where \textbf{tick} denotes the smallest unit of time in the game.
%
% This section describes temporal features' components and illustrates the distinctions in temporal features between normal players and cheaters.
\subsubsection{Temporal Features Construction} \label{sec:temporalfeatures}
% The temporal features were extracted directly from \textit{demo}. 
%
According to different in-game behaviors, temporal features are segmented into three main categories and subdivided into seven specific types. 
%
% This overview provides only categories and types due to the less-expressive nature of temporal features. 
%
Supplemental Material D lists elaborate descriptions.
\begin{itemize}[topsep=6pt, itemsep=2pt, parsep=0pt, partopsep=0pt]
    \item \textbf{Engagement features} describes in-game engagement events, which are subcategorized into five types as follows. 
    \begin{itemize}[topsep=6pt, itemsep=2pt, parsep=0pt, partopsep=0pt]
        \item \text{\textbf{Damage features}} indicate damage-inflicting events, including the occurring \textit{tick}, the attacker's and victim's locations, view directions, weapons, damage value, etc.
        \item \text{\textbf{Auxiliary props features}} related to auxiliary props-utilization events (e.g., flashbang), including the deploy \textit{tick}, victim's affected duration, etc.
        \item \text{\textbf{Offensive props features}} related with events involving props that aid the attack (e.g., grenade, incendiary, etc.) including the deploy and destroy \textit{ticks}, the props' type and coordinates, etc.
        \item \text{\textbf{Elimination features}} describes kill-related events, including the occurring \textit{tick}, attacker's weapon and location, victim's location and view direction, etc.
        \item \text{\textbf{Weapon fire features}} denotes firing-related occurrences, including the occurring \textit{tick}, the weapon information, the shooter's location and view direction, etc.
    \end{itemize}
    \item \textbf{Movement features} represents in-game positioning, mobility, and boolean flags per \textit{tick}. 
    For instance, coordinates, view directions, velocities in three dimensions, flags that indicate if the player is ducking, blinded, reloading, etc. 
    \item \textbf{Economy features} reflects in-game financial aspects. 
    For example, they include the equipment value and balance per \textit{round} for each player. 
\end{itemize}
\subsubsection{Temporal Data Comparative Visualization}
\begin{figure}[htbp]
% \vspace{-0.5em}
\centering
\includegraphics[width=0.47\textwidth]{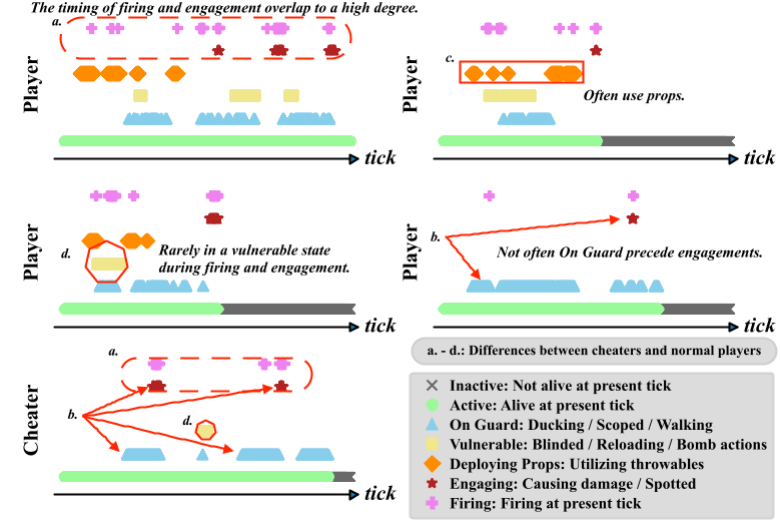}
% \vspace{-0.5em}
\caption{Comparative behavioral visualization between honest players and cheaters with respect to time scales.}
% \Description{Comparative behavioral visualization between honest players and cheaters with respect to time scales.}
% \vspace{-1.5em}
\label{fig:temporal_vis}
\end{figure}

\autoref{fig:temporal_vis} comparatively analyzes the temporal features between cheaters and normal players across one \textit{round}.
Where \textbf{round} denotes the unit determining each win-loss outcome. One match has multiple rounds, and the players who win more rounds are the winners.
In this example, there are four differences between cheaters and normal players. 
First, since the cheaters illegally acquire their opponents' position, the timing of cheaters' firing (pink) and engaging (red) highly overlap (a). 
Second, normal players stay on guard (blue) irregularly to check for potential enemy locations, whereas cheaters are often on guard just before engagement (b). 
Third, normal players deploy props (orange) to gain a gunfighting or positional advantage, whereas cheaters rarely do so because they already have illegal advantages (c). 
Fourth, cheaters are rarely vulnerable (yellow) during the fire or engagement because they know the enemy's location and therefore can circumvent it in advance (d).

\subsection{Sense and Performance Features}
\label{sec:Sense and Performance Features Classification}
Sense and performance features are derived from the previous subsections. 
Supplemental Material E includes detailed classification.
\begin{itemize}[topsep=6pt, itemsep=2pt, parsep=0pt, partopsep=0pt]
    \item \textbf{Sense features} denote a player's in-game cognition, reflecting strategic understanding, and tactical anticipation. 
    \begin{itemize}[topsep=6pt, itemsep=2pt, parsep=0pt, partopsep=0pt]
        \item \textbf{Temporal sense features} include temporal features that reflect a player's gaming sense such as economic considerations, movement dynamics, grenade usage, etc.
        \item \textbf{Structured sense features} contain structured features that describe in-game understanding, e.g., flash efficiency, prop utilization, etc.
    \end{itemize}
    \item \textbf{Performance features} demonstrate in-game efficacy, indicating engagement capabilities, precision in aiming, and mastery over game mechanisms. 
    \begin{itemize}[topsep=6pt, itemsep=2pt, parsep=0pt, partopsep=0pt]
        \item \textbf{Temporal performance features} include temporal features that indicate continuous in-game performances, e.g., eliminations, damage, weapon proficiency, etc.
        \item \textbf{Structured performance features} contain the remaining structured features that statistically describe the player's in-game performances, e.g., elimination-related features.
    \end{itemize}
\end{itemize}
\begin{figure*}[htbp]
\centering
\includegraphics[width=1.0\textwidth]{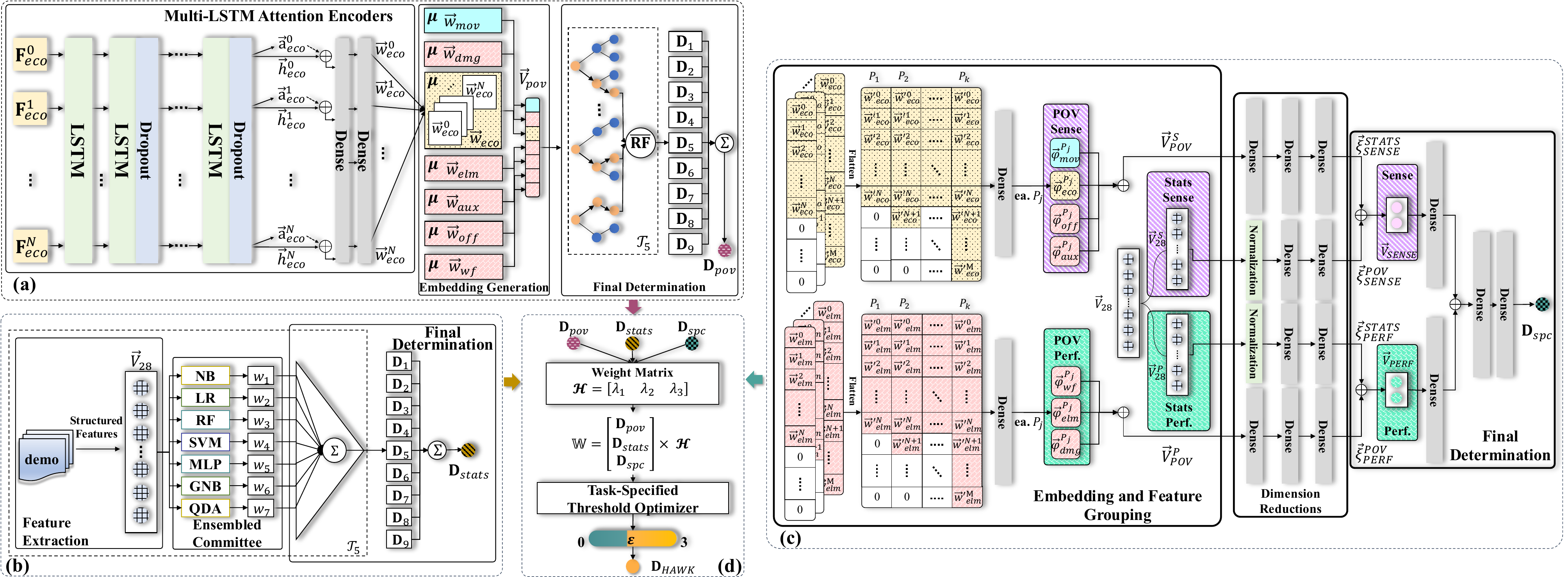}
\caption{\sys framework illustration. (a)-\syspov, (b)-\sysstats, (c)-\sysspc are corresponding to three observations in \autoref{sec: Fact OBS}, (d)-\sysinteg is an integration network for combining the determinations from the aforementioned three subsystems.}
% \Description{\sys framework illustration. (a)-\syspov, (b)-\sysstats, (c)-\sysspc are corresponding to three observations in \autoref{sec: Fact OBS}, (d)-\sysinteg is an integration network for combining the determinations from the aforementioned three subsystems.}
% \vspace{-1.5em}
\label{fig:framework}
\end{figure*} 
\section{\sys}
%
%
% \Jiayi{add technical contributions this section TBA}
%
To mimic human experts' identification process, we propose \sys which consists of four subsystems: Review Point-of-View (\syspov), Review Statistics (\sysstats), Examine Sense-Performance Consistency (\sysspc), and Multi-View Integration (\sysinteg). 
The first three subsystems are associated with the three observations we discovered regarding the inherent nature of the cheat identification process in \autoref{sec: Fact OBS}. 
\sysinteg integrates the results derived from the above three subsystems, just as humans integrate the three aspects collectively.
%
%The overarching framework, encompassing all these subsystems, is illustrated in \autoref{fig:framework}.

\subsection{\syspov} \label{sec:RevPOV}
Aligned with \textbf{OBS.1}, \syspov thoroughly examines a player's time-series operations within a match. 

\noindent\textbf{Multi-LSTM Attention Encoders:} In \autoref{fig:framework}\textcolor{Maroon}{a}, we feed the seven sets of temporal features (\autoref{sec:temporalfeatures}) as input to seven independent Long Short-Term Memory (LSTM) \cite{LSTM1997Hochreiter} networks to acquire the embeddings. 
%
% In \autoref{fig:framework} (a), the various sets of features are transformed into vectors and employed as input to acquire the embeddings. 
%
For explanatory purposes, we illustrate the process using the example of the \textit{Economy} features. 
$\mathbf{F_{eco}^{n}}$ denotes the input feature vector.
Dropout layers \cite{10.5555/2627435.2670313} are employed 
% with different parameters within the domain from 0 to 1, indicating the dropping ratio between each layer of neurons. This dropout layer is implemented 
to mitigate the risk of overfitting. 
% \CQ{There are multiple Dropout layers.}
%
$\vec{h}_{G}^{n}$ denotes the output of the encoder, consists of one set of the seven temporal features mentioned in \autoref{sec:temporalfeatures}, where $G$ represents the set and $n$ indicates the sequence number among all outputs. 
%
% Distinct categories are equipped with different numbers of LSTM layers to make the dimension reduction smoother.
%
% Different numbers of LSTM layers are used in different feature categories to make the dimension reduction smoother.
%
\adddiscuss{
For each time step, we introduce the Luong-style attention \cite{luong2015effective} afterward to help the model better understand the importance of different positions in the input sequence and aggregate different parts of the input weights, overcoming the challenge of identifying varying patterns in temporal data for a large number of samples.
}
In \autoref{eq:attention}, $a_G^i$ represents the attention weight for the time step $i$. 
$\mathbf{W}_q$ and $\mathbf{W}_k$ denotes the learned weight matrices of \textit{query} and \textit{key} in attention mechanism. 
$i$, $j$ and $k$ indicate different time steps. 
Concatenation operation is executed afterward, yielding $[\vec{a}_{G}^{n}||\vec{h}_{G}^{n}]$, where $||$ denotes concatenation. 
%
% Then, it is sequentially computed by two dense layers with \textit{ReLU} and \textit{Sigmoid} activation functions respectively for further dimension reduction. 
%
After two dense layers, $\vec{w}_{G}^{n}$ represents the final tensor for binary classification.
% \vspace{-0.4em}
{\normalsize
\begin{equation}\label{eq:attention}
\vec{a}_G^i = \sum_{j=0}^{N} \frac{\exp{(\mathbf{W}_q \cdot \vec{h}_G^i) \cdot (\mathbf{W}_k \cdot \vec{h}_G^j)}}{\sum_{k=0}^{N} \exp{(\mathbf{W}_q \cdot \vec{h}_G^i) \cdot (\mathbf{W}_k \cdot \vec{h}_G^k)}} \cdot \vec{h}_G^j
\end{equation}
}
% \vspace{-0.4em}

\noindent\textbf{Embedding Generation:} 
% Since LSTM outputs sequentially, 
An average pooling in \autoref{eq:outputSeq} is conducted to flatten the output sequence, where $N$ denotes the number of output sequences. 
$\vec{V}_{pov}$ is the final determination embedding.
% \vspace{-0.5em}
{\normalsize
\begin{equation}\label{eq:outputSeq}\vec{w}_{G}=\frac{
\sum_{n=0}^{N}\vec{w}_{G}^{n}}
{N+1}\end{equation}
}
% \vspace{-0.5em}
{\normalsize\begin{equation}\label{eq:avgOutput}
\vec{V}_{pov}=[\vec{w}_{eco}||\vec{w}_{mov}||\vec{w}_{dmg}||\vec{w}_{elm}||\vec{w}_{aux}||\vec{w}_{off}||\vec{w}_{wf}]
\end{equation}
}
% \vspace{-1.5em}

\noindent\textbf{Final Determination:} 
We find that only depending on the fully connected layers with different activation functions can not effectively classify.
Therefore, we conduct the following strategies to further improve \syspov.
%The process of optimizing the \syspov subsystem primarily hinges on maintaining an equitable distribution of classes in the training set. 
%
In the context of cheat detection, the dataset often exhibits a class imbalance, where the majority of players are honest, inadvertently leading to biased learning towards the dominant class. 
Therefore, we create a \textit{multi-subsampling} strategy to level the playing field for the minority class, creating balanced representations of both cheating and honest behaviors without dropping any samples.
%
% Suppose $\mathcal{T}$ is the initial training set.
% where the subset of honest players significantly outweighs the cheaters. 
%
We perform a selective sampling to divide the honest players in the initial training set $\mathcal{T}$ into nine sets in accordance with the ratio of normal to cheating players, generating nine training sets containing the same set of cheaters and distinct normal players, each denoted as $\mathcal{T}_k$, where $k$ indicates the iteration count. 
%
% Since the ratio of honest to dishonest players is approximately nine to one in our dataset, 
% We perform this operation nine times in accordance with the ratio of honest to dishonest players, generating nine distinct training sets, each denoted as $\mathcal{T}_k$, where $k$ indicates the iteration count. 
%
% A pivotal aspect of $\mathcal{T}_k$ is ensuring diversity across instances. 
%
% Conduct each iteration of the sampling process to select unique honest player samples, thereby adding varied examples to the learning process. 
%
% This enables the model to gain extensive insights into the different representations of honest player behaviors, facilitating more comprehensive learning.
%
Subsequently, after adding varied examples to the learning process, we obtain nine independent Random Forest models, each corresponding to one of the diversified and balanced training sets $\mathcal{T}_k$, we named this strategy as \textit{multi-forest}.
%
% This strategy, namely \textit{multi-forest} strategy, not only provides the model with a broad learning perspective but also enhances the robustness of our cheat detection system. 
%
The averaging method in \autoref{eq:RF} is selected for \syspov final class prediction, where random forest consisting of $N$ decision trees, each producing a prediction $h_i(\vec{V}_{pov}^{k})$, where $i$ indicates the $i$-th tree, $k$ denotes the $k$-th sampled set, and $\vec{V}_{pov}^{k}$ represents the input feature vector. 
%
% The prediction of each tree is a probability distribution, indicating the likelihood of the sample $\vec{V}_{pov}^{k}$ belonging to each class. 
%
$p_{i}$ represents the probability that the $i$-th tree predicts sample $\vec{V}_{pov}^{k}$ to be in class $j$. 
$\mathbf{\overline{RF}}(\vec{V}_{pov}^{k})$ denotes the average prediction output of the random forest, which is the average of the prediction probability vectors of all decision trees. 
%
% \autoref{eq:RF} is the average of the prediction probability vectors of all decision trees. 
%
The prediction for each $\vec{D}_{k}$ is obtained by selecting the class with the highest predicted probability shown in \autoref{eq:RF_prob}, where $(\cdot)_j$ represents the probability of class $j$ in $\mathbf{\overline{RF}}(\vec{V}_{pov}^{k})$. 
After the \textit{multi-forest}, we acquire the corresponding determination $\mathbf{D}_{k}$.
% , and the final prediction can be determined through majority voting. 
% Within the majority voting scheme, the ultimate prediction is determined by selecting the class that has received the highest frequency of predictions across all models. 
Lastly, as in \autoref{eq:majority_vote}, we obtain the final classification determination $\mathbf{D}_{pov}$ for \syspov.
% \vspace{-0.5em}
{\normalsize
\begin{equation}\label{eq:RF}
\mathbf{\overline{RF}}(\vec{V}_{pov}^{k}) = \frac{1}{N} \sum_{i=1}^{N} h_i(\vec{V}_{pov}^{k}) = \frac{1}{N} \sum_{i=1}^{N} \mathbf{p}_i
\end{equation}
}
% \vspace{-0.8em}
{\normalsize
\begin{equation}\label{eq:RF_prob}
\mathbf{D}_{k} = \arg \max_{j} \left(\mathbf{\overline{RF}}(\vec{V}_{pov}^{k})\right)_j
\end{equation}
}
% \vspace{-0.9em}
{\normalsize
\begin{equation}\label{eq:majority_vote}
\mathbf{D}_{pov} = \arg \max_{j} \sum_{k=1}^{9} \mathbf{D}_k^{(j)}
\end{equation} 
}
% \vspace{-1.5em}
\subsection{\sysstats}\label{sec:RevStats}
Aligned with \textbf{OBS.2}, \sysstats subsystem is designed to perform an in-depth analysis of structured features to capture the cheaters. 
%
% To capture these anomalies, we deploy machine-learning techniques that facilitate the effective segregation of cheaters from normal players based on their gameplay statistics.

\noindent\textbf{Feature Extraction:} 
%\sysstats aims to facilitate machine learning in a manner similar to human discernment. 
% Thus, we have defined features that are typically employed by human players in cheat detection. An exhaustive description of these features is provided in \autoref{appx:structuredfeatures}. 
%
In \autoref{fig:framework}\textcolor{Maroon}{b}, $\vec{V}_{28}$ denotes the vector representation of the structured features (\autoref{sec:Statistical Features Construction}).

\noindent\textbf{Ensembled Committee:} 
% \sysstats uses $\vec{V}_{28}$ as input.
% After extracting features, we input them into an ensemble classifier system. 
%
% Various machine learning algorithms can be employed to distinguish between normal and cheating behavior concerning the extracted features. 
%
% In this respect, we exhaustively experimented with classification models, and have ultimately chosen seven: 
%
During preliminary testing, our selected features showed promising results across most classifiers, demonstrating their effectiveness.
However, despite similar performance in terms of metrics alone, the testing revealed that different classifiers produced varying classifications on some individual samples. 
This suggests that relying on a single classifier, as done in prior works, can lead to unreliable outcomes.
However, the majority of classifiers correctly classified the samples, leading to the development of the Ensembled Committee.
Among hundreds of classifiers, we identify seven superior classifiers \cite{haykin1994neural,cramer2002origins,breiman2001random,cortes1995support,murphy2006naive,john2013estimating,srivastava2007bayesian} after the preliminary testing and use them in an ensembled way.
%
% Multilayer Perceptron (MLP) \cite{haykin1994neural}, 
% % Decision Tree \cite{myles2004introduction}, 
% % Gradient Boosting Decision Tree (GBDT) \cite{friedman2001greedy}, 
% % AdaBoost \cite{freund1997decision}, 
% Logistic Regression \cite{cramer2002origins}, Random Forest \cite{breiman2001random}, Support Vector Machines (SVM) \cite{cortes1995support}, Naive Bayes \cite{murphy2006naive}, 
% % k-Nearest Neighbors (KNN) \cite{altman1992introduction}, 
% Gaussian Naive Bayes \cite{john2013estimating}, and Quadratic Discriminant Analysis (QDA) \cite{srivastava2007bayesian}. 
%
% Each model is independently trained and provides an individual prediction. 
%
% Upon completion, each model independently adjudicates each test sample. 
%
% The outputs are binary signifying the presence or absence of cheating.

\noindent\textbf{Final Determination:} Due to imbalanced dataset,
% where the ratio of honest to cheating players is approximately 9:1, 
a similar \textit{multi-subsampling} strategy as employed in the \syspov subsystem is adopted for the \sysstats subsystem to fully leverage the available data without losing valuable information. 
For each classification model, we generate a distinct model for each subsampled set. 
%
% Each model within a particular classification algorithm is independently trained on one of these subsampled training sets. 
%
Each model is independently trained and provides an individual prediction. 
A majority voting scheme is utilized within each classification algorithm to aggregate the results of the seven models. 
The final determination is also achieved through majority voting, which collates the determinations from all nine subsampled sets. 
In \autoref{eq:MajorityVoting}, where $\mathbf{D}_{stats}$ signifies the binary final determination. 
Where $w_j$ denotes the decision from the $j$th classification algorithm, and $I(\cdot)$ is the indicator function, returning 1 if the condition within the brackets is true, and 0 otherwise. 
The decision, $i$, which receives the majority of votes from all seven classification algorithms, is accepted as the final determination.
% \vspace{-0.5em}
{\normalsize
\begin{equation}\label{eq:MajorityVoting}
\mathbf{D}_{stats}=\arg\max_{i\in \{0,1\}} \sum_{j=1}^{7} I(w_j = i)
\end{equation}
}
% \vspace{-1.5em}
\subsection{\sysspc}\label{sec:ExSPC}
Aligned with \textbf{OBS.3}, \sysspc is designed from scratch to inspect the consistency of a player's in-game senses and overall performances. 
%
% \sysspc utilizes both the aforementioned temporal and structured features.

\noindent\textbf{Embedding and Feature Grouping:} 
% %
% There are seven embeddings and two groups of vectors as input (\autoref{sec:Sense and Performance Features Classification}) in \sysspc, in which seven embeddings are generated by using the trained models in \syspov and two vectors are the subsets of $\vec{V}_{28}$ in \sysstats. 
% %
% Each embedding is padded to the maximum length of that in the whole dataset. 
% %
% $\vec{w}^{k}_{G}$ is the intermediary result from \syspov, where $k$ represents the numbering in the sequence, ${G}$ denotes the type of the temporal features. 
%
\sysspc takes seven embeddings and two groups of vectors as input (see \autoref{sec:Sense and Performance Features Classification}).
The seven embeddings are generated using trained models from \syspov, while the two vectors are subsets of $\vec{V}_{28}$ from \sysstats. 
Each embedding is padded to the maximum length observed in the dataset. 
The intermediary result from \syspov, denoted as $\vec{w}^{k}_{G}$, represents temporal features, where $k$ is the sequence number and ${G}$ denotes the type of the temporal features.
All embeddings are structured as two-dimensional arrays.
As illustrated in \autoref{fig:framework}\textcolor{Maroon}{c}, where $P_{1}$ to $P_{k}$ represent different players, we exemplify with one player $P_{j}$.
The embeddings are first flattened and dimensionally reduced.
After processing through seven groups of similar layers, we obtain $\varphi_{G}^{{P}_{j}}$.
These tensors are then divided into seven types and concatenated for each player.
In which, $\vec{V}_{POV}^{S}$ denotes the temporal sense features vector and $\vec{V}_{POV}^{P}$ denotes the temporal performance features vector. 
%
% In \autoref{fig:framework}c, $P_{1}$ to $P_{k}$ indicates different players, we exemplify with one player $P_{j}$. 
% %
% % We flatten the embeddings into one-dimensional arrays and use a dense layer with an \textit{ELU} activation function to shrink the dimensions. 
% %
% We first flatten the embeddings and shrink the dimensions. 
% %
% After the seven different similar groups of layers, we obtain $\varphi_{G}^{{P}_{j}}$. 
% %
% Subsequently, we divide seven different types of tensors as shown in \autoref{appx:spcfeatures}, and concatenate each group of tensors for each player. 
% %
% In which, $\vec{V}_{POV}^{S}$ denotes the temporal sense features vector and $\vec{V}_{POV}^{P}$ denotes the temporal performance features vector. 
%
% = [\varphi_{mov}^{{P}_{j}}||\varphi_{eco}^{{P}_{j}}||\varphi_{off}^{{P}_{j}}||\varphi_{aux}^{{P}_{j}}]
% = [\varphi_{wf}^{{P}_{j}}||\varphi_{elm}^{{P}_{j}}||\varphi_{dmg}^{{P}_{j}}]
% In terms of structured features vector $\vec{V}_{28}$, we also divided into two groups according to \autoref{appx:spcfeatures}, namely the sense vector of structured features $\vec{V}_{28}^{S}$ and the performance vector of structured features $\vec{V}_{28}^{P}$.
%
Similarly, the structured features vector $\vec{V}_{28}$ is divided into two groups: the sense vector $\vec{V}_{28}^{S}$ and the performance vector $\vec{V}_{28}^{P}$, as detailed in Supplemental Material E.

\noindent\textbf{Dimension Reduction:} $\vec{V}_{28}^{S}$ and $\vec{V}_{28}^{P}$ are normalized. 
Afterward, each group of the four vectors goes through different groups of dense layers to reduce the dimensions. 
Next, concatenate four vectors and yield two vectors eventually representing sense $\vec{V}_{SENSE}$ and performance $\vec{V}_{PREF}$ respectively.
% = [\vec{\xi}_{SENSE}^{STATS}||\vec{\xi}_{SENSE}^{POV}]
% = [\vec{\xi}_{PREF}^{STATS}||\vec{\xi}_{PREF}^{POV}]

\noindent\textbf{Final Determination:} When acquiring $\vec{V}_{SENSE}$ and $\vec{V}_{PERF}$, we further reduce their dimensions and deepen the network for better learning performances. 
Then, we concatenate two vectors and go through one final dense layer for reducing the dimension and another dense layer with \text{sigmoid} activation function for binary classification. 
Eventually, the final classification determination $\mathbf{D}_{spc}$ for \sysspc subsystem is obtained. 
The class weight is set as one to nine like previous subsystems.
Both \syspov and \sysspc use \textit{Binary Cross-Entropy} as loss function shown in \autoref{eq:BCE}, where \( N \) represents the number of samples, \( y_i \) is the true label of the \(i\)-th sample, \( \hat{y}_i \) is the predicted value for the \(i\)-th sample.
% \vspace{-0.5em}
{\normalsize
\begin{equation}\label{eq:BCE}
\text{Loss} = -\frac{1}{N} \sum_{i=1}^{N} \left[ y_i \log(\hat{y}_i) + (1 - y_i) \log(1 - \hat{y}_i) \right]
\end{equation}
}
% \vspace{-1.5em}
\subsection{\sysinteg}\label{sec:MVIN}
% By embracing the strengths of each preceding subsystem and mitigating their individual limitations, \sysinteg emerges as an integral and sophisticated cheat detection tool. Future endeavors might involve refining the deep learning model or exploring other integration techniques to further bolster the system's accuracy and reliability.
In \autoref{fig:framework}\textcolor{Maroon}{d}, \sysinteg is to dynamically integrate determinations from \syspov, \sysstats, and \sysspc.
%
% By integrating the predictions from these subsystems, \sysinteg aims to deliver a final verdict regarding a player's cheating disposition. 
%
% By leveraging deep learning techniques, it accentuates the synergy among the subsystems to elevate cheat detection performance.

\noindent\textbf{Data Integration:} 
The subsystem integrates the outcomes from \syspov, \sysstats, and \sysspc, symbolized as $\mathbf{D}_{pov}$, $\mathbf{D}_{stats}$, and $\mathbf{D}_{spc}$. Given the diverse nature of these outputs, a structured preprocessing phase becomes indispensable. This step is designed to merge the individual outputs and synchronize them for the succeeding integration process.
% To adapt to the dynamic requirements of the user and the game environment, the system introduces a "Task-Specified Threshold Optimizer." Users can specify their priority, be it recall-oriented or accuracy-driven. For instance, a recall-focused approach would strive to identify every potential cheater, even at the risk of some false positives. On the other hand, an accuracy-centered approach aims to minimize false positives, albeit with the possibility of some cheaters being overlooked. The threshold optimizer adapts the decision boundary accordingly, ensuring the desired outcome.

\noindent\textbf{Model Optimization:} 
\sysinteg adopts a weight matrix for each subsystem's determination, where weights are dynamically assigned based on the subsystem's importance and reliability.
The weighted output, \( \mathbb{W} \), is calculated as \autoref{eq:D_HAWK_Computation}. 
Where \( \lambda_1 \), \( \lambda_2 \), and \( \lambda_3 \) represent the weights associated with the outputs of the subsystems \syspov, \sysstats, and \sysspc, respectively. 
% \vspace{-0.5em}
{\normalsize
\begin{equation}\label{eq:D_HAWK_Computation}
\mathbb{W} = \lambda_1 \cdot \mathbf{D}_{pov} + \lambda_2 \cdot \mathbf{D}_{stats} + \lambda_3 \cdot \mathbf{D}_{spc}
\end{equation}
}
% \vspace{-1em}

\noindent\textbf{Final Determination:} \( \mathbb{W} \) undergoes a learnable thresholding via the \textit{Task-Specified Threshold Optimizer} (TSTO), which provides a binary determination based on a dynamic threshold \( \varepsilon \). 
The threshold optimizer can be dynamically set based on anti-cheat requirements.
For instance, in \textit{Recall} mode, the optimizer would aim to identify every potential cheater, accepting the risk of higher false positives. 
%
% Conversely, in an accuracy-driven mode, the optimizer would aim to reduce false positives, even if it means potentially missing some cheaters. 
%
This threshold \( \varepsilon \) therefore acts as a decision boundary, converting the continuous weighted value \( \mathbb{W} \) into a discrete binary outcome \( \mathbf{D}_{HAWK} \).
TSTO offers \sys a degree of selectivity, enabling flexible tuning and offering promising performance without requiring retraining any prior subsystems, overcoming the challenge of lack of system adaptability.
By harnessing the strengths of each subsystem and compensating for their limitations, \sysinteg is crucial to integrate determinations from different subsystems. 
%
% One of its distinctive features is the ability to adapt to user preferences by selecting the appropriate \textit{Task-Specified Threshold Optimizer}. 

\section{Experiments}\label{sec:eval}
% This section includes the evaluation criteria and presents the resultant findings.  
%
%As stated in \autoref{sec:incomparability}, there are no precedents in terms of evaluation nor extant state-of-the-art designs for direct comparison with the current datasets. 
%
%We explained the reasons for the incomparability to prior works, especially the reason for not conducting the comparison analysis with the client-side SOTA designs.
%
%Nevertheless, we undertook a comprehensive assessment of all the subsystems encompassed in this study and executed an ablation analysis to verify the efficacy of each subsystem. 
%
%We also found some interesting results that not merely indicate the inefficacy of the current industrial solutions and \sys's superior performance but also demonstrate the bonus suspects and illegal actions among the false positives.
%
% Besides, we explained the reasons for the incomparability to prior works, especially the reason for not conducting the comparison analysis with the client-side SOTA designs.
%
We conduct experiments to answer seven research questions (\textbf{RQ.1}$-$\textbf{RQ.7}):
% \Jiayi{TODO: modify this.}
\begin{itemize}[topsep=6pt, itemsep=2pt, parsep=0pt, partopsep=0pt]
    \item \textbf{RQ.1:} What are the performances of \sys and each subsystem? (\autoref{sec:ablationstudy})
    % \item \textbf{RQ.2:} What is the effectiveness of each subsystem? 
    \item \textbf{RQ.2:} To what extent has \sys improved effectiveness and efficiency compared to current official inspections? (\autoref{sec:officalcomparison})
    \item \textbf{RQ.3:} What are the performances of \sys under different settings of the \textit{Task-Specified Threshold Optimizer}? (\autoref{appx:MVINOpt})
    % \item \textbf{RQ.5:} What are the performance of \sys compared with the SOTA design \textbf{BotScreen} \cite{choi2023botscreen}?
    \item \textbf{RQ.4:} What is the robustness of \sys and its subsystems against cheat evolution? (\autoref{appx:robustness})
    \item \textbf{RQ.5:} Is there any suspicious activity that still exists but has never been detected by officials? (\autoref{sec:case_study_FPs})
    % \item \textbf{RQ.6:} What are the overheads of \sys's subsystems? (\autoref{sec:overheads})
    \item \textbf{RQ.6:} What is the overhead of \sys? (\autoref{sec:overheads})
    \item \textbf{RQ.7:} Is \sys performing better than prior works? (\autoref{sec:prior_comp})
\end{itemize}
\begin{figure}[htbp]
\centering
% \vspace{-1em}
\includegraphics[width=0.48\textwidth]{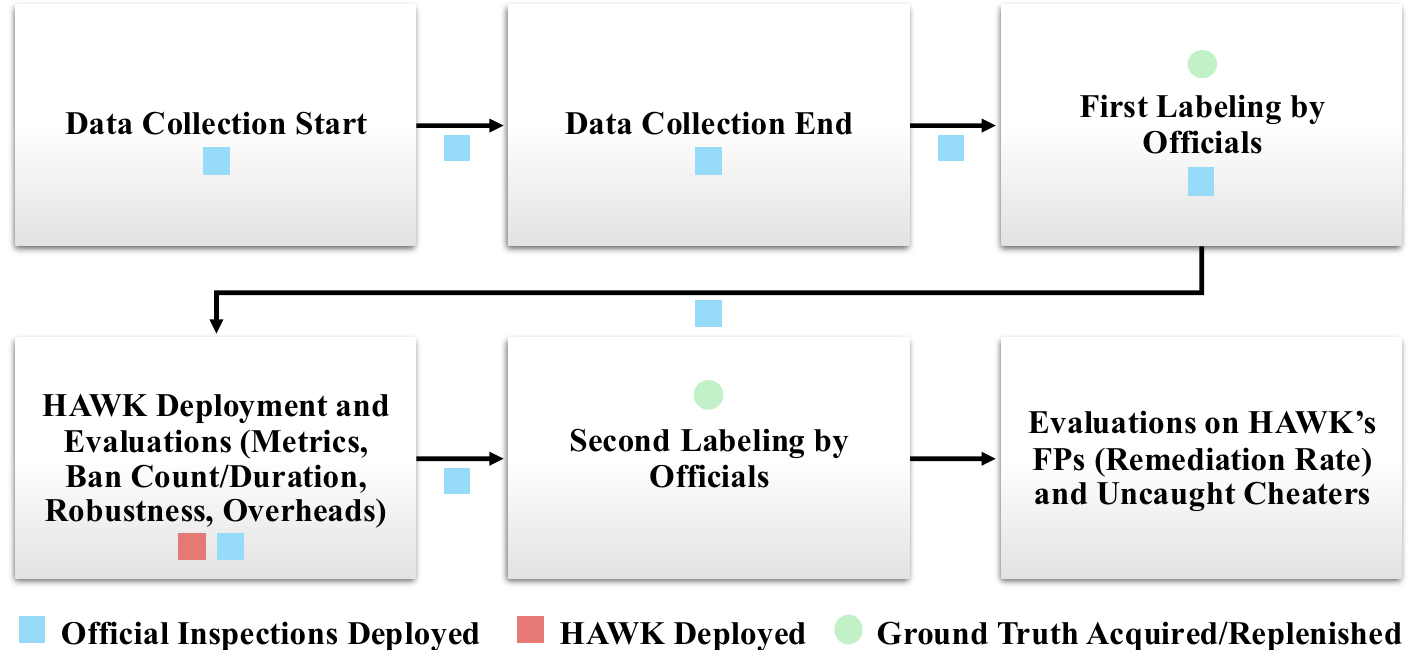}
% \vspace{-1em}
\caption{Experiments' process and labeling details.}
% \Description{Experiments' process and labeling details.}
% \vspace{-1em}
\label{fig:groundtruth}
\end{figure}
\begin{table}[htbp]
\centering
% \vspace{-1em}
\caption{Dataset separation and description.}
% \vspace{-0.5em}
\label{tb:dataset}
\begin{threeparttable}
\resizebox{0.48\textwidth}{!}{%
\begin{tabular}{@{}ccccclcccclcccc@{}}
\toprule
\multirow{2}{*}{\textbf{CT\tnote{1}}} & \multicolumn{4}{c}{\textbf{Train}}                        &  & \multicolumn{4}{c}{\textbf{Validation}}                   &  & \multicolumn{4}{c}{\textbf{Test}}                         \\ \cmidrule(lr){2-5} \cmidrule(lr){7-10} \cmidrule(l){12-15} 
                             & \text{\#M\tnote{2}} & \text{\#P\tnote{3}} & \text{\#N\tnote{4}} & \text{\#C\tnote{5}} &  & \text{\#M} & \text{\#P} & \text{\#N} & \text{\#C} &  & \text{\#M} & \text{\#P} & \text{\#N} & \text{\#C} \\ \midrule
\textbf{Aimbot}                       & 1,009        & 10,127            & 9,150            & 977            &  & 975          & 8,561            & 7,697            & 864            &  & 995          & 9,047            & 8,163            & 884            \\ \midrule
\textbf{Wallhack}                     & 833          & 8,320            & 7,500            & 820            &  & 1,037        & 9,712            & 8,745            & 967            &  & 1,101        & 10,274            & 9,249            & 1,025            \\ \bottomrule
\end{tabular}
}
\raggedright \footnotesize
\textsuperscript{1}\textbf{CT}: Cheat type; 
\textsuperscript{2}\textbf{\text{\#M}}: Number of matches; 
\textsuperscript{3}\textbf{\text{\#P}}: Number of players in total; 
\textsuperscript{4}\textbf{\text{\#N}}: Number of normal (honest) players; 
\textsuperscript{5}\textbf{\text{\#C}}: Number of cheaters.
\end{threeparttable}
% \vspace{-1.5em}
\end{table}

% \begin{table}[htbp]
% \centering
% % \vspace{-1em}
% \caption{Dataset separation and description.}
% % \vspace{-0.5em}
% \label{tb:dataset}
% \begin{threeparttable}
% \resizebox{0.48\textwidth}{!}{%
% \begin{tabular}{@{}ccccclcccclcccc@{}}
% \toprule
% \multirow{2}{*}{\textbf{CT}} & \multicolumn{4}{c}{\textbf{Train}}                        &  & \multicolumn{4}{c}{\textbf{Validation}}                   &  & \multicolumn{4}{c}{\textbf{Test}}                         \\ \cmidrule(lr){2-5} \cmidrule(lr){7-10} \cmidrule(l){12-15} 
%                              & \text{\#Match} & \text{\#Player} & \text{\#Normal} & \text{\#Cheater} &  & \text{\#Match} & \text{\#Player} & \text{\#Normal} & \text{\#Cheater} &  & \text{\#Match} & \text{\#Player} & \text{\#Normal} & \text{\#Cheater} \\ \midrule
% \textbf{Aimbot}                       & 1,009        & 10,127            & 9,150            & 977            &  & 975          & 8,561            & 7,697            & 864            &  & 995          & 9,047            & 8,163            & 884            \\ \midrule
% \textbf{Wallhack}                     & 833          & 8,320            & 7,500            & 820            &  & 1,037        & 9,712            & 8,745            & 967            &  & 1,101        & 10,274            & 9,249            & 1,025            \\ \bottomrule
% \end{tabular}
% }
% \end{threeparttable}
% % \vspace{-2em}
% \end{table}
% \usepackage{booktabs}
% \usepackage{multirow}
% \usepackage{graphicx}
% \usepackage{xcolor} % 添加颜色支持包
\begin{table*}[htbp]
\centering
\caption{Incomparability to prior works.}
\label{tab:incomp}
% \vspace{-1em}
\resizebox{\textwidth}{!}{%
\begin{tabular}{@{}lcccccccccc@{}}
\toprule
\multirow{2}{*}{\textbf{Work}} & \multicolumn{2}{c}{\textbf{Task}} & & \multicolumn{2}{c}{\textbf{Data Extraction}} & & \multicolumn{2}{c}{\textbf{Targeting}} & \multirow{2}{*}{\textbf{Hardware Constraint}} & \multirow{2}{*}{\textbf{Code Availability}} \\ \cmidrule(lr){2-3} \cmidrule(lr){5-6} \cmidrule(lr){8-9}
 & \text{Detection} & \text{Prevention} & & \text{Runtime} & \text{Replay file} & & \text{Aimbot} & \text{Wallhack} &  &  \\ \midrule
BotScreen \cite{choi2023botscreen} & \ding{51} &  & & \ding{51} (deprecated) &  & & \ding{51} &  & Intel SGX (deprecated) & \ding{51} \\
BlackMirror \cite{10.1145/3372297.3417890} &  & \ding{51} & & \ding{51} &  & &  & \ding{51} & Intel SGX (deprecated) & \ding{51} \\
Invisibility Cloak \cite{invisibilitycloak} &  & \ding{51} & & \ding{51} &  & & \ding{51} &  & None & \ding{51} \\
Prior works \cite{yeung2006detecting, yu2012statistical,6633617, liu2017detecting} & \ding{51} &  & & \ding{51} (unavailable) &  & & \ding{51} &  & None & \ding{55} \\ \midrule
\sys & \ding{51} &  & &  & \ding{51} & & \ding{51} & \ding{51} & None & \ding{51} \\ \bottomrule
\end{tabular}%
}
% \vspace{-1em}
\end{table*}
\begin{table*}[htbp]
\centering
\caption{Ablation study results on each of the subsystems.}
% \vspace{-1em}
\label{tb:result}
\begin{threeparttable}
\resizebox{\textwidth}{!}{%
\begin{tabular}{@{}ccccccccccccccccccccccc@{}}
\toprule
\multirow{2}{*}{\textbf{CT}} & & \multirow{2}{*}{\textbf{System}} & & \multicolumn{9}{c}{\textbf{Validation}} & & \multicolumn{9}{c}{\textbf{Test}} \\ \cmidrule(lr){5-13} \cmidrule(l){15-23}
& & & & TP & TN & FP & FN & Accuracy & Recall & NPV & AUC-ROC & FPR & & TP & TN & FP & FN & Accuracy & Recall & NPV & AUC-ROC & FPR \\ \midrule
\multirow{7}{*}{\textbf{Aimbot}} & & \syspov & & 458 & 4,796 & 2,901 & 406 & 0.614 & 0.530 & 0.922 & 0.577 & 0.378 & & 481 & 5,088 & 3,075 & 403 & 0.616 & 0.544 & 0.927 & 0.584 & 0.377 \\
& & \sysstats & & 453 & 7,290 & 407 & 411 & 0.904 & 0.524 & 0.947 & 0.736 & 0.053 & & 566 & 6,797 & 1,366 & 318 & 0.814 & 0.640 & 0.955 & 0.736 & 0.167 \\
& & \sysspc & & 553 & 6,187 & 1,510 & 311 & 0.787 & 0.640 & 0.952 & 0.779 & 0.196 & & 558 & 6,616 & 1,547 & 326 & 0.793 & 0.631 & 0.953 & 0.786 & 0.190 \\
& & \textsc{RevPov}+\sysstats & & 648 & 4,603 & 3,094 & 216 & 0.613 & 0.750 & 0.955 & 0.760 & 0.402 & & 691 & 4,364 & 3,799 & 193 & 0.559 & 0.782 & 0.958 & 0.750 & 0.465 \\
& & \textsc{RevPov}+\sysspc & & 611 & 5,733 & 1,964 & 253 & 0.741 & 0.707 & 0.958 & 0.781 & 0.255 & & 608 & 6,082 & 2,081 & 276 & 0.739 & 0.688 & 0.957 & 0.784 & 0.255 \\
& & \textsc{RevStats}+\sysspc & & 605 & 5,828 & 1,869 & 259 & 0.751 & 0.700 & 0.957 & 0.798 & 0.243 & & 639 & 5,877 & 2,286 & 245 & 0.720 & 0.723 & 0.960 & 0.796 & 0.280 \\ \cline{3-23}
& & \sys & & 614 & 5,803 & 1,894 & 250 & 0.750 & 0.711 & 0.959 & 0.732 & 0.246 & & 642 & 5,837 & 2,326 & 242 & 0.716 & 0.726 & 0.960 & 0.721 & 0.285 \\ \midrule
\multirow{7}{*}{\textbf{Wallhack}} & & \syspov & & 573 & 6,081 & 2,664 & 394 & 0.685 & 0.593 & 0.939 & 0.644 & 0.305 & & 627 & 6,364 & 2,885 & 398 & 0.680 & 0.612 & 0.941 & 0.650 & 0.312 \\
& & \sysstats & & 604 & 8,241 & 504 & 363 & 0.911 & 0.625 & 0.958 & 0.783 & 0.058 & & 855 & 6,988 & 2,261 & 170 & 0.763 & 0.834 & 0.976 & 0.795 & 0.245 \\
& & \sysspc & & 760 & 7,258 & 1,487 & 207 & 0.826 & 0.786 & 0.972 & 0.885 & 0.170 & & 501 & 8,484 & 765 & 524 & 0.875 & 0.489 & 0.942 & 0.840 & 0.083 \\
& & \textsc{RevPov}+\sysstats & & 604 & 8,241 & 504 & 363 & 0.911 & 0.625 & 0.958 & 0.798 & 0.058 & & 855 & 6,988 & 2,261 & 170 & 0.763 & 0.834 & 0.976 & 0.816 & 0.244 \\
& & \textsc{RevPov}+\sysspc & & 806 & 6,900 & 1,845 & 161 & 0.793 & 0.834 & 0.977 & 0.885 & 0.211 & & 573 & 8,257 & 992 & 452 & 0.859 & 0.559 & 0.948 & 0.840 & 0.107 \\
& & \textsc{RevStats}+\sysspc & & 791 & 7,110 & 1,635 & 176 & 0.814 & 0.818 & 0.976 & 0.889 & 0.187 & & 868 & 6,907 & 2,342 & 157 & 0.757 & 0.847 & 0.978 & 0.859 & 0.253 \\ \cline{3-23}
& & \sys & & 798 & 7,033 & 1,712 & 169 & 0.806 & 0.825 & 0.977 & 0.815 & 0.196 & & 869 & 6,902 & 2,347 & 156 & 0.756 & 0.848 & 0.978 & 0.797 & 0.254 \\ \bottomrule
\end{tabular}
}%
\end{threeparttable}
% \vspace{-0.5em}
\end{table*}
\begin{table*}[htbp]
\centering
\caption{Selected comparison result between different superior baseline models and \sysstats.} 
% \vspace{-1em}
\label{tb:stats}
\begin{threeparttable}
\resizebox{\textwidth}{!}{%
\begin{tabular}{@{}cclcccccccccccccccccccc@{}}
\toprule
\multirow{2}{*}{\textbf{CT}}       &  & \multirow{2}{*}{\textbf{Model}} &  & \multicolumn{9}{c}{\textbf{Validation}}                                 &  & \multicolumn{9}{c}{\textbf{Test}}                                        \\ \cmidrule(lr){5-13} \cmidrule(l){15-23} 
                                   &  &                                  &  & TP  & TN    & FP    & FN  & Accuracy & Recall & NPV   & AUC-ROC & FPR &  & TP  & TN    & FP    & FN  & Accuracy & Recall & NPV   & AUC-ROC  & FPR    \\ \midrule
\multirow{8}{*}{\textbf{Aimbot}}   
&& LogisticRegression \cite{cramer2002origins} & & 570 & 6,204 & 1,493 & 294 & 0.791 & 0.660 & 0.955 & 0.733 & 0.194 && 713 & 4,897 & 3,266 & 171 & 0.620 & 0.807 & 0.966 & 0.703 & 0.400 \\
&& Naive Bayes \cite{murphy2006naive} & & 467 & 6,740 & 957 & 397 & 0.842 & 0.874 & 0.944 & 0.708 & 0.124 && 625 & 5,928 & 2,235 & 259 & 0.724 & 0.707 & 0.958 & 0.717 & 0.274 \\
&& RandomForest \cite{breiman2001random} & & 415 & 7,694 & 3 & 449 & 0.947 & 0.480 & 0.945 & 0.740 & 0.000 && 427 & 7,634 & 529 & 457 & 0.891 & 0.483 & 0.944 & 0.709 & 0.065 \\
&& SVM \cite{cortes1995support} & & 309 & 7,615 & 82 & 555 & 0.926 & 0.358 & 0.932 & 0.673 & 0.011 && 405 & 7,665 & 498 & 479 & 0.892 & 0.458 & 0.941 & 0.699 & 0.061 \\
&& MLP \cite{haykin1994neural} & & 858 & 7,697 & 0 & 6 & 0.999 & 0.993 & 0.999 & 0.997 & 0.000 && 508 & 7,325 & 838 & 376 & 0.866 & 0.575 & 0.951 & 0.736 & 0.103 \\
&& GaussianNB \cite{john2013estimating} & & 467 & 6,740 & 957 & 397 & 0.842 & 0.541 & 0.944 & 0.708 & 0.124 && 625 & 5,928 & 2,235 & 259 & 0.724 & 0.707 & 0.958 & 0.717 & 0.274 \\
&& QDA \cite{srivastava2007bayesian} & & 350 & 7,181 & 516 & 514 & 0.880 & 0.405 & 0.933 & 0.669 & 0.067 && 534 & 6,770 & 1,393 & 350 & 0.807 & 0.604 & 0.951 & 0.717 & 0.171 \\ \cline{3-23}
&& \text{\sysstats} & & \text{453} & \text{7,290} & \text{407} & \text{411} & \text{0.904} & \text{0.524} & \text{0.947} & \text{0.736} & \text{0.053} && \text{566} & \text{6,797} & \text{1,366} & \text{318} & \text{0.814} & \text{0.640} & \text{0.955} & \text{0.736} & \text{0.167} \\
\midrule
\multirow{8}{*}{\textbf{Wallhack}}
&& LogisticRegression & & 714 & 7,629 & 1,116 & 253 & 0.859 & 0.738 & 0.968 & 0.805 & 0.128 && 943 & 5,622 & 3,627 & 82 & 0.639 & 0.920 & 0.986 & 0.764 & 0.392 \\
&& Naive Bayes & & 604 & 7,677 & 1,068 & 363 & 0.853 & 0.625 & 0.955 & 0.751 & 0.122 && 891 & 6,103 & 3,146 & 134 & 0.681 & 0.869 & 0.979 & 0.765 & 0.340 \\
&& RandomForest & & 967 & 8,745 & 0 & 0 & 1.000 & 1.000 & 1.000 & 1.000 & 0.000 && 751 & 8,083 & 1,166 & 274 & 0.860 & 0.733 & 0.967 & 0.803 & 0.126 \\
&& SVM & & 548 & 8,323 & 422 & 419 & 0.913 & 0.567 & 0.952 & 0.759 & 0.048 && 817 & 7,227 & 2,022 & 208 & 0.783 & 0.797 & 0.972 & 0.789 & 0.219 \\
&& MLP & & 472 & 8,489 & 256 & 495 & 0.923 & 0.488 & 0.945 & 0.729 & 0.029 && 788 & 7,726 & 1,523 & 237 & 0.829 & 0.769 & 0.970 & 0.802 & 0.165 \\
&& GaussianNB & & 604 & 7,677 & 1,068 & 363 & 0.853 & 0.625 & 0.955 & 0.751 & 0.122 && 891 & 6,103 & 3,146 & 134 & 0.681 & 0.869 & 0.979 & 0.765 & 0.340 \\
&& QDA & & 493 & 8,276 & 469 & 474 & 0.903 & 0.510 & 0.946 & 0.728 & 0.054 && 815 & 7,270 & 1,979 & 210 & 0.787 & 0.795 & 0.972 & 0.791 & 0.214 \\ \cline{3-23}
&& \text{\sysstats} & & \text{604} & \text{8,241} & \text{504} & \text{363} & \text{0.911} & \text{0.625} & \text{0.958} & \text{0.783} & \text{0.058} && \text{855} & \text{6,988} & \text{2,261} & \text{170} & \text{0.763} & \text{0.834} & \text{0.976} & \text{0.795} & \text{0.245} \\
\bottomrule
\end{tabular}
}%
% \newline
% \raggedright \footnotesize
% \textsuperscript{1}\textbf{CT}: Cheat type.
\end{threeparttable}
% \vspace{-1.5em}
\end{table*}
\subsection{Incomparability to Prior Works}\label{sec:incomparability}
% \Jiayi{DELETE WORDS}
First, most prior works are not open-sourced as listed in \autoref{tab:work_comp}.
% Only client-side anti-cheat approach BotScreen \cite{choi2023botscreen} and BlackMirror \cite{10.1145/3372297.3417890} have open-sourced their systems, the remaining outdated client-side or server-side approaches \cite{yeung2006detecting,inproceedings,yu2012statistical,6633617,liu2017detecting,zheng2016efficient} have not published their source code or datasets.
% %
% Besides, to the best of our knowledge, none of the previous client-side and server-side approaches has been evaluated on the real-world dataset in the form of \textit{replay files} extracted from the official platforms or the vendors directly. 
% %
% All previous works are evaluated under the same conditions: building a private server, modifying the client or server engine to log the desired data, and inviting volunteers to participate in the experiment. 
% %
% This brings two aspects of incomparability explained below, both in theoretical and practical foundations.
%
Second, the data collection method is incompatible. 
\sys relies on post-game \textit{replay files}, whereas prior studies modified the game engine or applied mods during gameplay.
Prior works are unable to extract data from the \textit{replay files}.
%
% The \textit{replay files} are unable to be modified as the prior works did to obtain the desired data while the game process is running simultaneously.
%
Third, installation on commercial clients or servers is infeasible. 
Only if the prior designs are implemented before the data collection, the data can then be gathered and evaluated. 
However, the partner platform prohibits installing unverified programs on clients or servers due to legal restrictions. 
%
% Consequently, it is impractical to install the programs from previous studies and conduct comparative analyses before collecting data from real-world matches within the scope of our dataset.
%
% \begin{itemize}
%     \item \textbf{Data collection method is distinct and incompatible.} Our data collection method, utilizing the \textit{replay file} post-game, differs fundamentally from prior studies that modified the game engine or applied mods during gameplay. The \textit{replay files} are unable to be modified as the prior works did to obtain the desired data while the game process is running simultaneously.
%     \item \textbf{Installation on commercial clients or servers is infeasible.} As long as the installation of the prior works is implemented before the commencement of data collection, the data can then be gathered and evaluated. However, the \textit{demos} we collected come directly from the game platform, which prohibits the installation of any unverified programs on clients or servers due to legal constraints. Consequently, it is impractical to install the programs from previous studies and conduct comparative analyses before collecting data from real-world matches within the scope of our dataset.
% \end{itemize}
%
In addition to the aforementioned reasons with commonality, we also further state below the specific reasons for the detailed incomparability between \sys and the SOTA designs.
There are three aspects of incomparability with BotScreen \cite{choi2023botscreen}.
First, BotScreen's data extraction is highly coupled with the game version and deprecated.
The authors of BotScreen confirmed through email that their data extraction does not apply to the current game versions.
%
%This further proves the significance of addressing \textbf{L3}.
%
Second, the deprecation of Intel SGX \cite{bleepingcomputer2024, intelcommunity2024} prevents us from conducting comparative experiments.
Third, it can only detect \textit{pure aimbot}, making the comparison less significant.
% (excluding \textit{wallhacks}, \textit{triggerbots}, \textit{micro-settings} and \textit{wallhack-mixed aimbots}, which all occurred in \sys's evaluation datasets)
% We should have conducted an additional experiment on a private server with BotScreen \cite{choi2023botscreen} configured on ten different clients installed with proper Intel Core Processors with Intel SGX TEE. 
% %
% And invite volunteers to simulate real-world matches with an open-sourced cheat Osiris \cite{KrupinskiOsiris} (the identical tool used in \cite{choi2023botscreen} for evaluation) to comparatively analyze the performance of \sys and the SOTA design.
% %
% However, as mentioned in \autoref{sec:intro}, only 6th to 10th generation Intel Core Processors released from May 2015 to August 2020 \cite{intelark} were equipped with SGX (10th generation discontinued in July 2021 \cite{comet_lake_wikipedia_2024}), and the following generations (i.e. from 11th until today's 14th generation) deprecated SGX \cite{bleepingcomputer2024, intelcommunity2024}. 
% %
% Therefore, we were unable to assemble or purchase enough equipment for this supplementary experiment.
% %
% Besides, \sys can detect multiple cheat genres, however, BotScreen can only detect \textit{pure aimbot} (excluding \textit{wallhacks}, \textit{triggerbots}, \textit{micro-settings} and \textit{wallhack-mixed aimbots}, which all occurred in \sys's evaluation datasets), making the comparison less significant.
%
Invisibility Cloak \cite{invisibilitycloak} and BlackMirror \cite{10.1145/3372297.3417890} are prevention designs only for counteracting \textit{computer-vision-based aimbots} and \textit{wallhacks}, respectively.
Both studies are only functional before the cheats happen.

\begin{figure}[t]
\centering
% \vspace{-0.5em}
\includegraphics[width=0.3\textwidth]{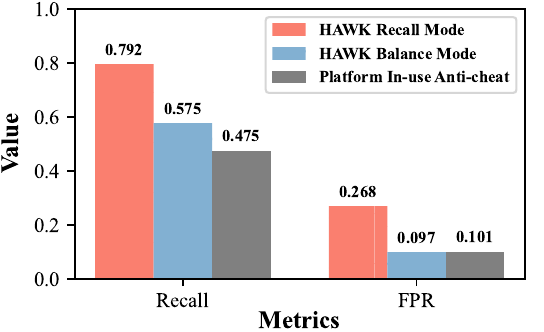}
% \vspace{-1em}
\caption{Recall and FPR performance comparison among \sys-Recall priority mode, \sys-Balance mode, and our partner in-use anti-cheat\protect\footnotemark. \sys's different modes are introduced and evaluated in \autoref{sec:MVIN} and \autoref{appx:MVINOpt}.}
% \Description{Recall and FPR performance comparison among \sys-Recall priority mode, \sys-Balance mode, and our partner in-use anti-cheat. \sys's different modes are introduced and evaluated in \autoref{sec:MVIN} and \autoref{appx:MVINOpt}.}
% \vspace{-1.5em}
\label{fig:comparison_metrics}
\end{figure}
\footnotetext{Commercial anti-cheat (detailed in \autoref{sec:AC_flaws}), not open-source, leverages reverse engineering on the known cheats to detect cheating signatures.}

Whereas \sys and other prior works are detection systems that function after cheating.
We detailed the incomparability to prior works in \autoref{tab:incomp}. We manage to compare \sys with our partner's industrial in-use anti-cheat in \autoref{fig:comparison_metrics} and reproduce two server-side works \cite{yu2012statistical,6633617} for further comparison in \autoref{tab:prior_comp}.
%
% Therefore it is also incomparable between \sys and BlackMirror.
%
% \subsection{Implementation}
% This research primarily focuses on CS:GO for practicality and detailed analysis. 
% %
% The selection of CS:GO stems from its widespread popularity, which makes it a prime target for cheaters, and its advanced replay system enables sophisticated analysis. 
% %
% By implementing through CS:GO, we ensure \sys and its evaluations are conducted under genuine conditions. 
\subsection{Experimental Settings and Ground Truth}\label{sec:dataset}
The \textit{demos} are open-sourced and acquired from a well-known CS:GO platform\footnote{5EPlay: \href{https://www.5eplay.com/}{https://www.5eplay.com/}.}.
The initial data extraction utilizes a public library \textit{awpy} \cite{awpy} to parse the \textit{demo} into a JSON file, which contains the raw data for \sys's feature constructions. 
%\footnote{https://www.5eplay.com/}

\autoref{tb:dataset} is the dataset description. 
\autoref{fig:groundtruth} illustrates our experimental process.
We evaluate \sys with two of the most prevalent cheat types, \textit{aimbot} and \textit{wallhack}. 
For \textit{aimbot}, data is collected over a single 16-day window, while for \textit{wallhack}, two collection windows totaling 27 days are used (the rationale for this inconsistency is explained in \autoref{appx:robustness}).
All \textit{demos} are from real-world ranked matches within these windows, ensuring that the cheats used by players remain black-box to the anti-cheat system. 
Both datasets naturally contain varying levels of cheating sophistication.
The cheaters' labels in the \textit{first labeling} are acquired on the last day of data collection for each set. 
These labels are used for training and most evaluations, as illustrated in \autoref{fig:groundtruth}.
Official inspections, which include proprietary closed-source anti-cheat systems and manual checks, are conducted continuously, with daily updates to the ban list.
However, due to the lengthy ban cycle, most cheaters are not banned immediately after cheating. 
To address this, a \textit{second labeling} is performed 65 days after the \textit{first labeling} to identify cheaters who were initially mislabeled. This allowed us to refine our analysis of false positives, with the updated cheater counts detailed in \autoref{tb:remedy}.

Official inspections employ a combination of proprietary anti-cheat software, developed by \textit{5Eplay}, and manual verification. 
The anti-cheat system operates by installing a kernel-level driver to scan system memory and active processes for known cheat signatures. 
A significant drawback of this method is its inherent reliance on a blacklist, making it one step behind the cheating developers. 
Modern cheats frequently update their code and employ advanced obfuscation techniques to evade signature-based detection. 
When a potential violation is identified, the system forwards the detection results to human reviewers for final determination.
Note that despite \textit{5EPlay} tried their best to annotate the samples as accurately as possible.
This process inherently and inevitably introduces a small amount of false negatives (i.e., the cheaters who escape the inspections) after the \textit{second labeling} due to the current limitation of anti-cheat low performance.
This is an unsolvable dilemma for this area.
All positive samples (cheaters) are accurately labeled.

The increase of cheaters between the first and second labeling is indicated in \autoref{tb:remedy}'s \textit{\#IncrPs}. 
Based on 5EPlay's statistics and the case study in \autoref{sec:case_study_FPs}, there are approximately 1.5\%-2\% and 3\%-4\%  uncaught cheaters in the negative samples after the \textit{second labeling} across different sets in the \textit{Aimbot} and \textit{Wallhack} datasets, respectively.

Although the two datasets include different cheat types, the official in-use anti-cheat systems cannot label specific subcategories; they only annotate broad cheat categories. 
Quantifying cheating sophistication is inherently subjective and challenging, making it infeasible to provide a more detailed breakdown of the datasets. 
Additionally, due to the poor performance of current anti-cheat systems, the ground truth inevitably contains noise (cheaters who evade detection) within the negative labels.
But to the best of our knowledge, this is the most realistic and well-annotated dataset.

To evaluate how effectively cheaters are identified, we use the metrics, accuracy, recall, NPV, AUC-ROC, and FPR, defined in Supplemental Material G. 
\begin{table}[htbp]
\centering
\caption{The manual average ban duration comparison and the remediation rate among \sys's false positives after the first labeling.}
% \vspace{-0.5em}
\label{tb:remedy}
\begin{threeparttable}
\resizebox{0.47\textwidth}{!}{%
\begin{tabular}{@{}clcccclccc@{}}
\toprule
\multirow{2}{*}{\textbf{CT}} & \multirow{2}{*}{\textbf{\sys}\tnote{1}}             & \multirow{2}{*}{\textbf{Off.}\tnote{2}} & \multicolumn{3}{c}{\textbf{Validation}} &           & \multicolumn{3}{c}{\textbf{Test}} \\ \cmidrule(lr){4-6} \cmidrule(l){8-10} 
                             &                                                  &                                & \#MTPs\tnote{3}     & Rmd\%\tnote{4}     & \#IncrPs\tnote{5}    & \textbf{} & \#MTPs     & Rmd\%     & \#IncrPs  \\ \midrule
\textbf{Aimbot}              & \multicolumn{1}{c}{\multirow{2}{*}{$\sim$4 min}} & 5.88 days                      & 93         & 4.91\%       & 237         &           & 97       & 4.17\%     & 233       \\
\textbf{Wallhack}            & \multicolumn{1}{c}{}                             & 4.47 days                      & 93         & 5.43\%       & 226         &           & 153      & 6.52\%     & 303       \\ \bottomrule
\end{tabular}
}
\raggedright \footnotesize
\textsuperscript{1}\textbf{\sys}: The process time of \sys varies with match length, \sys's detailed overheads are evaluated in \autoref{sec:overheads}; 
\textsuperscript{2}\textbf{Off.}: The average ban duration of official manual inspections per match; 
\textsuperscript{3}\textbf{\#MTPs}: The number of missed true positives among the false positives; 
\textsuperscript{4}\textbf{Rmd\%}: The percentage of \#MTPs in the number of false positives; 
\textsuperscript{5}\textbf{\#IncrPs}: The number of increased positives (cheaters) labeled by the officials after the second labeling.
\end{threeparttable}
% \vspace{-1em}
\end{table}

\begin{table*}[htbp]
\centering
\caption{Comparison result on the different baseline models of \syspov.} 
% \vspace{-1em}
\label{tb:pov}
\begin{threeparttable}
\resizebox{\textwidth}{!}{%
\begin{tabular}{@{}cclcccccccccccccccccccc@{}}
\toprule
\multirow{2}{*}{\textbf{CT}}       &  & \multirow{2}{*}{\textbf{Model}} &  & \multicolumn{9}{c}{\textbf{Validation}}                                 &  & \multicolumn{9}{c}{\textbf{Test}}                                        \\ \cmidrule(lr){5-13} \cmidrule(l){15-23} 
                                   &  &                                  &  & TP  & TN    & FP    & FN  & Accuracy & Recall & NPV   & AUC-ROC  & FPR  &  & TP  & TN    & FP    & FN  & Accuracy & Recall & NPV   & AUC-ROC  & FPR   \\ \midrule
\multirow{12}{*}{\textbf{Aimbot}}   
&& REPTree \cite{elomaa2001analysis} && 19 & 7,660 & 37 & 845 & 0.897 & 0.022 & 0.901 & 0.509 & 0.005 && 33 & 8,118 & 45 & 851 & 0.901 & 0.037 & 0.905 & 0.516 & 0.006 \\
% && RIONIDA \cite{rseslib2019} && 0 & 7,697 & 0 & 864 & 0.899 & 0.000 & 0.899 & 0.500 & N/A && 0 & 8,163 & 0 & 884 & 0.902 & 0.000 & 0.902 & 0.500 & N/A \\
% && MLP && 0 & 7,697 & 0 & 864 & 0.899 & 0.000 & 0.899 & 0.500 & N/A && 0 & 8,163 & 0 & 884 & 0.902 & 0.000 & 0.902 & 0.500 & N/A \\
&& RandomTree && 123 & 6,914 & 783 & 741 & 0.822 & 0.142 & 0.903 & 0.520 & 0.102 && 125 & 7,399 & 764 & 759 & 0.832 & 0.141 & 0.907 & 0.524 & 0.094 \\
&& RseslibKNN \cite{gora2002riona} && 40 & 7,460 & 237 & 824 & 0.876 & 0.046 & 0.901 & 0.508 & 0.031 && 33 & 7,868 & 295 & 851 & 0.873 & 0.037 & 0.902 & 0.501 & 0.036 \\
% && RBF Classifier \cite{frank2014fully} && 0 & 7,697 & 0 & 864 & 0.899 & 0.000 & 0.899 & 0.500 & N/A && 0 & 8,163 & 0 & 884 & 0.902 & 0.000 & 0.902 & 0.500 & N/A \\
&& SPAARC \cite{yates2019spaarc} && 0 & 7,697 & 0 & 864 & 0.899 & 0.000 & 0.899 & 0.500 & 0.000 && 0 & 8,163 & 0 & 884 & 0.902 & 0.000 & 0.902 & 0.500 & 0.000 \\
% && J48 \cite{Quinlan1993} && 0 & 7,697 & 0 & 864 & 0.899 & 0.000 & 0.899 & 0.500 & N/A && 0 & 8,163 & 0 & 884 & 0.902 & 0.000 & 0.902 & 0.500 & N/A \\
&& RandomForest \cite{breiman2001random} && 16 & 7,665 & 32 & 848 & 0.897 & 0.019 & 0.900 & 0.507 & 0.004 && 33 & 8,128 & 35 & 851 & 0.902 & 0.037 & 0.905 & 0.517 & 0.004 \\
% && AdaBoostM1 \cite{Freund1996} && 0 & 7,697 & 0 & 864 & 0.899 & 0.000 & 0.899 & 0.500 & N/A && 0 & 8,163 & 0 & 884 & 0.902 & 0.000 & 0.902 & 0.500 & N/A \\
&& Bagging \cite{Breiman1996} && 18 & 7,666 & 31 & 846 & 0.898 & 0.021 & 0.901 & 0.508 & 0.004 && 30 & 8,129 & 34 & 854 & 0.902 & 0.034 & 0.905 & 0.515 & 0.004 \\
&& VFI \cite{demiroz1997classification} && 822 & 445 & 7,252 & 42 & 0.148 & 0.951 & 0.914 & 0.505 & 0.942 && 848 & 472 & 7,691 & 36 & 0.146 & 0.959 & 0.929 & 0.509 & 0.942 \\
&& JRip \cite{Cohen1995} && 21 & 7,663 & 34 & 843 & 0.898 & 0.024 & 0.901 & 0.510 & 0.004 && 42 & 8,116 & 47 & 842 & 0.902 & 0.048 & 0.906 & 0.521 & 0.006 \\
&& MODLEM \cite{stefanowski1998rough} && 19 & 7,588 & 109 & 845 & 0.889 & 0.022 & 0.900 & 0.504 & 0.014 && 34 & 8,029 & 134 & 850 & 0.891 & 0.038 & 0.904 & 0.511 & 0.016 \\
% && CHIRP \cite{wilkinson2011chirp} && 0 & 7,697 & 0 & 864 & 0.899 & 0.000 & 0.899 & 0.500 & 0.000 && 0 & 8,163 & 0 & 884 & 0.902 & 0.000 & 0.902 & 0.500 & 0.000 \\
&& CSForest \cite{siers2015software} && 50 & 7,292 & 405 & 814 & 0.858 & 0.058 & 0.900 & 0.503 & 0.053 && 58 & 7,755 & 408 & 826 & 0.864 & 0.066 & 0.904 & 0.508 & 0.050
\\ \cline{3-23}
& & \text{\syspov} & & \text{458} & \text{4,796} & \text{2,901} & \text{406} & \text{0.614} & \text{0.530} & \text{0.922} & \text{0.577} & \text{0.377} & & \text{481} & \text{5,088} & \text{3,075} & \text{403} & \text{0.616} & \text{0.544} & \text{0.927} & \text{0.584} & \text{0.377} \\
\midrule
\multirow{12}{*}{\textbf{Wallhack}}
&& REPTree && 47 & 8,641 & 104 & 920 & 0.895 & 0.049 & 0.904 & 0.518 & 0.012 && 66 & 9,140 & 109 & 959 & 0.896 & 0.064 & 0.905 & 0.526 & 0.012 \\
% && RIONIDA && 0 & 8,745 & 0 & 967 & 0.900 & 0.000 & 0.900 & 0.500 & N/A && 0 & 9,249 & 0 & 1,025 & 0.900 & 0.000 & 0.900 & 0.500 & N/A \\
% && MLP && 0 & 8,745 & 0 & 967 & 0.900 & 0.000 & 0.900 & 0.500 & N/A && 0 & 9,249 & 0 & 1,025 & 0.900 & 0.000 & 0.900 & 0.500 & N/A \\
&& RandomTree && 178 & 8,006 & 738 & 789 & 0.843 & 0.184 & 0.910 & 0.550 & 0.084 && 198 & 8,433 & 816 & 827 & 0.840 & 0.193 & 0.911 & 0.552 & 0.088 \\
&& RseslibKNN && 89 & 8,515 & 230 & 878 & 0.886 & 0.092 & 0.907 & 0.533 & 0.026 && 120 & 8,939 & 310 & 905 & 0.882 & 0.117 & 0.908 & 0.542 & 0.034 \\
% && RBF Classifier && 0 & 8,745 & 0 & 967 & 0.900 & 0.000 & 0.900 & 0.500 & N/A && 0 & 9,249 & 0 & 1,025 & 0.900 & 0.000 & 0.900 & 0.500 & N/A \\
&& SPAARC && 69 & 8,679 & 66 & 898 & 0.901 & 0.071 & 0.906 & 0.532 & 0.008 && 87 & 9,164 & 85 & 938 & 0.900 & 0.085 & 0.907 & 0.538 & 0.009 \\
% && J48 && 0 & 8,745 & 0 & 967 & 0.900 & 0.000 & 0.900 & 0.500 & N/A && 0 & 9,249 & 0 & 1,025 & 0.900 & 0.000 & 0.900 & 0.500 & N/A \\
&& RandomForest && 27 & 8,713 & 32 & 940 & 0.900 & 0.028 & 0.903 & 0.512 & 0.004 && 35 & 9,212 & 37 & 990 & 0.900 & 0.034 & 0.903 & 0.515 & 0.004 \\
% && AdaBoostM1 && 0 & 8,745 & 0 & 967 & 0.900 & 0.000 & 0.900 & 0.500 & N/A && 0 & 9,249 & 0 & 1,025 & 0.900 & 0.000 & 0.900 & 0.500 & N/A \\
&& Bagging && 26 & 8,705 & 40 & 941 & 0.899 & 0.027 & 0.902 & 0.511 & 0.005 && 37 & 9,201 & 48 & 988 & 0.899 & 0.036 & 0.903 & 0.515 & 0.005 \\
&& VFI && 950 & 563 & 8,182 & 17 & 0.156 & 0.982 & 0.971 & 0.523 & 0.936 && 1,008 & 689 & 8,560 & 17 & 0.166 & 0.983 & 0.976 & 0.529 & 0.925 \\
&& JRip && 36 & 8,670 & 75 & 931 & 0.896 & 0.037 & 0.903 & 0.514 & 0.009 && 71 & 9,179 & 70 & 954 & 0.900 & 0.069 & 0.906 & 0.531 & 0.008 \\
&& MODLEM && 35 & 8,687 & 58 & 932 & 0.898 & 0.036 & 0.903 & 0.515 & 0.007 && 56 & 9,171 & 78 & 969 & 0.898 & 0.055 & 0.904 & 0.523 & 0.008 \\
% && CHIRP && 6 & 8,738 & 7 & 961 & 0.900 & 0.006 & 0.901 & 0.503 & 0.001 && 8 & 9,242 & 7 & 1,017 & 0.900 & 0.008 & 0.901 & 0.504 & 0.001 \\
&& CSForest && 0 & 8,745 & 0 & 967 & 0.900 & 0.000 & 0.900 & 0.500 & 0.000 && 0 & 9,249 & 0 & 1,025 & 0.900 & 0.000 & 0.900 & 0.500 & 0.000 \\ \cline{3-23}
& & \text{\syspov} & & \text{573} & \text{6,081} & \text{2,664} & \text{394} & \text{0.685} & \text{0.593} & \text{0.939} & \text{0.644} & \text{0.305} & & \text{627} & \text{6,364} & \text{2,885} & \text{398} & \text{0.680} & \text{0,612} & \text{0.941} & \text{0.650} & \text{0.312} \\
\bottomrule
\end{tabular}
}%
% \newline
% \raggedright \footnotesize
% \textsuperscript{1}\textbf{CT}: Cheat type.
\end{threeparttable}
% \vspace{-0.5em}
\end{table*}
\begin{table*}[ht]
\centering
\caption{\sys detection result with different \textit{Task-Specified Threshold Optimizer} (TSTO) settings.}
% \vspace{-1em}
\label{tab:optresult}
\begin{threeparttable}
\resizebox{1\textwidth}{!}{%
\begin{tabular}{@{}cclcccccccccccccccccccc@{}}
\toprule
\multirow{2}{*}{\textbf{CT}}       & \textbf{} & \multirow{2}{*}{\textbf{TSTO}}    & \textbf{} & \multicolumn{9}{c}{\textbf{Validation}}                                 &  & \multicolumn{9}{c}{\textbf{Test}}                                       \\ \cmidrule(lr){5-13} \cmidrule(l){15-23}
                                   &           &                                        &           & TP  & TN    & FP    & FN  & Accuracy & Recall & AUC-ROC & NPV & FPR  &  & TP  & TN    & FP    & FN  & Accuracy & Recall & AUC-ROC & NPV & FPR \\ \midrule
\multirow{8}{*}{\textbf{Aimbot}}   &           & Balance\tnote{1}                               &           & 389 & 7,390 & 307   & 475 & 0.909    & 0.450  & 0.705   & 0.940 & 0.040 && 457 & 7,444 & 719  & 427 & 0.873    &  0.517  & 0.714   & 0.946 & 0.088 \\
&           & F1-Score                               &           & 448 & 7,305 & 392   & 416 & 0.906    & 0.519  & 0.734   & 0.946  & 0.051 && 548 & 6,928 & 1235  & 336 & 0.826    & 0.620  & 0.734   & 0.954 & 0.151 \\
                                   &           & Accuracy                               &           & 269 & 7,528 & 169   & 595 & 0.911    & 0.311  & 0.645   & 0.927 & 0.022 && 305 & 7,888 & 275   & 579 & 0.906    & 0.345  & 0.656   & 0.932 & 0.034 \\
                                   &           & AUC-ROC                                &           & 648 & 4,603 & 3,094 & 216 & 0.613    & 0.750  & 0.674   & 0.955 & 0.402 && 691 & 4,364 & 3,799 & 193 & 0.559    & 0.782  & 0.658   & 0.958 & 0.465 \\
                                   &           & $\text{Acc}_{\textit{rc}>0.75}$\tnote{2}              &           & 652 & 5,321 & 2,376 & 212 & 0.698    & 0.755  & 0.723   & 0.962  & 0.309 && 672 & 5,365 & 2,798 & 212 & 0.667    & 0.760  & 0.709   & 0.962  & 0.343 \\
                                   &           & $\text{Acc}_{\textit{rc}>0.8}$               &           & 700 & 4,611 & 3,086 & 164 & 0.620    & 0.810  & 0.705   & 0.966  & 0.401 && 716 & 4,700 & 3,463 & 168 & 0.599    & 0.810  & 0.693   & 0.965  & 0.424 \\
                                   &           & $\text{Acc}_{\textit{rc}>0.85}$              &           & 736 & 3,790 & 3,907 & 128 & 0.529    & 0.852  & 0.672   & 0.967  & 0.508 && 761 & 3,875 & 4,288 & 123 & 0.512    & 0.861  & 0.668   & 0.969  & 0.525 \\
                                   &           & $\text{Acc}_{\textit{rc}>0.9}$ &           & 778 & 2,859 & 4,838 & 86  & 0.425    & 0.900  & 0.636   & 0.971  & 0.628 && 787 & 2,955 & 5,208 & 97  & 0.414    & 0.890  & 0.626   & 0.968  & 0.638 \\ \midrule
\multirow{8}{*}{\textbf{Wallhack}} &           & Balance                               &           & 592 & 8,256 & 489   & 375 & 0.911    & 0.612  & 0.778   & 0.957 & 0.056 && 616 & 8,344 & 905  & 409 & 0.872    &  0.601  & 0.752   & 0.953  & 0.098 \\
&           & F1-Score                               &           & 576 & 8,313 & 432   & 391 & 0.915    & 0.596  & 0.773   & 0.955  & 0.049 && 651 & 8,040 & 1,209 & 374 & 0.846    & 0.635  & 0.752   & 0.956 & 0.131 \\
                                   &           & Accuracy                               &           & 374 & 8,582 & 163   & 593 & 0.922    & 0.387  & 0.684   & 0.935  & 0.019 && 167 & 9,158 & 91    & 858 & 0.908    & 0.163  & 0.577   & 0.914  & 0.010 \\
                                   &           & $\text{Acc}_{\textit{rc}>0.7}$               &           & 677 & 7,898 & 847   & 290 & 0.883    & 0.700  & 0.802   & 0.965  & 0.097 && 803 & 7,506 & 1,743 & 222 & 0.809    & 0.783  & 0.797   & 0.971  & 0.188 \\
                                   &           & $\text{Acc}_{\textit{rc}>0.75}$              &           & 727 & 7,589 & 1,156 & 240 & 0.856    & 0.752  & 0.810   & 0.969  & 0.132 && 859 & 6,964 & 2,285 & 166 & 0.761    & 0.838  & 0.795   & 0.977  & 0.247 \\
                                   &           & $\text{Acc}_{\textit{rc}>0.8}$               &           & 776 & 7,186 & 1,559 & 191 & 0.820    & 0.802  & 0.812   & 0.974  & 0.178 && 583 & 8,280 & 969   & 442 & 0.863    & 0.569  & 0.732   & 0.949  & 0.105 \\
                                   &           & $\text{Acc}_{\textit{rc}>0.85}$              &           & 822 & 6,712 & 2,033 & 145 & 0.776    & 0.850  & 0.809   & 0.979  & 0.232 && 871 & 6,810 & 2,439 & 154 & 0.748    & 0.850  & 0.793   & 0.978  & 0.264 \\
                                   &           & $\text{Acc}_{\textit{rc}>0.9}$ &           & 871 & 5,755 & 2,990 & 96  & 0.682    & 0.901  & 0.779   & 0.984 & 0.342 && 855 & 6,885 & 2,364 & 170 & 0.753    & 0.834  & 0.789   & 0.976 & 0.256 \\ \bottomrule
\end{tabular}
}%
\newline
\raggedright \footnotesize
\textsuperscript{1}\textbf{Balance}: maintain industry-standard FPR (shown in \autoref{fig:comparison_metrics}) while optimizing recall;
\textsuperscript{2}\textbf{\textit{$\text{Acc}_{\textit{rc}>n}$}}: highest accuracy while maintaining recall above $n$.
\end{threeparttable}
% \vspace{-1.5em}
\end{table*}
\subsection{Overall Performance, Ablation and Baseline Comparison}\label{sec:ablationstudy}
To answer \textbf{RQ.1}, we conduct an ablation study, shown in \autoref{tb:result}. 
%
% The evaluation embarks on a meticulous analysis of each subsystem within \sys, evaluating them against two prominent cheat types: \textit{aimbot} and \textit{wallhack}. 
%
We set the TSTO to achieve the highest \textit{accuracy} with at least 70\% \textit{recall} for detecting \textit{aimbot}, and the highest \textit{AUC-ROC} for detecting \textit{wallhack} (same below unless otherwise stated), illustrating different anti-cheat scenarios (further evaluated in \autoref{appx:MVINOpt}).
Overall, \sys excels in identifying authentic cheaters, evidenced by the highest \textit{recall} and \textit{NPV} across all evaluations.
For \textit{aimbot} detection, \sysstats enhances \textit{accuracy} by effectively filtering normal players in both validation and test sets. 
In \textit{wallhack} detection, \sysstats excels in the validation set, while \sysspc leads in the test set.
Although some components like \sysstats and \sysspc outperform \sys in \textit{accuracy} and \textit{FPR}, their lower \textit{recall} (up to 64\%) limits their effectiveness, affirming \sys as the optimal choice.
Notably, the \textit{Wallhack} dataset yields better overall performance than \textit{Aimbot}.
We believe this is due to the latter's diverse subcategories of \textit{aimbots}. 
We conduct the pair-wise ablation study to evaluate the performance among \sys and different combinations of subsystems.
In \autoref{tb:result}, we observe that \syspov and \sysstats combination has significantly higher \textit{FPR} on the \textit{Aimbot} set and slightly lower \textit{recall} on the \textit{Wallhack} set. 
The other two combinations both suffer from lower \textit{recall}.
While evaluating the performance, we prioritize \textit{recall}, \textit{accuracy} and \textit{FPR} on the test set over other metrics, because they can more directly reflect the ability to identify cheaters or the overall performance.
\sys's framework is designed with both detection effectiveness and efficiency in mind. We select models with acceptable detection performance, and further select models with low training and prediction costs based on the overhead described in the relevant articles.
\autoref{tb:pov} and \autoref{tb:stats} show \syspov's and \sysstats's promising model selection results among hundreds of tested baselines.
\syspov outperforms other models with balanced, consistent accuracy and recall.
For \sysstats, we found that Random Forest outperformed other models. 
Yet, it sometimes failed to overcome the overfitting problem (e.g., wallhack validation set).
Thus, we adopt \textit{multi-forest} and \textit{multi-subsampling} with Random Forest.
As a result, \sysstats outperforms individual classifiers in balancing recall and accuracy by capitalizing on the strengths of each classifier.
We also prioritize metrics \textit{recall}, \textit{accuracy} and \textit{FPR} on the test set over other metrics.
For example, although Logistic Regression has a higher \textit{recall}, its \textit{accuracy} and \textit{FPR} are significantly lower. 
The contribution of the \textit{recall}'s increase can not make up for the decrease in accuracy and \textit{FPR}.

The current framework has optimized overall performance compared with more innovative models. For example, in the preliminary test of \syspov's encoder selection, we obtain lower detection performance with extended training and testing time while using the Transformer.
%
% The training and validation loss for \sysspc in \autoref{fig:loss_wallhack} and \autoref{fig:loss_aimbot} (discussed in \autoref{appx:ExSPC_loss}).
\begin{figure}[htbp]
\centering
% \vspace{-0.5em}
\includegraphics[width=0.47\textwidth]{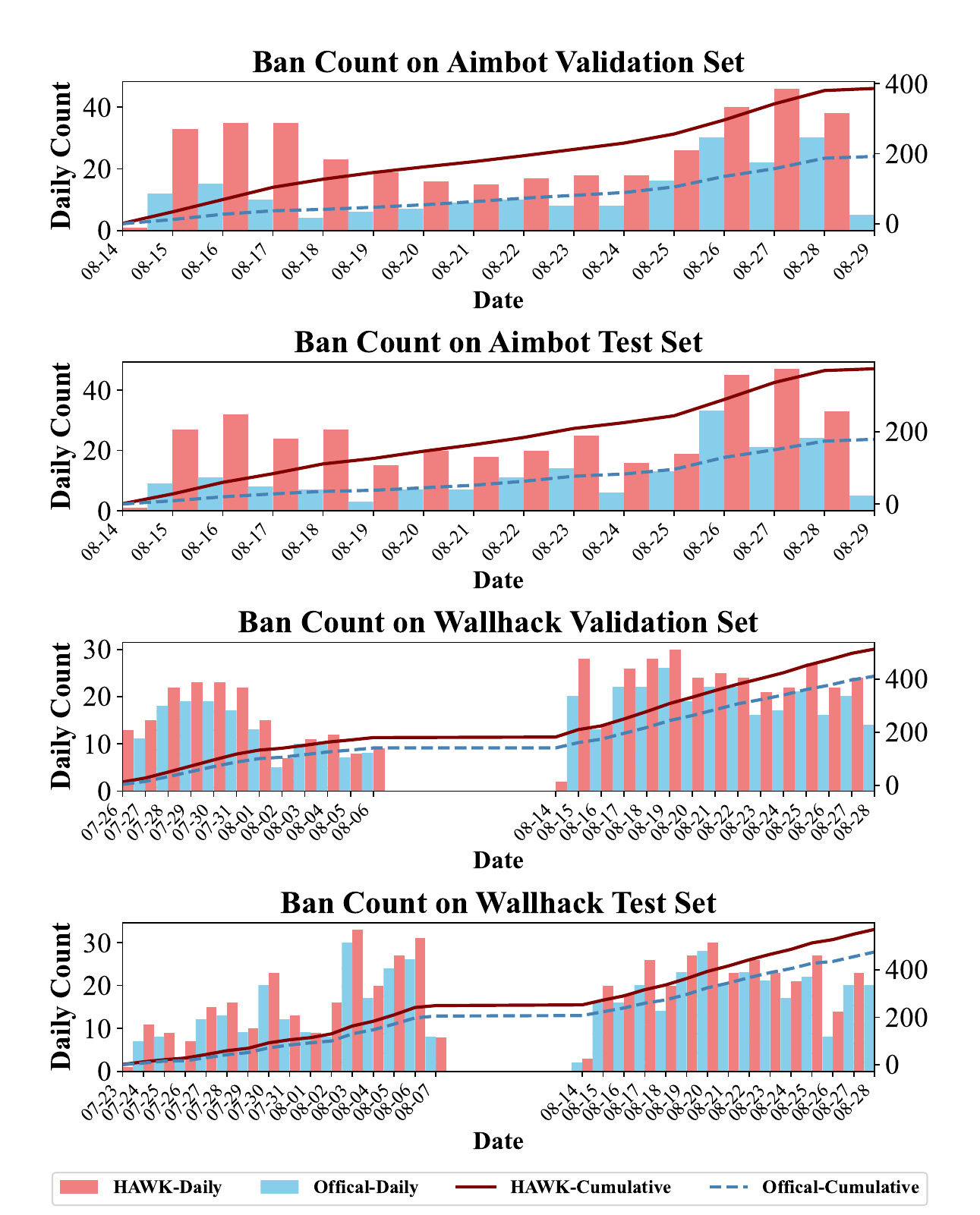}
% \vspace{-1em}
\caption{Daily number of bans for the day and cumulative ban-sums by \sys versus the official manual inspections.}
% \Description{Daily number of bans for the day and cumulative ban-sums by \sys versus the official manual inspections.}
\label{fig:bancycle}
% \vspace{-1.5em}
\end{figure}
\subsection{\sys and Official Inspection Comparison}\label{sec:officalcomparison}
To answer \textbf{RQ.2}, we calculate the daily number of bans for the day and cumulative ban-sums by \sys and the official inspections illustrated in \autoref{fig:bancycle}. 
The figure indicates that \sys's identification effectiveness (daily count in red bars) and efficiency (cumulative count in solid red lines) outrank those of the official manual inspections (hatched blue bars and dashed blue lines) on both cheat types.
Additionally, \autoref{tb:remedy} compares the average ban duration per match, revealing that \sys's seconds-scale processing time is exponentially more efficient than official inspections.
The remediation rate indicates the percentage of cheaters not banned in the first labeling but caught in the second labeling.
\autoref{tb:remedy} lists the remediation rate after the second labeling among \sys's false positives (which are obtained after the first labeling).
It demonstrates that \sys can capture the authentic cheaters who escape from the official anti-cheat system.
Moreover, this indicates that \sys is robust against mislabeled data.
% using a purer dataset would further improve results.
%
We also compare the \sys's performance with the official in-use anti-cheat in the next section, as it is relevant to \sysinteg.

% \usepackage{booktabs}
% \usepackage{graphicx}
% \usepackage{xcolor} % 添加颜色支持包
\begin{table}[htbp]
\centering
\caption{\sys comparative study results to prior works on the \textit{Aimbot} test set. The optimal results are bolded.}
\label{tab:prior_comp}
\resizebox{\columnwidth}{!}{%
\begin{tabular}{@{}lcccccc@{}}
\toprule
\textbf{Method} & & \textbf{Accuracy} & \textbf{Recall} & \textbf{NPV} & \textbf{AUC-ROC} & \textbf{FPR} \\ \midrule
AccA\cite{yu2012statistical} & & 0.103 & \textbf{0.997} & 0.897 & 0.558 & 0.997 \\
HitAcc\cite{6633617} & & 0.675 & 0.545 & 0.931 & 0.658 & 0.311 \\
\textbf{\sys} & & \textbf{0.716} & 0.726 & \textbf{0.960} & \textbf{0.721} & \textbf{0.285} \\ \bottomrule
\end{tabular}%
}
\end{table}
\subsection{Comparative Analysis Among \sys and Prior Works}
\label{sec:prior_comp}

To answer \textbf{RQ.7}, we compare \sys with two other reproduced methods using \sys's \textit{Aimbot} dataset. 
Due to the lack of available open-source solutions, we write our own code to reproduce both \textit{aimbot} detection methods based on statistical features related to aiming or elimination used in prior works. 
Both methods are related to \sys but use naive statistical features, data, and approach. 
Since prior work only targeted \textit{aimbots}, we only used the \textit{Aimbot} dataset for comparison.
The results are listed in \autoref{tab:prior_comp}.
\sys outperforms prior works across all metrics but \textsc{AccA}'s recall, because \textsc{AccA} classifies most of the samples as cheaters.
Below is a summary of each method:

\begin{itemize}[topsep=2pt, itemsep=2pt, parsep=0pt, partopsep=0pt]
\item \textsc{AccA}: Filters cheaters when the player's field-of-view acceleration exceeds a threshold \cite{yu2012statistical}.
\item \textsc{HitAcc}: Filters cheaters based on the weapon accuracy of the player, using SVM \cite{6633617}.
\end{itemize}

\subsection{Analysis on Different \sysinteg Settings}
\label{appx:MVINOpt}
To address \textbf{RQ.3}, we conduct a comparative analysis of \sys under different \sysinteg settings, as shown in \autoref{tab:optresult} and \autoref{fig:comparison_metrics}.
In \autoref{tab:optresult}, except \sysinteg, other subsystems are trained and validated with recall priority.
By only adjusting the TSTO in \sysinteg, \sys achieves rapid dynamic tuning ability, by using pre-trained \sysinteg or retraining \sysinteg to generate weight matrices and thresholds in about one minute.
%
% An inevitable trade-off exists between capturing more cheaters and minimizing false positives (discuss in \autoref{sec:discussion}).
% %
% As discussed in \autoref{sec:MVIN}, GMs can dynamically adjust optimizers by retraining \sysinteg to generate weight matrices and thresholds in about one minute. 
% %
% This adaptability aligns with the varying intensity of required anti-cheat measures, offering flexibility in balancing performance and efficiency.
% %
% GMs can use different settings to meet specific anti-cheat needs.
%
\adddiscuss{For example, the TSTO settings can be adjusted to achieve a lower FPR if necessary, avoiding the potential surge of excessive manual verification workload.
\autoref{tab:optresult} shows that leveraging different optimizer settings, e.g., \textit{Accuracy} or \textit{F1-Score} optimization modes, can obtain low FPR (1\%-3\% or 5\%-15\%, respectively). 
Alternatively, the \textit{Balance} mode maintains a lower FPR than current industry standards (grey bar in \autoref{fig:comparison_metrics}) while optimizing recall, resulting in slightly lower FPR and significantly higher recall compared to industry anti-cheat system (see blue bars in \autoref{fig:comparison_metrics}, industry anti-cheats are introduced in \autoref{sec:AC_flaws}).
}
%
% Interestingly, GMs do not always aim to catch all cheaters. 
% %
% They sometimes balance catching cheaters with permissiveness, as certain cheaters may bring short-term profits (e.g., in buyout games, where bans may be delayed until after refund periods). 
% %
% When complaints rise, GMs may then conduct a mass ban (a regularly organized operation to clean up the gaming environment) to restore community confidence.
% %
% This also partially draws to the creation of \sysinteg.
%
\subsection{Robustness to Cheat Evolution}
\label{appx:robustness}
\begin{figure}[htbp]
\centering
\begin{subfigure}[htbp]{0.19\textwidth}
    \centering
    \includegraphics[width=\textwidth]{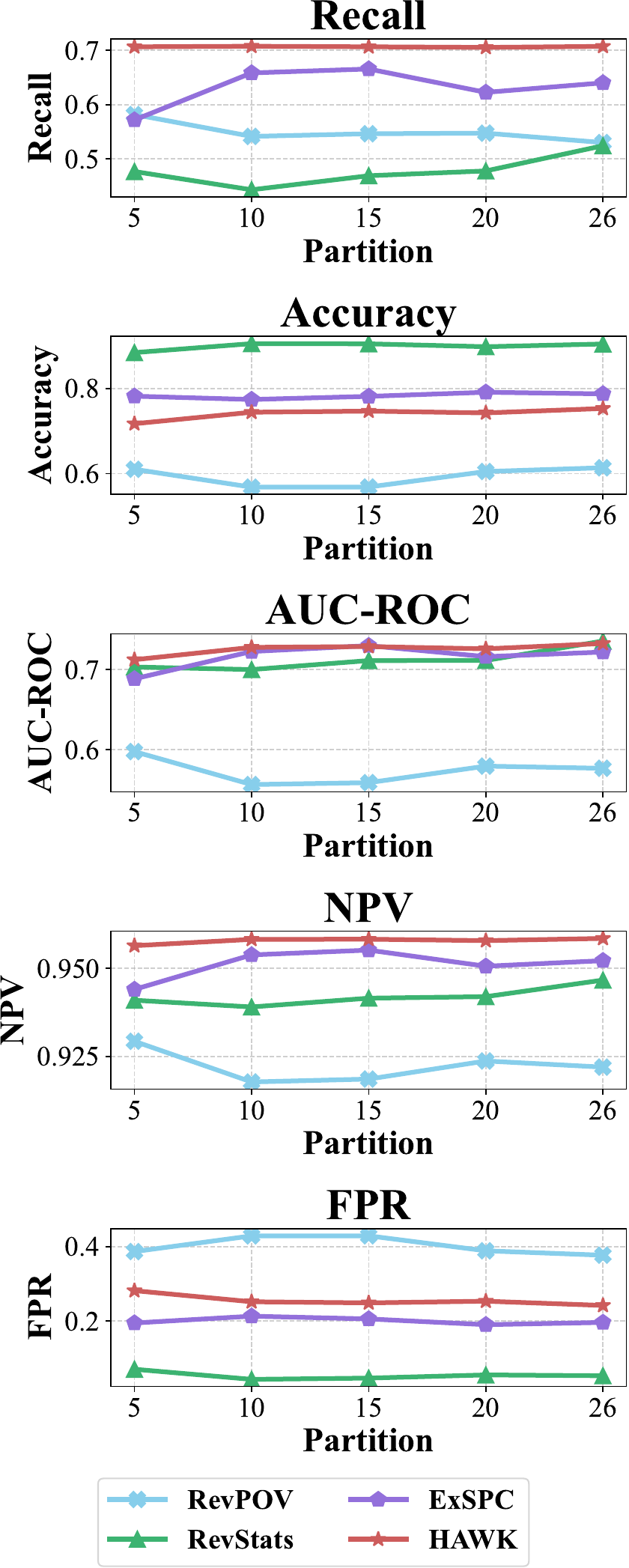}
    \caption{Validation set}
    \label{fig:robust_val}
\end{subfigure}
% \hfill
\begin{subfigure}[htbp]{0.19\textwidth}
    \centering
    \includegraphics[width=\textwidth]{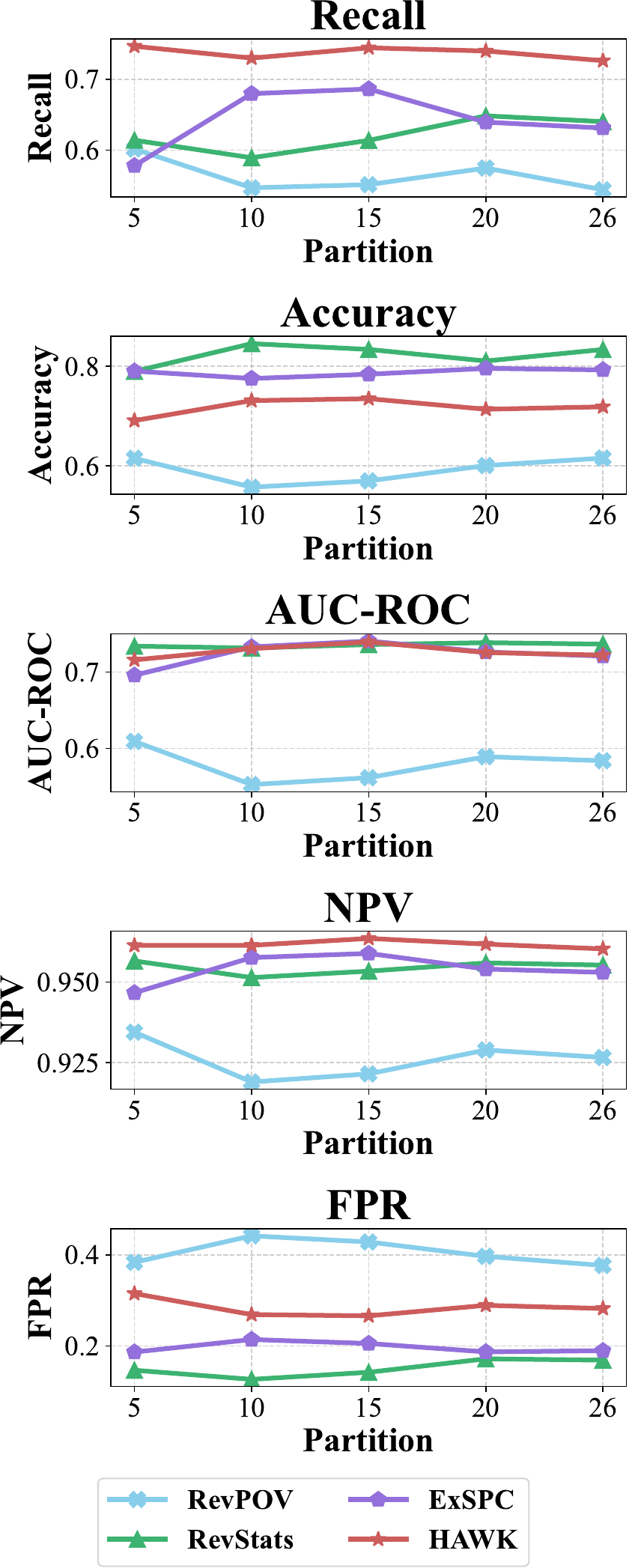}
    \caption{Test set}
    \label{fig:robust_test}
\end{subfigure}
% \vspace{-0.5em}
\caption{Ablated robustness against cheat evolution under different metrics for validation (a) and test (b) sets.}
% \Description{Ablated Robustness on \textit{Aimbot} Validation and Test Sets}
% \vspace{-1.5em}
\label{fig:robust_combined}
\end{figure}

Although anti-cheat measures are often opaque to prevent circumvention, cheat developers treat them as black boxes, adapting and camouflaging their cheats on a timely basis to evade detection.
\textit{Cheat evolution} refers to the diminishing effectiveness of specific anti-cheat over time.
Commercially used anti-cheat methods are susceptible to this evolution (explained in \autoref{sec:AC_flaws}).
However, \sys starts the analysis from the behavior, thus cheaters must appear the data-level distinctions mentioned in \autoref{sec:CT}.
Hence, no matter how the cheats adapt, as long as the cheaters still want to gain unfair advantages, it is hard to bypass the detection of \sys.
To address \textbf{RQ.4}, we test \sys's robustness to cheat evolution using the \textit{Aimbot} dataset. 
Classified through date, the dataset is divided into partitions (40 \textit{matches} per partition).
The training and validation are performed on all data up to the current partition. 
We compared its performance on the same test set with different partitions’ trained models. 
The reason for selecting the \textit{Aimbot} dataset is that during its collection window, an official mass ban on \textit{aimbots} took place (at the 10th partition), and the cheats are usually updated within a few days after such an incident.
\autoref{fig:robust_val} and \autoref{fig:robust_test} demonstrate the ablated performances of subsets under different metrics of \sys across different numbers of partitions for training.
According to the average ban duration of the official bans in \autoref{tb:remedy}, approximately three sequestration ban cycles have passed during the collection window, and at least two ban cycles after the mass ban occurred.
This offers more than enough time for cheat to evolve.
The ablated results of \sys on both sets have been fluctuating to maintain within a small range over partitions (time) and with the cheat evolution.
This indicates even without using the latter models of partitioned samples after the mass ban, \sys’s performance remains stable on the test set before and after the mass ban occurred.
Besides, in \autoref{fig:bancycle}, we observe the effectiveness of \sys has not continuously decreased over time on both datasets (we separated and extended the collection windows in \textit{Wallhack} dataset for further proof).
Thus, both experiments indicate \sys is robust to cheat evolution.
%
%
% \begin{figure}[htbp]
% \centering
% % \vspace{-1em}
% \includegraphics[width=0.47\textwidth]{Figures/aimbot_val.pdf}
% % \vspace{-1em}
% \caption{Ablated robustness against cheat evolution over partitions on \textit{Aimbot}'s validation set under different metrics.}
% \Description{Ablated Robustness on \textit{Aimbot} Validation Set}
% % \vspace{-0.5em}
% \label{fig:robust_val}
% \end{figure}
% \begin{figure}[htbp]
% \centering
% % \vspace{-1em}
% \includegraphics[width=0.47\textwidth]{Figures/aimbot_test.pdf}
% % \vspace{-1em}
% \caption{Ablated robustness against cheat evolution over partitions on \textit{Aimbot}'s test set under different metrics.}
% \Description{Ablated Robustness on \textit{Aimbot} Test Set}
% % \vspace{-0.5em}
% \label{fig:robust_test}
% \end{figure}
\subsection{Detecting Uncaught Cheaters in the Wild}
\label{sec:case_study_FPs}
To answer \textbf{RQ.5}, given the above-known presence of cheaters in the false positives in \autoref{tb:remedy} after the \textit{first labeling}, we conduct a use case analysis to explore if there are additional illegal activities that exist after the \textit{second labeling}. 
Given the lengthy average duration of a match, which demands significant manpower, we select the top 30 \sys's false positives based on the decreasing order of the \textit{akpr} from the \textit{Aimbot} and \textit{Wallhack} validation sets for manual review, respectively (experts' skill level is listed in Supplemental Material Table 1). 
With the assistance of three experts in FPS games, we identify 27 suspects, who are labeled as \textit{normal players} after the second labeling, to be highly suspicious within these 60 false positives. 
Of these 27 suspects, 14 originate from the \textit{Aimbot} dataset (4 are considered using \textit{aimbots} and 10 are considered \textit{boosting}, i.e., elevate the account rank by other higher-rank players playing the account), while 13 are from the \textit{Wallhack} dataset (8 are considered using \textit{wallhacks} and 5 are considered \textit{boosting}). 
We reported these suspicions to the official personnel. 
Upon their review, they confirm all 12 of the aforementioned \textit{aimbots} and \textit{wallhacks} suspects are committed to cheats, subsequently issuing bans for these cheaters.
This reaffirms the poor performance of current in-use anti-cheats.
The website mentioned in \autoref{sec:intro} demonstrates their illegal actions.
In terms of \textit{boosting} suspects, there are no existing solutions for the officials to evidently verify their illegal actions. 
Therefore, there are no bans issued on the suspicious \textit{boosting} accounts.
This experiment further proves the limited performance of current industrial anti-cheats, while demonstrating \sys's superior anti-cheat capabilities and high robustness to noise.
\subsection{Overheads}\label{sec:overheads}
\sys is designed as an asynchronous, load-balancing system that analyzes post-game \textit{replay files} during periods of low server activity.
It operates independently of active game servers and can be deployed on a separate server.
However, we still conduct overhead assessments for reference.
To answer \textbf{RQ.6}, we utilize an Ubuntu server equipped with 128 cores AMD EPYC 7713P 64-Core Processor and 256 GB RAM, a real dedicated game server configuration \cite{forum_serverreq}.
Leveraging the pre-trained model in the artifact, \sys requires only CPU resources to function efficiently. 
\autoref{tab:overhead} shows the maximum overhead statistics per match, with each match analyzed concurrently.
Additionally, we estimate the \sys-to-server overheads in terms of a real deployment scenario.
Our partner platform reports that at most 150,000 matches are played daily, with each server needing to handle up to 573 matches a day.
According to our platform’s report and public server statistics \cite{steamdb2024}, player activity is lowest between 1-5 AM local time, requiring analysis of up to 144 matches per hour. 
Based on the elapsed time in \autoref{tab:overhead}, \sys needs to process up to 12 matches concurrently to complete all analyses within the timeframe per server.
We further conduct concurrent processing evaluation, it occupies at most 17.95\% memory and 35.45\% CPU usage for processing 12 matches concurrently.
Therefore, each server has more than sufficient resources for deploying \sys, even when hosting several matches simultaneously.
It is vital for a design to maintain stable performance in high-concurrency environments \cite{ParliRobo}.
To further validate the performance of \sys under varying concurrency levels, we employed a game server equipped with a 24-core Intel® Xeon® Gold 5418Y CPUs and 503 GB RAM. 
During idle periods on this server, we concurrently request \sys 10, 20, and 50 times, respectively. 
The results are listed in Table \autoref{tab:concurrent_overhead}. 
The system maintains stable performance when handling different levels of concurrent requests. 
Furthermore, based on the aforementioned server load data, each server requires running \sys concurrently 12 times per hour at most. 
Our tests reached a maximum concurrency of 50 requests, exceeding the data upper bound by over four times. 
But \sys requires less than 7 minutes to complete all 50 requests. 
Therefore, we conclude that using \sys under high concurrency conditions does not significantly affect the system overhead.

\begin{table}[htbp]
\caption{Maximum overheads of \sys's subsystems per match. The numbers are retrieved by the Time \cite{time_linux} in Linux.}
% \vspace{-0.5em}
\label{tab:overhead}
\begin{threeparttable}
\resizebox{\columnwidth}{!}{%
\begin{tabular}{@{}lcccc@{}}
\toprule
\textbf{Max. Overheads per Match}                & \multicolumn{1}{c}{\syspov} & \multicolumn{1}{c}{\sysstats} & \multicolumn{1}{c}{\sysspc} & \multicolumn{1}{c}{\sysinteg} \\ \midrule
User time (sec)               & 415.72                     & 178.12                       & 36.89                     & 3.97                                  \\
System time (sec)             & 69.32                      & 436.61                       & 27.12                     & 13.53                                \\
Elapsed time (mm:ss)            & 03:40                   & 00:14                     & 00:45                  & 00:01                               \\
Percent of CPU this job got     & 220\%                    & 4371\%                       & 140\%                   & 2860\%                                \\
Maximum resident set size (GB)  & 3.56                    & 0.27                       & 2.57                   & 0.10                                   \\ \bottomrule
% File system inputs\tnote{6}            & 6,261,818                    & 3,050,561                      & 786,996                    & 30,902                    & 10,130,277                 \\
% File system outputs\tnote{7}            & 1,068                       & 59,615                        & 3,065                      & 1,457                    & 65,205                    \\ 
\end{tabular}
}
\end{threeparttable}
% \vspace{-1em}
\end{table}
% \usepackage{booktabs}
% \usepackage{xcolor}
\begin{table}[htbp]
\centering
\caption{Performance under varying concurrency levels.}
\vspace{-0.5em}
\label{tab:concurrent_overhead}
\resizebox{\columnwidth}{!}{%
\begin{tabular}{@{}lccccccc@{}}
\toprule
\textbf{\# Concurrent Request} & & & \text{10} & & \text{20} & & \text{50} \\ \midrule
\text{Average time (sec)} & & & 384.512 & & 365.267 & & 377.084 \\
\text{Minimum time (sec)} & & & 380.343 & & 357.570 & & 368.070 \\
\text{Maximum time (sec)} & & & 393.472 & & 391.536 & & 386.522 \\
\text{Standard deviation} & & & 4.502 & & 7.793 & & 5.282 \\ \bottomrule
\end{tabular}%
}%
\end{table}
\section{Discussions}
\label{sec:discussion}
% This section discusses the generalizability of \sys, the dataset noise, detailed labeling, optimizable temporal features, and workflow manual involvement.
\noindent\textbf{Generalizability.}
Narrowed generalizability is a meta-problem in the anti-cheat field. Since cheats are game-specific, creating a universal anti-cheat system is challenging. 
Although we successfully implemented \sys on a leading tactical shooter game \cite{tactical_shooter} (a subgenre of FPS games like CS:GO), similar to prior works \cite{choi2023botscreen, yeung2006detecting, yu2012statistical,6633617, liu2017detecting} discussed in \autoref{sec:AC_flaws}, it requires further adaptation for other FPS subgenres with distinct gameplay.
%
% \sys's generalizability concerns the core observations, methodologies, and workflow rather than specific features. 
%
The proposed features are applicable to tactical shooters with different engines in FPS games, e.g., CS2.
However, adaptation is needed for other subgenres, where certain elements like flashes and grenades may not apply. 
%
% For instance, in Hero Shooters (e.g., Overwatch, Valorant), such features could be substituted with character-specific abilities, or in Battle Royale (BR) games (e.g., Fortnite, PUBG), where elements like angles and velocities may not be vital due to long time-to-kill.
%
We summarize two core principles for adapting \sys to other subgenres.
(a) \textit{Generalized anti-cheat perspectives transform from three observations.} 
The fundamental structure of transforming three key observations into three corresponding anti-cheat subsystems is a generalized framework applicable to any FPS subgenre. 
While the specific features or model architectures within each subsystem may need to be optimized for different games, the principle of using these three complementary perspectives to identify cheating behavior is universally necessary and applicable.
(b) \textit{Adaptable feature extraction strategy}.
We propose that feature extraction for any FPS game that uses \sys should be divided into two distinct categories.
\textit{Core FPS features}. 
These are fundamental features inherent to all FPS games and should be retained across all subgenres. 
For example, this includes data like player position, distance to target, and view direction (in \syspov); or various aiming speeds, angles, and precision/hit rates (in \sysstats).
While generalizing \sys to other subgenres, these features are applicable without any modification.
\textit{Subgenre-specific features}.
These features are designed around the unique mechanics that define a specific subgenre. 
The design of subgenre-specific features, as well as the classification rules in \sysspc, should be led by senior anti-cheat developers with deep expertise in the specific game. 
This ensures the system is tailored to detect cheats that exploit that subgenre's unique mechanics. For instance, in a Hero Shooter (e.g., \textit{Overwatch}, \textit{Valorant}), the ability to use unique character skills is a key differentiator. 
To adapt \sys to other subgenres, one would: i. augment \syspov with basic temporal information related to skill activation and targeting; ii. extend \sysstats to record metrics like skill accuracy, the mean and variance of skill usage timing, and their effectiveness.
%
% Since \textit{replay files} and labeling are significant for \sys, and due to dataset scarcity, further research on different subgenres to prove \sys’s generalizability requires cooperation with game vendors to obtain data.
% %
% We have contacted several vendors, and one specializing in BR games has shown interest in cooperating. 
% %
% However, implementation and evaluation have to be left for future work.
%

\noindent\textbf{Manual involvement.} 
Although \sys has significantly increased the detection performance and efficiency, as mentioned in \autoref{sec:workflow}, \sys-after-use manual verification is unavoidable. 
%
% The presence of real false positives undeniably indicates the requirement for manual checks to avoid false bans.
%
\adddiscuss{However, platform statistics indicate their anti-cheat recall (47\%) is significantly lower than \sys's (79\% on average) under recall-first mode. 
Although \sys’s false FPR is around 25\%, the high recall outweighs the FPR increase (about 15\%). 
A higher FNR indicates more cheaters undetected, leading to user attrition with greater losses than those caused by higher FPR \cite{9772248}.
\sys also offers Balance mode whose FPR is lower and recall is higher than the current in-use anti-cheat (\autoref{appx:MVINOpt} and \autoref{fig:comparison_metrics}).
Under this mode, \sys outperforms current anti-cheat and obtains lower manual involvement, due to lower FPR.
}
Note that for any anti-cheat system, all results must undergo manual review and confirmation before any bans are issued (i.e., the GM verification stage). 
To provide an intuitive demonstration of \sys's performance, unless otherwise specified, all results presented for this system are shown before manual review. 
Therefore, all positive samples mentioned in the evaluation require manual review before any bans can be enforced in actual operation. 
No player will be mistakenly banned or inconvenienced before the manual revision. 
Regarding the trade-off between \textit{recall} and \textit{FPR}, we have demonstrated the system's performance under different anti-cheating requirements in \autoref{tab:optresult}. 
It can be observed that for every approximately 5\% increase in \textit{recall}, \textit{FPR} increases by 10\% and 20.00\% and 20.84\% increase of the manpower required on average in terms of the \textit{aimbot} and the \textit{wallhack}, respectively.
Therefore, if human resources are relatively scarce, the default or balanced mode of \sys should be used.
%
% Therefore, we believe \sys plays an important role in the FPS games' server-side anti-cheat research field. 

\noindent\textbf{Dataset imbalance and result significance.}
To statistically validate the performance stability of our models, we employed bootstrap resampling \cite{bootstrap_resample} with 5,000 iterations to estimate 95\% confidence intervals for all evaluation metrics in \autoref{tab:confidence_interval}. 
The bootstrap analysis confirmed the statistical significance and robustness of our models under class-imbalanced conditions. 
All performance metrics showed narrow 95\% confidence intervals. 
This demonstrates that both \textit{aimbot} and \textit{wallhack} detection performances are not only practically meaningful but also statistically reliable, validating the effectiveness of our approach even with inherent dataset imbalances in the context of game anti-cheating.
% \usepackage{booktabs}
% \usepackage{multirow}
% \usepackage{graphicx}
% \usepackage{xcolor} % 添加颜色支持包
\begin{table}[htbp]
\centering
\caption{Performance metrics with 95\% confidence intervals from bootstrap analysis (5,000 resamples).}
\label{tab:confidence_interval}
\resizebox{\columnwidth}{!}{%
\begin{tabular}{@{}llcccccc@{}}
\toprule
\textbf{Dataset} & \textbf{Metric} & \textbf{Original} & \textbf{Bootstrap\_Mean} & \textbf{Std\_Dev} & \textbf{Std\_Error} & \textbf{CI} & \textbf{CI\_Width} \\ \midrule
\multirow{5}{*}{\textbf{Aimbot}} & \textbf{Accuracy} & 0.716149 & 0.716142 & 0.00467 & 0.000066 & [0.706972, 0.725323] & 0.018351 \\
 & \textbf{Recall} & 0.726244 & 0.726073 & 0.014762 & 0.000209 & [0.697646, 0.755435] & 0.057789 \\
 & \textbf{NPV} & 0.960191 & 0.960179 & 0.00249 & 0.000035 & [0.955263, 0.964983] & 0.00972 \\
 & \textbf{AUC-ROC} & 0.72065 & 0.720535 & 0.007924 & 0.000112 & [0.704752, 0.736172] & 0.03142 \\
 & \textbf{FPR} & 0.284944 & 0.284831 & 0.004967 & 0.00007 & [0.274992, 0.294452] & 0.01946 \\ \midrule
\multirow{5}{*}{\textbf{Wallhack}} & \textbf{Accuracy} & 0.756375 & 0.756422 & 0.004182 & 0.000059 & [0.748394, 0.764651] & 0.016257 \\
 & \textbf{Recall} & 0.847805 & 0.847965 & 0.011178 & 0.000158 & [0.826003, 0.869812] & 0.043808 \\
 & \textbf{NPV} & 0.977897 & 0.977882 & 0.001741 & 0.000025 & [0.97437, 0.981204] & 0.006834 \\
 & \textbf{AUC-ROC} & 0.797024 & 0.797097 & 0.006064 & 0.000086 & [0.785037, 0.80896] & 0.023923 \\
 & \textbf{FPR} & 0.253757 & 0.253778 & 0.004606 & 0.000065 & [0.244804, 0.262858] & 0.018054 \\ \bottomrule
\end{tabular}%
}
\end{table}
\section{Related Works}\label{sec:AC_flaws}

\noindent\textbf{Server-side anti-cheats}.
%
% \Jiayi{TODO: more intro on prior works' features}
%
Yeung et al. \cite{yeung2006detecting} initially proposed a Bayesian model to tune the probability threshold with naive features (e.g., whether the player is moving, the distance between the attacker and victim) in 2006, but its performance lacks real-world validation and the features are outdated for today’s advanced cheating techniques and sophistication.
Subsequent works \cite{yu2012statistical,6633617,liu2017detecting, han2015online,orlova2024estimation} employed statistical tests with basic features (e.g., time-on-target) to examine data distribution difference \cite{yu2012statistical}, set fixed threshold with limited features (e.g., extra shots count on an invisible bait) \cite{liu2017detecting,orlova2024estimation}, or used one machine learning model (e.g., SVM) with basic non-temporal features (e.g., whether the target's position changed) \cite{6633617}, which couldn't monitor sophisticated behaviors or describe entire matches \cite{choi2023botscreen}, or ineffective to certain type of cheat.
Due to the scarcity of real-world datasets, these server-side studies relied on private servers and simulated data for evaluation, with the largest dataset containing only 47 cheaters \cite{liu2017detecting}.
Such small-scale datasets are insufficient to reflect real-world anti-cheat performance and possibly lead to inconsistent performance between the reported results and real-world performance. 
Additionally, works like \cite{yu2012statistical,6633617,liu2017detecting, han2015online} require game engine modifications for concurrent data collection, necessitating significant adaptation efforts after each update and limiting generalizability to other games.
Importantly, prior works leverage a single model to classify and could result in biased predictions on the complex real-world dataset as discussed in \autoref{sec:RevStats}.
CUPID \cite{cupid} is a matchmaking design to optimize the user’s gameplay experience. In terms of anti-cheat, it may be modified to arrange cheaters within the same match so that they would not appear in the normal matches. 
However, its applicability remains unknown since it does not target identifying cheaters.

\noindent\textbf{Client-side anti-cheats}.
%
% \Jiayi{TODO: more intro on in-use IDS/anti-cheat}
%
BotScreen \cite{choi2023botscreen}, the latest \textit{aimbot} detector, uses SGX to securely store and run its detection model with temporal aiming features.
It constructs its dataset using a self-setup server with 14 players in Deathmatch (gameplay is different from Rank) and evaluates with the deprecated open-source \textit{aimbot} Osiris \cite{KrupinskiOsiris}.
This setup may not accurately reflect performance against more sophisticated cheats or common cheating scenarios \cite{miketendo64aimbots2021,mediumaimbotcomplaints2021}, leaving its real-world functionality unclear.
It is also troubled by the deprecated TEE constraint \cite{bleepingcomputer2024, intelcommunity2024}. 
Invisibility Cloak \cite{invisibilitycloak} and BlackMirror \cite{10.1145/3372297.3417890} only prevent without detecting and are limited to hardware constraints or client overheads, and targeting on a specific cheat subcategory. 
Commercial anti-cheats like VAC \cite{SteamSupport}, Vanguard \cite{RiotGamesInc}, EAC \cite{EasyAC} and BattlEye \cite{BattlEye} offer both detection and prevention \cite{Silva_2022} for various games through customization. 
They monitor and manage process privileges, scan DNS caches and cookies for browsing history and cheat purchase records \cite{lehtonen2020comparative,SteamSupport}, and may use kernel-level drivers \cite{pushtotalk_anticheat}.
However, these solutions are limited by security and privacy concerns mentioned in \autoref{sec:intro}.
The common approach of creating blacklists for known cheats and issuing bans based on detected malware signatures \cite{9774028, pushtotalk_anticheat, choi2023botscreen, liu2017detecting} is ineffective against new or updated cheats with different signatures \cite{choi2023botscreen, liu2017detecting}, lacking robustness to cheat evolutions.
Most cheaters avoid using legacy cheats as paid cheats often offer auto-updates to evade bans, leading to low recall in industry anti-cheats, which are always one step behind.
These anti-cheats are integrated into games before release and must run concurrently with gameplay \cite{pushtotalk_anticheat}, causing client overheads and limited generalizability.
They also combine multiple methods for cross-checking, extending the banning cycle \cite{pushtotalk_anticheat}.
Importantly, these commercial anti-cheats are proprietary and closed-source \cite{pushtotalk_anticheat}, making it infeasible for researchers to conduct comparison studies without collaborating with game vendors.
%
% A high-level comparison among \sys and the related detection works is shown in \autoref{tab:work_comp}.
%
VESPA \cite{vespa} combines computer vision techniques and human inspection to capture the user's screen and detect the \textit{wallhack}'s UI overlays and thereby identify cheaters.
This may raise privacy concerns and can be easily bypassed by managing process privileges or disabling the visible overlays.
Therefore, it is less practical for real-world deployment.
FCDGame \cite{fcd} leverages few-shot learning to learn highly repetitive gesture trajectory patterns on the mobile phone screen to identify cheaters.
This pattern targets the bot identification in MMORPGs rather than FPS games. 
Because the former only requires fixed navigation and simple gesture operations to achieve repetitive tasks (e.g., collect mines), but the latter requires real-time and complex inputs to adjust aiming and perform shooting.
The FPS trajectory patterns are more complex and irregular.
Therefore, the screen gesture trajectory feature may be less functional in FPS games.

\section{Conclusion}

We proposed an FPS games' server-side anti-cheat framework \sys, drawing inspiration from how humans identify cheaters. 
%
% Drawing inspiration from how humans identify cheaters, \sys includes three subsystems, \syspov, \sysstats, \sysspc, and one integration network \sysinteg. 
%
With promising experimental results, we have demonstrated \sys's real-world effectiveness, efficiency, and robustness in identifying cheaters in a modern FPS game.
%
% Our objective is to ensure a fair FPS competitive environment and save the loss brought by cheating activities.
%
Ultimately, we hope \sys can be a pathfinder for next-generation FPS anti-cheat, providing a valuable dataset and a reference for the follow-up research.

\section*{Acknowledgments}
\noindent We thank 5EPlay for sharing and labeling the dataset. We appreciate Associate Editor Prof. Abderrahim Benslimane for arranging the review process, and three anonymous reviewers for their professional reviews. We also acknowledge the anonymous reviewers from IEEE S\&P, ACM CCS, USENIX Security and NDSS for their constructive reviews on the past versions of this research paper. This work is partially supported by the NSFC for Young Scientists of China (No.62202400) and the
RGC for Early Career Scheme (No.27210024). Any opinions, findings, or conclusions expressed in this material are those of the authors and do not necessarily reflect the views of NSFC and RGC.
% \input{Sections/Ethical_Consideration}
% \clearpage
\bibliographystyle{IEEEtran}
\bibliography{mybibliography}

\begin{thebibliography}{10}
\providecommand{\url}[1]{#1}
\csname url@samestyle\endcsname
\providecommand{\newblock}{\relax}
\providecommand{\bibinfo}[2]{#2}
\providecommand{\BIBentrySTDinterwordspacing}{\spaceskip=0pt\relax}
\providecommand{\BIBentryALTinterwordstretchfactor}{4}
\providecommand{\BIBentryALTinterwordspacing}{\spaceskip=\fontdimen2\font plus
\BIBentryALTinterwordstretchfactor\fontdimen3\font minus \fontdimen4\font\relax}
\providecommand{\BIBforeignlanguage}[2]{{%
\expandafter\ifx\csname l@#1\endcsname\relax
\typeout{** WARNING: IEEEtran.bst: No hyphenation pattern has been}%
\typeout{** loaded for the language `#1'. Using the pattern for}%
\typeout{** the default language instead.}%
\else
\language=\csname l@#1\endcsname
\fi
#2}}
\providecommand{\BIBdecl}{\relax}
\BIBdecl

\bibitem{newzoo2024}
Newzoo, ``{The global games market will generate \$187.7 billion in 2024},'' \url{https://newzoo.com/resources/blog/global-games-market-revenue-estimates-and-forecasts-in-2024}, August 2024.

\bibitem{irdeto2023}
Irdeto, ``{Cheating in video games: The A to Z},'' \url{https://blog.irdeto.com/video-gaming/cheating-in-games-everything-you-always-wanted-to-know-about-it/}, 2023.

\bibitem{wepc2023}
WePC, ``{Video Game Industry Statistics, Trends and Data In 2023},'' \url{https://www.wepc.com/news/video-game-statistics/}, 2023.

\bibitem{belyaeva2022stakeholder}
Z.~Belyaeva, A.~Petrosyan, and R.~Shams, ``Stakeholder data analysis in the video gaming industry: Implications for regional development,'' \emph{EuroMed Journal of Business}, vol.~17, 05 2022.

\bibitem{liu2021lower}
\BIBentryALTinterwordspacing
S.~Liu, M.~Claypool, A.~Kuwahara, J.~Sherman, and J.~J. Scovell, ``Lower is better? the effects of local latencies on competitive first-person shooter game players,'' in \emph{Proceedings of the 2021 CHI Conference on Human Factors in Computing Systems}, ser. CHI '21.\hskip 1em plus 0.5em minus 0.4em\relax New York, NY, USA: Association for Computing Machinery, 2021. [Online]. Available: \url{https://doi.org/10.1145/3411764.3445245}
\BIBentrySTDinterwordspacing

\bibitem{choi2023botscreen}
M.~Choi, G.~Ko, and S.~K. Cha, ``Botscreen: trust everybody, but cut the aimbots yourself,'' in \emph{Proceedings of the 32nd USENIX Conference on Security Symposium}, ser. SEC '23.\hskip 1em plus 0.5em minus 0.4em\relax USA: USENIX Association, 2023.

\bibitem{gianvecchio2009battle}
\BIBentryALTinterwordspacing
S.~Gianvecchio, Z.~Wu, M.~Xie, and H.~Wang, ``Battle of botcraft: fighting bots in online games with human observational proofs,'' in \emph{Proceedings of the 16th ACM Conference on Computer and Communications Security}, ser. CCS '09.\hskip 1em plus 0.5em minus 0.4em\relax New York, NY, USA: Association for Computing Machinery, 2009, p. 256–268. [Online]. Available: \url{https://doi.org/10.1145/1653662.1653694}
\BIBentrySTDinterwordspacing

\bibitem{invisibilitycloak}
\BIBentryALTinterwordspacing
C.~Sun, K.~Ye, L.~Su, J.~Zhang, and C.~Qian, ``Invisibility cloak: Proactive defense against visual game cheating,'' in \emph{33rd USENIX Security Symposium (USENIX Security 24)}.\hskip 1em plus 0.5em minus 0.4em\relax Philadelphia, PA: USENIX Association, Aug. 2024, pp. 3045--3061. [Online]. Available: \url{https://www.usenix.org/conference/usenixsecurity24/presentation/sun-chenxin}
\BIBentrySTDinterwordspacing

\bibitem{10.1145/3372297.3417890}
\BIBentryALTinterwordspacing
S.~Park, A.~Ahmad, and B.~Lee, ``Blackmirror: Preventing wallhacks in 3d online fps games,'' in \emph{Proceedings of the 2020 ACM SIGSAC Conference on Computer and Communications Security}, ser. CCS '20.\hskip 1em plus 0.5em minus 0.4em\relax New York, NY, USA: Association for Computing Machinery, 2020, p. 987–1000. [Online]. Available: \url{https://doi.org/10.1145/3372297.3417890}
\BIBentrySTDinterwordspacing

\bibitem{yeung2006detecting}
S.~Yeung, J.~Lui, J.~Liu, and J.~Yan, ``Detecting cheaters for multiplayer games: theory, design and implementation,'' in \emph{CCNC 2006. 2006 3rd IEEE Consumer Communications and Networking Conference, 2006.}, vol.~2.\hskip 1em plus 0.5em minus 0.4em\relax Las Vegas, NV, USA: {IEEE}, 2006, pp. 1178--1182.

\bibitem{yu2012statistical}
S.-Y. Yu, N.~Hammerla, J.~Yan, and P.~Andras, ``A statistical aimbot detection method for online fps games,'' in \emph{The 2012 International Joint Conference on Neural Networks (IJCNN)}.\hskip 1em plus 0.5em minus 0.4em\relax Brisbane, QLD, Australia: {IEEE}, 2012, pp. 1--8.

\bibitem{6633617}
H.~Alayed, F.~Frangoudes, and C.~Neuman, ``Behavioral-based cheating detection in online first person shooters using machine learning techniques,'' in \emph{2013 IEEE Conference on Computational Inteligence in Games (CIG)}.\hskip 1em plus 0.5em minus 0.4em\relax Niagara Falls, ON, Canada: {IEEE}, Aug 2013, pp. 1--8.

\bibitem{liu2017detecting}
D.~Liu, X.~Gao, M.~Zhang, H.~Wang, and A.~Stavrou, ``Detecting passive cheats in online games via performance-skillfulness inconsistency,'' in \emph{2017 47th Annual IEEE/IFIP International Conference on Dependable Systems and Networks (DSN)}.\hskip 1em plus 0.5em minus 0.4em\relax Denver, CO, USA: {IEEE}, 2017, pp. 615--626.

\bibitem{bleepingcomputer2024}
{BleepingComputer}, ``New intel chips won't play blu-ray disks due to sgx deprecation,'' \url{https://www.bleepingcomputer.com/news/security/new-intel-chips-wont-play-blu-ray-disks-due-to-sgx-deprecation/}, {Bleeping Computer LLC}, 2022, {Accessed: 2024-02-27}.

\bibitem{intelcommunity2024}
{Intel Community}, ``Rising to the challenge: Data security with intel confidential,'' \url{https://community.intel.com/t5/Blogs/Products-and-Solutions/Security/Rising-to-the-Challenge-Data-Security-with-Intel-Confidential/post/1353141}, 2022, {Accessed: 2024-02-27}.

\bibitem{EasyAC}
{Epic Games, Inc.}, ``{Easy Anti-Cheat},'' \url{https://www.easy.ac/}, 2024.

\bibitem{BattlEye}
{BattlEye Innovations}, ``{Interested in using BattlEye?}'' \url{https://www.battleye.com/}, 2024.

\bibitem{SteamSupport}
V.~Corporation, ``Valve anti-cheat (vac) system,'' \url{https://help.steampowered.com/en/faqs/view/571A-97DA-70E9-FF74}, 2023.

\bibitem{mikkelsen2017information}
K.~K. Mikkelsen, ``Information security as a countermeasure against cheating in video games,'' Master's thesis, NTNU, 2017.

\bibitem{9566108}
A.~Maario, V.~K. Shukla, A.~Ambikapathy, and P.~Sharma, ``Redefining the risks of kernel-level anti-cheat in online gaming,'' in \emph{2021 8th International Conference on Signal Processing and Integrated Networks (SPIN)}.\hskip 1em plus 0.5em minus 0.4em\relax Noida, India: {IEEE}, 2021, pp. 676--680.

\bibitem{10.1145/3380786.3391397}
\BIBentryALTinterwordspacing
T.~Witschel and C.~Wressnegger, ``Aim low, shoot high: Evading aimbot detectors by mimicking user behavior,'' in \emph{Proceedings of the 13th European Workshop on Systems Security}, ser. EuroSec '20.\hskip 1em plus 0.5em minus 0.4em\relax New York, NY, USA: Association for Computing Machinery, 2020, p. 19–24. [Online]. Available: \url{https://doi.org/10.1145/3380786.3391397}
\BIBentrySTDinterwordspacing

\bibitem{BackEngineering2021}
{bright, IDontCode, irql0}, ``{EasyAntiCheat Exploit to inject unsigned code into protected processes},'' \url{https://blog.back.engineering/10/08/2021/}, August 2021, {Accessed: 2024-02-29}.

\bibitem{CVE-2019-16098}
CVE-2019-16098, ``{CVE-2019-16098: Vulnerability in MSI Afterburner},'' \url{https://nvd.nist.gov/vuln/detail/CVE-2019-16098}, 2019, {Accessed: 2023-12-07}.

\bibitem{CVE-2020-36603}
CVE-2020-36603, ``{CVE-2020-36603: Vulnerability in Genshin Impact Anti-Cheat Software},'' \url{https://nvd.nist.gov/vuln/detail/CVE-2020-36603}, 2022, {Accessed: 2023-12-07}.

\bibitem{CVE-2021-3437}
CVE-2021-3437, ``{CVE-2021-3437: HP OMEN Gaming Hub Privilege Escalation Bug Hits Millions of Gaming Devices},'' \url{https://www.sentinelone.com/labs/cve-2021-3437-hp-omen-gaming-hub-privilege-escalation-bug-hits-millions-of-gaming-devices/}, 2021, {Accessed: 2023-12-07}.

\bibitem{CVE-2023-38817}
CVE-2023-38817, ``{CVE-2023-38817: Vulnerability in Minecraft Anti-Cheat Tool},'' \url{https://nvd.nist.gov/vuln/detail/CVE-2023-38817}, 2023, {Accessed: 2023-12-07}.

\bibitem{gaffer2016nevertrust}
\BIBentryALTinterwordspacing
G.~on~Games, ``Never trust the client,'' 2016, accessed: 2024-12-20. [Online]. Available: \url{https://web.archive.org/web/20170510170653/https://gafferongames.com/2016/04/25/never-trust-the-client/}
\BIBentrySTDinterwordspacing

\bibitem{guardian2016hackerscheats}
\BIBentryALTinterwordspacing
K.~Stuart, ``How hackers and cheats ruined the division on pc,'' \emph{The Guardian}, 2016, accessed: 2024-12-20. [Online]. Available: \url{https://www.theguardian.com/technology/2016/apr/26/hackers-cheats-ruined-the-division-pc-ubisoft}
\BIBentrySTDinterwordspacing

\bibitem{i3d2024counteringcheating}
\BIBentryALTinterwordspacing
i3D.net, ``Countering the scourge of cheating in games,'' 2024, accessed: 2024-12-20. [Online]. Available: \url{https://www.i3d.net/countering-scourge-of-cheating-in-games/}
\BIBentrySTDinterwordspacing

\bibitem{orlova2024estimation}
J.~Orlova, A.~Stepanov, A.~Vinogradov, L.~Orlova, A.~Baldycheva, and A.~Somov, ``Estimation of in-game player behavior by measuring key gameplay parameters,'' in \emph{2024 IEEE International Instrumentation and Measurement Technology Conference (I2MTC)}.\hskip 1em plus 0.5em minus 0.4em\relax IEEE, 2024, pp. 1--6.

\bibitem{han2015online}
M.~L. Han, J.~K. Park, and H.~K. Kim, ``Online game bot detection in fps game,'' in \emph{Proceedings of the 18th Asia Pacific Symposium on Intelligent and Evolutionary Systems-Volume 2}.\hskip 1em plus 0.5em minus 0.4em\relax Springer, 2015, pp. 479--491.

\bibitem{zleague_warzone_anticheat}
{JARVIS THE NPC}, ``{Warzone Reddit Outcry: Anti-Cheat Failures Fuel Frustration Among Players},'' \url{https://www.zleague.gg/theportal/warzone-reddit-outcry-anti-cheat-failures-fuel-frustration-among-players/}, April 2024, {Accessed: 2024-04-24}.

\bibitem{pushtotalk_anticheat}
R.~K. RIGNEY, ``The gamers do not understand anti-cheat,'' \url{https://www.pushtotalk.gg/p/the-gamers-do-not-understand-anti-cheat}, Febuary 2024, {Accessed: 2024-04-24}.

\bibitem{algshack2024}
B.~Toulas, ``{Apex Legends players worried about RCE flaw after ALGS hacks},'' \url{https://www.bleepingcomputer.com/news/security/apex-legends-players-worried-about-rce-flaw-after-algs-hacks/}, March 2024, {Accessed: 2024-04-08}.

\bibitem{RiotGamesInc}
R.~G. Inc., ``What is vanguard?'' \url{https://support-valorant.riotgames.com/hc/en-us/articles/360046160933-What-is-Vanguard-}, Jun 2022.

\bibitem{WELLBIA}
L.~WELLBIA.COM~CO., ``Wellbia,'' \url{https://wellbia.com/}, 2023.

\bibitem{10.1145/3456631}
\BIBentryALTinterwordspacing
S.~Fei, Z.~Yan, W.~Ding, and H.~Xie, ``Security vulnerabilities of sgx and countermeasures: A survey,'' \emph{ACM Comput. Surv.}, vol.~54, no.~6, jul 2021. [Online]. Available: \url{https://doi.org/10.1145/3456631}
\BIBentrySTDinterwordspacing

\bibitem{9774028}
A.~Kanervisto, T.~Kinnunen, and V.~Hautamäki, ``Gan-aimbots: Using machine learning for cheating in first person shooters,'' \emph{IEEE Transactions on Games}, vol.~15, no.~4, pp. 566--579, 2023.

\bibitem{inproceedings}
L.~Galli, D.~Loiacono, L.~Cardamone, and P.~L. Lanzi, ``A cheating detection framework for unreal tournament iii: A machine learning approach,'' in \emph{2011 IEEE Conference on Computational Intelligence and Games (CIG'11)}.\hskip 1em plus 0.5em minus 0.4em\relax Seoul, Korea (South): {IEEE}, 2011, pp. 266--272.

\bibitem{afkgaming}
{AFK Gaming}, ``{8 types of hacks and cheats in CS:GO},'' \url{https://afkgaming.com/csgo/news/2619-8-types-of-hacks-and-cheats-in-csgo}, 2019.

\bibitem{consalvo2007cheating}
M.~Consalvo, \emph{Cheating: Gaining Advantage in Videogames}.\hskip 1em plus 0.5em minus 0.4em\relax Cambridge, MA: MIT Press, 2007.

\bibitem{LSTM1997Hochreiter}
S.~Hochreiter and J.~Schmidhuber, ``Long short-term memory,'' \emph{Neural Computation}, vol.~9, no.~8, pp. 1735--1780, 1997.

\bibitem{10.5555/2627435.2670313}
N.~Srivastava, G.~Hinton, A.~Krizhevsky, I.~Sutskever, and R.~Salakhutdinov, ``Dropout: A simple way to prevent neural networks from overfitting,'' \emph{J. Mach. Learn. Res.}, vol.~15, no.~1, p. 1929–1958, jan 2014.

\bibitem{luong2015effective}
M.-T. Luong, H.~Pham, and C.~D. Manning, ``Effective approaches to attention-based neural machine translation,'' 2015.

\bibitem{haykin1994neural}
S.~Haykin, \emph{Neural networks: a comprehensive foundation}.\hskip 1em plus 0.5em minus 0.4em\relax USA: Prentice Hall PTR, 1994.

\bibitem{cramer2002origins}
J.~S. Cramer, ``The origins of logistic regression,'' Tinbergen Institute, Tinbergen Institute Discussion Papers 02-119/4, 2002.

\bibitem{breiman2001random}
L.~Breiman, ``Random forests,'' \emph{Machine learning}, vol.~45, pp. 5--32, 2001.

\bibitem{cortes1995support}
C.~Cortes and V.~Vapnik, ``Support-vector networks,'' \emph{Machine learning}, vol.~20, pp. 273--297, 1995.

\bibitem{murphy2006naive}
K.~P. Murphy \emph{et~al.}, ``Naive bayes classifiers,'' \emph{University of British Columbia}, vol.~18, no.~60, pp. 1--8, 2006.

\bibitem{john2013estimating}
G.~H. John and P.~Langley, ``Estimating continuous distributions in bayesian classifiers,'' in \emph{Proceedings of the Eleventh Conference on Uncertainty in Artificial Intelligence}, ser. UAI'95.\hskip 1em plus 0.5em minus 0.4em\relax San Francisco, CA, USA: Morgan Kaufmann Publishers Inc., 1995, p. 338–345.

\bibitem{srivastava2007bayesian}
\BIBentryALTinterwordspacing
S.~Srivastava, M.~R. Gupta, and B.~A. Frigyik, ``Bayesian quadratic discriminant analysis,'' \emph{Journal of Machine Learning Research}, vol.~8, no.~46, pp. 1277--1305, 2007. [Online]. Available: \url{http://jmlr.org/papers/v8/srivastava07a.html}
\BIBentrySTDinterwordspacing

\bibitem{awpy}
P.~Xenopoulos, ``{awpy: A Python package for Counter-Strike: Global Offensive data parsing, analytics, and visualization},'' \url{https://awpy.readthedocs.io/en/latest/}, 2023, {Accessed: 2025-02-21}.

\bibitem{elomaa2001analysis}
T.~Elomaa and M.~Kaariainen, ``An analysis of reduced error pruning,'' \emph{Journal of Artificial Intelligence Research}, vol.~15, pp. 163--187, 2001.

\bibitem{gora2002riona}
G.~G{\'o}ra and A.~Wojna, ``Riona: A classifier combining rule induction and k-nn method with automated selection of optimal neighbourhood,'' in \emph{Machine Learning: ECML 2002: 13th European Conference on Machine Learning Helsinki, Finland, August 19--23, 2002 Proceedings 13}.\hskip 1em plus 0.5em minus 0.4em\relax Berlin, Heidelberg: Springer, 2002, pp. 111--123.

\bibitem{yates2019spaarc}
D.~Yates, M.~Z. Islam, and J.~Gao, ``Spaarc: a fast decision tree algorithm,'' in \emph{Data Mining: 16th Australasian Conference, AusDM 2018, Bahrurst, NSW, Australia, November 28--30, 2018, Revised Selected Papers 16}.\hskip 1em plus 0.5em minus 0.4em\relax Singapore: Springer, 2019, pp. 43--55.

\bibitem{Breiman1996}
L.~Breiman, ``Bagging predictors,'' \emph{Machine Learning}, vol.~24, no.~2, pp. 123--140, 1996.

\bibitem{demiroz1997classification}
G.~Demir{\"o}z and H.~A. G{\"u}venir, ``Classification by voting feature intervals,'' in \emph{European Conference on Machine Learning}.\hskip 1em plus 0.5em minus 0.4em\relax Berlin, Heidelberg: Springer, 1997, pp. 85--92.

\bibitem{Cohen1995}
W.~W. Cohen, ``Fast effective rule induction,'' in \emph{Proceedings of the Twelfth International Conference on International Conference on Machine Learning}, ser. ICML'95.\hskip 1em plus 0.5em minus 0.4em\relax San Francisco, CA, USA: Morgan Kaufmann Publishers Inc., 1995, p. 115–123.

\bibitem{stefanowski1998rough}
\BIBentryALTinterwordspacing
J.~Stefanowski, \emph{On Combined Classifiers, Rule Induction and Rough Sets}.\hskip 1em plus 0.5em minus 0.4em\relax Berlin, Heidelberg: Springer, 2007, pp. 329--350. [Online]. Available: \url{https://doi.org/10.1007/978-3-540-71200-8_18}
\BIBentrySTDinterwordspacing

\bibitem{siers2015software}
M.~J. Siers and M.~Z. Islam, ``Software defect prediction using a cost sensitive decision forest and voting, and a potential solution to the class imbalance problem,'' \emph{Information Systems}, vol.~51, pp. 62--71, 2015.

\bibitem{forum_serverreq}
DatHost, ``{CS2 Server Hosting},'' \url{https://dathost.net/cs2-server-hosting}, 2024, {Accessed: 2024-06-05}.

\bibitem{steamdb2024}
SteamDB, ``{Counter-Strike 2 Steam Charts},'' \url{https://steamdb.info/app/730/charts/#48h}, 2024, {Accessed: 2024-06-05}.

\bibitem{ParliRobo}
\BIBentryALTinterwordspacing
J.~Zheng, C.~Xiao, M.~Li, Z.~Li, F.~Qian, W.~Liu, and X.~Wu, ``Parlirobo: Participant lightweight ai robots for massively multiplayer online games (mmogs),'' in \emph{Proceedings of the 31st ACM International Conference on Multimedia}, ser. MM '23.\hskip 1em plus 0.5em minus 0.4em\relax New York, NY, USA: Association for Computing Machinery, 2023, p. 9093–9102. [Online]. Available: \url{https://doi.org/10.1145/3581783.3613764}
\BIBentrySTDinterwordspacing

\bibitem{time_linux}
M.~Kerrisk, ``{time(1) — Linux manual page},'' \url{https://man7.org/linux/man-pages/man1/time.1.html}, 2024, {Accessed: 2024-06-05}.

\bibitem{tactical_shooter}
Wikipedia, ``Tactical shooter,'' \url{https://en.wikipedia.org/wiki/Tactical_shooter}, 2024, {Accessed: 2024-06-14}.

\bibitem{9772248}
J.~Tao, Y.~Xiong, S.~Zhao, R.~Wu, X.~Shen, T.~Lyu, C.~Fan, Z.~Hu, S.~Zhao, and G.~Pan, ``Explainable ai for cheating detection and churn prediction in online games,'' \emph{IEEE Transactions on Games}, vol.~15, no.~2, pp. 242--251, 2023.

\bibitem{bootstrap_resample}
\BIBentryALTinterwordspacing
B.~Efron, ``Bootstrap methods: Another look at the jackknife,'' \emph{The Annals of Statistics}, vol.~7, no.~1, pp. 1--26, 1979. [Online]. Available: \url{http://www.jstor.org/stable/2958830}
\BIBentrySTDinterwordspacing

\bibitem{cupid}
\BIBentryALTinterwordspacing
G.~Fan, C.~Zhang, K.~Wang, Y.~Li, J.~Chen, and Z.~Xu, ``Cupid: Improving battle fairness and position satisfaction in online moba games with a re-matchmaking system,'' \emph{Proc. ACM Hum.-Comput. Interact.}, vol.~8, no. CSCW2, Nov. 2024. [Online]. Available: \url{https://doi.org/10.1145/3686978}
\BIBentrySTDinterwordspacing

\bibitem{KrupinskiOsiris}
D.~Krupiński, ``Osiris,'' \url{https://github.com/danielkrupinski/Osiris}, 2023, {Accessed: 2024-02-27}.

\bibitem{miketendo64aimbots2021}
{Mike Scorpio}, ``{What You Should Know About Aimbots Before Using Them in a Video Game},'' \url{https://miketendo64.com/2021/01/07/what-you-should-know-about-aimbots-before-using-them-in-a-video-game/}, January 2021, {Accessed: 2024-06-13}.

\bibitem{mediumaimbotcomplaints2021}
m2qchqo110, ``{The Most Hilarious Complaints We've Heard About PUBG Aimbot},'' \url{https://medium.com/@p5hvlpr124/the-most-hilarious-complaints-weve-heard-about-pubg-aimbot-b596bbcb1ce5}, August 2019, {Accessed: 2024-06-13}.

\bibitem{Silva_2022}
J.~N. Silva, ``Towards automated server-side video game cheat detection,'' \url{https://repositorio-aberto.up.pt/bitstream/10216/142935/2/572983.pdf}, July 2022.

\bibitem{lehtonen2020comparative}
S.~Lehtonen \emph{et~al.}, ``Comparative study of anti-cheat methods in video games,'' \emph{University of Helsinki, Faculty of Science}, 2020.

\bibitem{vespa}
S.~Zhao, J.~Qi, Z.~Hu, H.~Yan, R.~Wu, X.~Shen, T.~Lv, and C.~Fan, ``Vespa: A general system for vision-based extrasensory perception anticheating in online fps games,'' \emph{IEEE Transactions on Games}, vol.~16, no.~3, pp. 611--620, 2024.

\bibitem{fcd}
\BIBentryALTinterwordspacing
Y.~Su, D.~Yao, X.~Chu, W.~Li, J.~Bi, S.~Zhao, R.~Wu, S.~Zhang, J.~Tao, and H.~Deng, ``Few-shot learning for trajectory-based mobile game cheating detection,'' in \emph{Proceedings of the 28th ACM SIGKDD Conference on Knowledge Discovery and Data Mining}, ser. KDD '22.\hskip 1em plus 0.5em minus 0.4em\relax New York, NY, USA: Association for Computing Machinery, 2022, p. 3941–3949. [Online]. Available: \url{https://doi-org.eproxy.lib.hku.hk/10.1145/3534678.3539157}
\BIBentrySTDinterwordspacing

\end{thebibliography}
% \newpage 
% \appendix
% % \appendixpage
% \pagenumbering{arabic}  % 重新设置页码格式为阿拉伯数字
% \setcounter{page}{1}  % 让页码从 1 开始
% \markboth{IEEE TRANSACTIONS ON INFORMATION FORENSICS AND SECURITY, Supplemental Material, March 2025}{Supplemental Material}
% \maketitle  % 重新插入标题
% \input{Appendix/ExSPC_Training_Loss}
% % \input{Appendix/RevPOV_Model_Selection_Preliminary_Test}
% % \input{Appendix/RevStats_Model_Selection}
% \input{Appendix/Structured_Features}
% \input{Appendix/Temporal_Features}
% \input{Appendix/Sense_and_Performance_Classification}
% \input{Appendix/PCA_Visualization}
% \input{Appendix/Metrics}
\end{document}

\title{Identify As A Human Does: A Pathfinder of Next-Generation Anti-Cheat Framework for First-Person Shooter Games}

\author{Jiayi Zhang\orcidlink{0009-0001-4452-6464}, Chenxin Sun\orcidlink{0009-0001-5057-6089}, Yue Gu\orcidlink{0009-0005-7827-7098}, Qingyu Zhang\orcidlink{0009-0009-4422-3971}, Jiayi Lin\orcidlink{0009-0004-6790-1302}, Xiaojiang Du\orcidlink{0000-0003-4235-9671}, Chenxiong Qian\orcidlink{0000-0002-6201-6011}}

\markboth{IEEE TRANSACTIONS ON INFORMATION FORENSICS AND SECURITY,~Vol.~X, No.~Y, March~2025}%
{Identify As A Human Does: A Pathfinder of Next-Generation Anti-Cheat Framework for First-Person Shooter Games}

\newcommand\CQ[1]{\textbf{\textcolor{purple}{CQ: #1}}}
\newcommand\Jiayi[1]{\textbf{\textcolor{orange}{Jiayi: #1}}}
\newcommand\Du[1]{\textbf{\textcolor{brown}{#1}}}

\newcommand{\syspov}{\textsc{RevPov}\xspace}
\newcommand{\sysstats}{\textsc{RevStats}\xspace}
\newcommand{\sysinteg}{\textsc{Mvin}\xspace}
\newcommand{\sysspc}{\textsc{ExSPC}\xspace}
\newcommand{\sys}{\textsc{Hawk}\xspace}

\maketitle
\markboth{IEEE TRANSACTIONS ON INFORMATION FORENSICS AND SECURITY, Supplemental Material, March 2025}{Supplemental Material}
\maketitle  % 重新插入标题
\subsection{Ethics Considerations}\label{sec:ethical}
The dataset only contains publicly accessible information including in-game operations and data, and the banning information.
Its publication ensures no direct or indirect harm to the game, its distributors, players, or the community. The data provider has approved its release and reuse for research purposes, and no private information will be disclosed. 
Still, we anonymized all information to minimize the probability of causing any unforeseen privacy concerns.
A data usage agreement will require users to commit to ethical, academic, and non-malicious use, with clear guidelines provided on ethical use and potential risks before using. 
The study adheres to local legislation and institutional requirements. 
\begin{table}[htbp]
\centering
\caption{Experts' skill level for sampled match review, all stats are retrieved from \cite{tracker}.}
\label{tab:expert_skill_level}
\resizebox{\columnwidth}{!}{%
\begin{tabular}{@{}ccl@{}}
\toprule
\multicolumn{1}{c}{\textbf{Expert}} & \multicolumn{1}{c}{\textbf{Playtime}}      & \multicolumn{1}{c}{\textbf{Skill Level}}                                                                                                              \\ \midrule
\#1               & \textgreater{}7,000hrs & \begin{tabular}[c]{@{}l@{}}CS:GO - \textit{Premier} 19,000 (Top 2.62\%)\\ Overwatch2 - \textit{Role} Grand Master (Top 1.46\%)\\ Apex Legend - \textit{BR} Master (Top 0.6\%)\end{tabular} \\
\#2               & \textgreater{}1,400hrs & CS:GO - \textit{Premier} 19,200 (Top 2.62\%) \\
\#3               & \textgreater{}5,000hrs & CS:GO - \textit{Premier} 18,000 (Top 4.59\%) \\ \bottomrule
\end{tabular}%
}
\end{table}
\subsection{Training and Validation Loss for \sysspc Model}
\label{appx:ExSPC_loss}
\begin{figure}[htbp]
\centering
\begin{subfigure}[b]{0.48\textwidth}
    \centering
    \includegraphics[width=\textwidth]{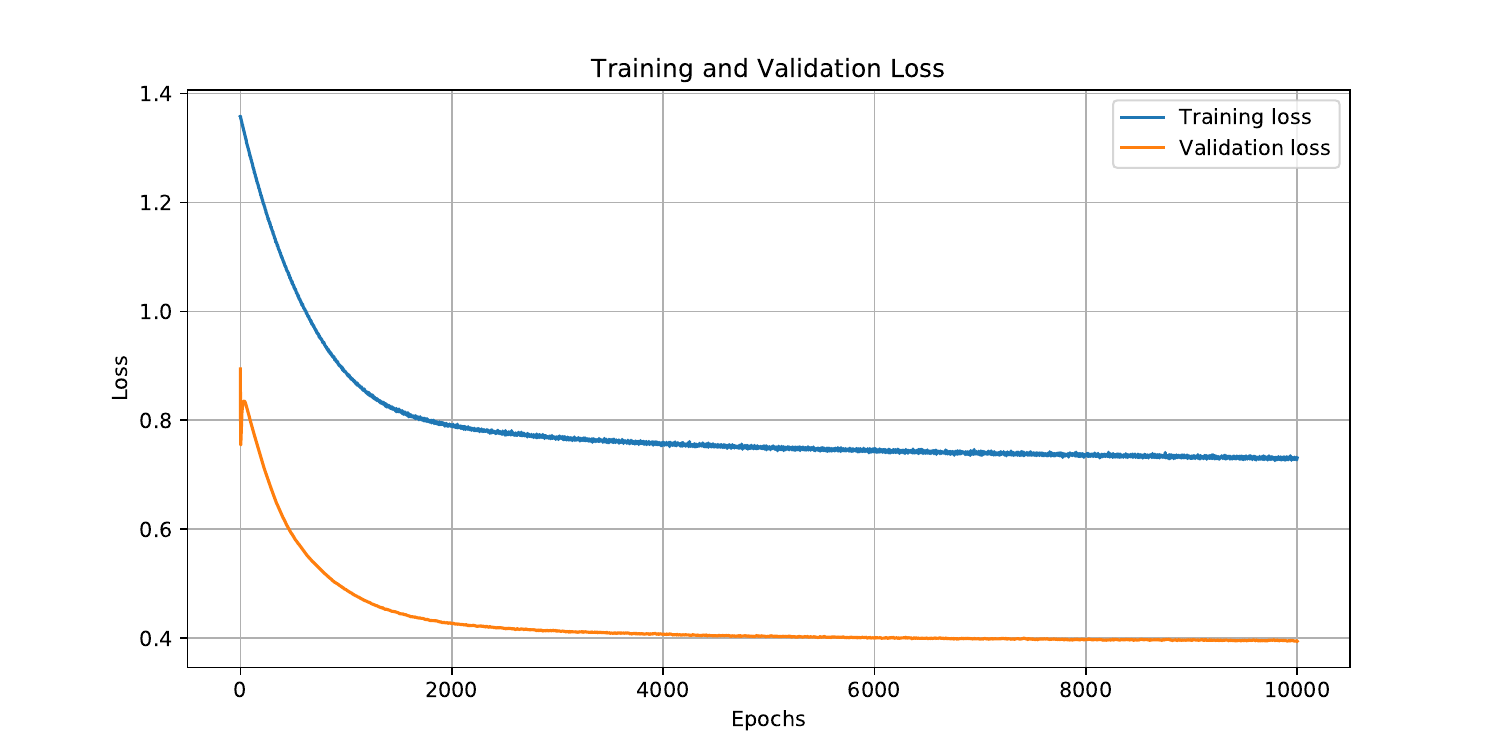}
    \caption{\textit{Wallhack} dataset}
    \label{fig:loss_wallhack}
\end{subfigure}
\hfill
\begin{subfigure}[b]{0.48\textwidth}
    \centering
    \includegraphics[width=\textwidth]{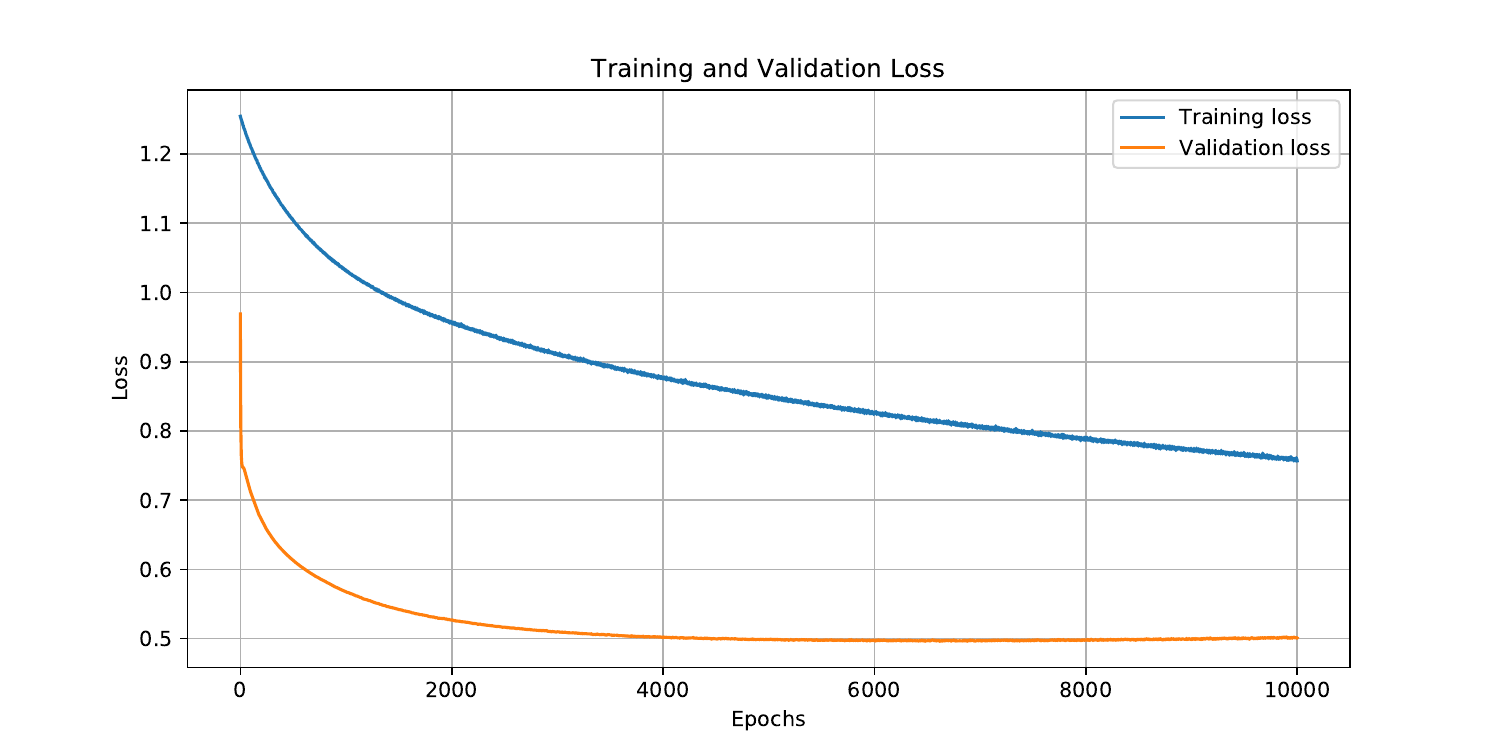}
    \caption{\textit{Aimbot} dataset}
    \label{fig:loss_aimbot}
\end{subfigure}
\caption{Training and validation loss of \sysspc on two datasets: \textit{Wallhack} (a) and \textit{Aimbot} (b).}
% \Description{Training and validation loss of \sysspc on two datasets: \textit{Wallhack} (a) and \textit{Aimbot} (b).}
\label{fig:loss_combined}
\end{figure}

The training process on each of the datasets in \sysspc utilizes the validation set for optimization.
\autoref{fig:loss_wallhack} and \autoref{fig:loss_aimbot} demonstrate the training and validation loss for each dataset, respectively.
It is noticeable that the performance result of \sysspc is not as ideal as \sys, although it utilizes both \syspov and \sysstats’s inputs. 
And thereby prove the need for all three subsystems as well as the \textit{TSTO}.
Thus, we use the multi-view design to complement one another of the subsystems.
% \begin{figure}[htbp]
% \centering
% \includegraphics[width=0.47\textwidth]{Figures/Training_Validation_Loss_Wallhack.pdf}
% \caption{Training and validation loss of \sysspc on \textit{Wallhack} dataset.}
% % \Description{Training and validation loss of \sysspc on \textit{Wallhack} dataset.}
% \label{fig:loss_wallhack}
% \end{figure}
% \begin{figure}[htbp]
% \centering
% \includegraphics[width=0.47\textwidth]{Figures/Training_Validation_Loss_Aimbot.pdf}
% \caption{Training and validation loss of \sysspc on \textit{Aimbot} dataset.}
% % \Description{Training and validation loss of \sysspc on \textit{Aimbot} dataset.}
% \label{fig:loss_aimbot}
% \end{figure}

\subsection{Structured Features}
\label{appx:structuredfeatures}
In this section, the extracted data from demos are further processed to create strong-expressiveness features that can indicate the sense and performance of a player clearly. For notations used in the formulas throughout this section, please refer to \autoref{tab:univ-compa}.
[S] denotes that the feature is a subgenre-specific feature, and [C] denotes that the feature is an FPS core feature metioned in Section VII.

\begin{table*}[t]
    \centering
    \small % or \footnotesize
    \caption{Notations and descriptions of structured features.}
    \label{tab:univ-compa}
    \begin{tabular}{ccccc}
    \toprule
        \textbf{Notation} & Description & \textbf{Notation} & Description \\
    \midrule
        $i$ & the current player &
        
        $\textbf{E}_{\textit{x}}^{a \rightarrow b}[k]$ & the $k^{th}$ type \textit{x} event conducted by player \textit{a} to \textit{b} \\
        
        $p, p^{*}$ & a particular player, any player &
        $\#(\bullet)$ & the number of counts of the parameter \\
        
        $o, o^{*}$ & a particular opponent, any opponent &
        $wp(\bullet)$ & the weapon used by a player \\
        
        $a, a^{*}$ & a particular ally, any ally &
        $hg(\bullet)$ & the body part hit by a player \\
        
        $ts$ & through smoke &
        $\textbf{\textit{t}}(\bullet)$ & the moment of the event occurs in second \\
        
        $p(\bullet)$ & the number of occluders penetrated &
        $mag(\bullet)$ & the number of bullets remain in a player's magazine \\
        
        $\Delta(\bullet)$ & the change in the parameter &
        $dist(x, y)$ & the distance range between entities x and y \\
        
        $\textbf{ELAPSE}(\bullet)$ & the duration from the start to the end of an event &
        $\textbf{HEG}(\bullet)$ & high-explosive grenade is applied by a player \\
        
        $\textbf{INC}(\bullet)$ & incendiary or Molotov is applied by a player &
        $\textbf{SMK}(\bullet)$ & smoke grenade is applied by a player \\
        
        $\textbf{FLS}(\bullet)$ & flash grenade is applied by a player &
        $\textbf{FK}(\bullet)$ & first kill is accomplished by a player \\
    \bottomrule
    \end{tabular}
\end{table*}

\begin{figure*}[htbp]
\centering
\includegraphics[width=1\textwidth]{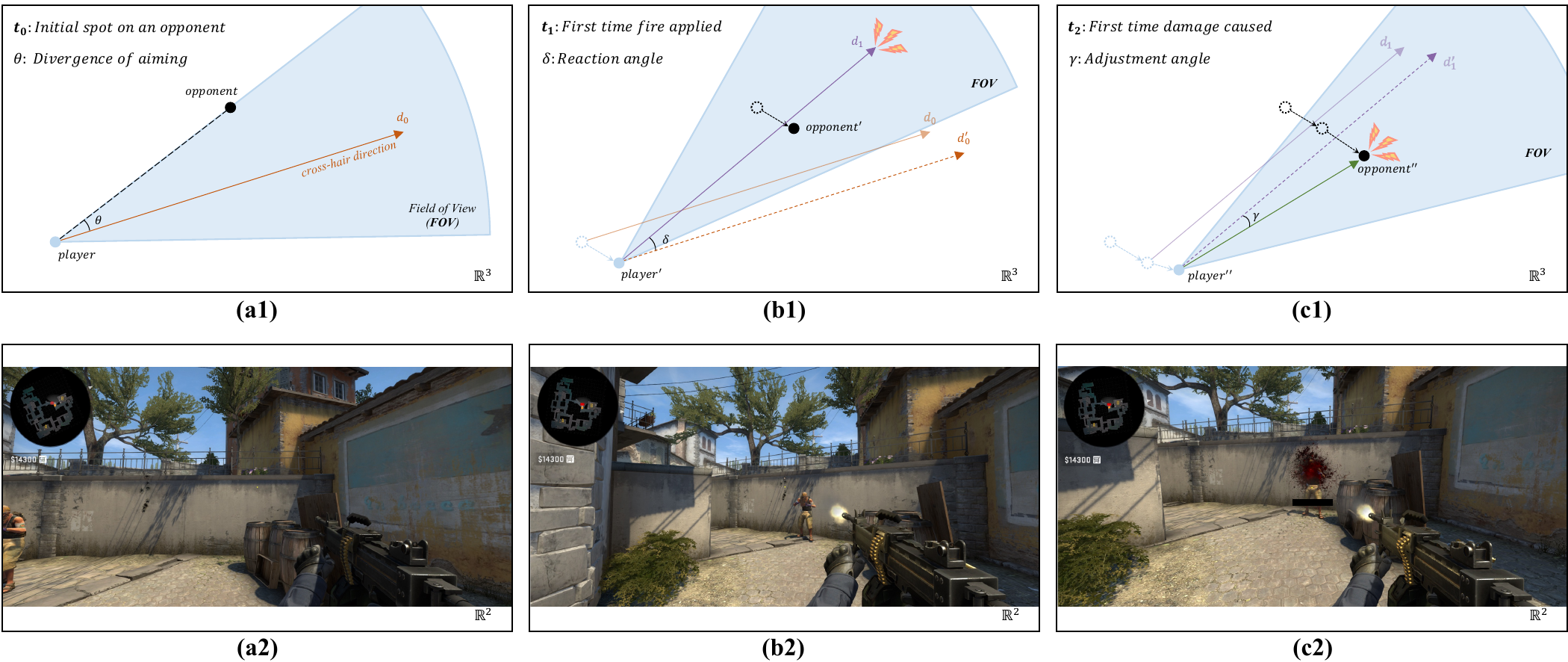}
\caption{Illustrations of reaction and adjustment features. The black rectangle overlay in c2 hides the player's ID.}
% \Description{Illustrations of reaction and adjustment features. The black rectangle overlay in c2 hides the player's ID.}
\label{fig:feature}
\end{figure*}

\subsubsection{Aiming features}

\paragraph*{Durations}

\autoref{fig:feature} illustrates the three distinct moments that form three different durations. In terms of the moments, $\textbf{\textit{t}}_{0}$ denotes the moment that the current player initially spots one opponent.  In \autoref{fig:feature}, a1 and a2 present a possible set of states at $\textbf{\textit{t}}_{0}$. In which a1 represents a 3-dimensional world in the game map, where the opponent happens to appear at the edge of the current player's field of view (FOV). And a2 demonstrates the 2-dimensional state correspondingly which is typically displayed through the player's monitor, where the opponent also happens to be spotted at the edge of the current player's POV. $\textbf{\textit{t}}_{1}$ denotes the moment that the current player notices the hostile movement and initiates responsive action (\textit{i.e.}, open fire). Note that at $\textbf{\textit{t}}_{1}$ the current firing event may or may not be equivalent to a damage event upon the opponent. $\textbf{\textit{t}}_{2}$ denotes the moment that the current player performs a damage event upon the opponent within one period of valid engagement for the first time. In our case, the threshold for resetting the valid engagement period detection (\textit{i.e.,} no opponent in sight, no damage or firing event) is set as 10 seconds. In \autoref{fig:feature}, it demonstrates a series of generic circumstances that often happen in games where the positions of the current player and the opponent vary along from $\textbf{\textit{t}}_{0}$ to $\textbf{\textit{t}}_{2}$. And it is common that sometimes $\textbf{\textit{t}}_{1}$ and $\textbf{\textit{t}}_{2}$ are overlapped. In terms of durations, the elapsed period from $\textbf{\textit{t}}_{0}$ to $\textbf{\textit{t}}_{1}$ is defined as \textit{reaction time (rat)}. The elapsed period from $\textbf{\textit{t}}_{1}$ to $\textbf{\textit{t}}_{2}$ is defined as \textit{adjustment time (ajt)}. The elapsed period from $\textbf{\textit{t}}_{0}$ to $\textbf{\textit{t}}_{2}$ is defined as \textit{duration time (drt)}. The average and variance calculations are applied after obtaining the data. The average performance of all samples is obtained through the average, and the degree of dispersion of the samples is represented by the variance. The integration of average and variance is able to represent the overall performance and stability of the current player within a match.

\textit{rat average.} [C] Denotes the average performance regarding \textit{rat} of the current player among one match. The lower this feature is, the faster the player reacts.

{\normalsize
\begin{equation}\mathbf{F_{\textbf{rat-avg}}(i)}=\overline{RAT}=\frac{\sum_{j=1}^{N}RAT_{j}}{N}\end{equation}
}
\begin{align*}
\text{where } N \text{ denotes } \#(RAT) \text{ within a match.}
\end{align*}

\textit{rat variance.} [C] Denotes the degree of dispersion of \textit{rat} distributions of the current player among one match. 
{\normalsize
\begin{equation}\mathbf{F_{\textbf{rat-var}}(i)}=Var(RAT)=\frac{\sum_{j=1}^{N}\left(RAT_{j}-\overline{RAT}\right)^{2}}{N}\end{equation}
}
\begin{align*}
\text{where } N \text{ denotes } \#(RAT) \text{ within a match.}
\end{align*}

\textit{ajt average.} [C] Denotes the average performance regarding \textit{ajt} of the current player among one match. The lower this feature is, the faster the player adjusts.

{\normalsize
\begin{equation}\mathbf{F_{\textbf{ajt-avg}}(i)}=\overline{AJT}=\frac{\sum_{j=1}^{N}AJT_{j}}{N}\end{equation}
}
\begin{align*}
\text{where } N \text{ denotes } \#(AJT) \text{ within a match.}
\end{align*}

\textit{ajt variance.} [C] Denotes the degree of dispersion of \textit{ajt} distributions of the current player among one match.

{\normalsize
\begin{equation}\mathbf{F_{\textbf{ajt-var}}(i)}=Var(AJT)=\frac{\sum_{j=1}^{N}\left(AJT_{j}-\overline{AJT}\right)^{2}}{N}\end{equation}
}
\begin{align*}
\text{where } N \text{ denotes } \#(AJT) \text{ within a match.}
\end{align*}

\textit{drt average.} [C] Denotes the average performance regarding \textit{drt} of the current player among one match. The lower this feature is, the faster the player hits.

{\normalsize
\begin{equation}
\mathbf{F_{\textbf{drt-avg}}(i)}=\overline{DRT}=\frac{\sum_{j=1}^{N}DRT_{j}}{N}
\end{equation}
}

\textit{drt variance.} [C] Denotes the degree of dispersion of \textit{drt} distributions of the current player among one match.

{\normalsize
\begin{equation}
\mathbf{F_{\textbf{drt-var}}(i)}=Var(DRT)=\frac{\sum_{j=1}^{N}\left(DRT_{j}-\overline{DRT}\right)^{2}}{N}
\end{equation}
}
\begin{align*}
\text{where } N \text{ denotes } \#(DRT) \text{ occurrences within a match.}
\end{align*}
\begin{figure*}[htbp]
\centering
\includegraphics[width=1\textwidth]{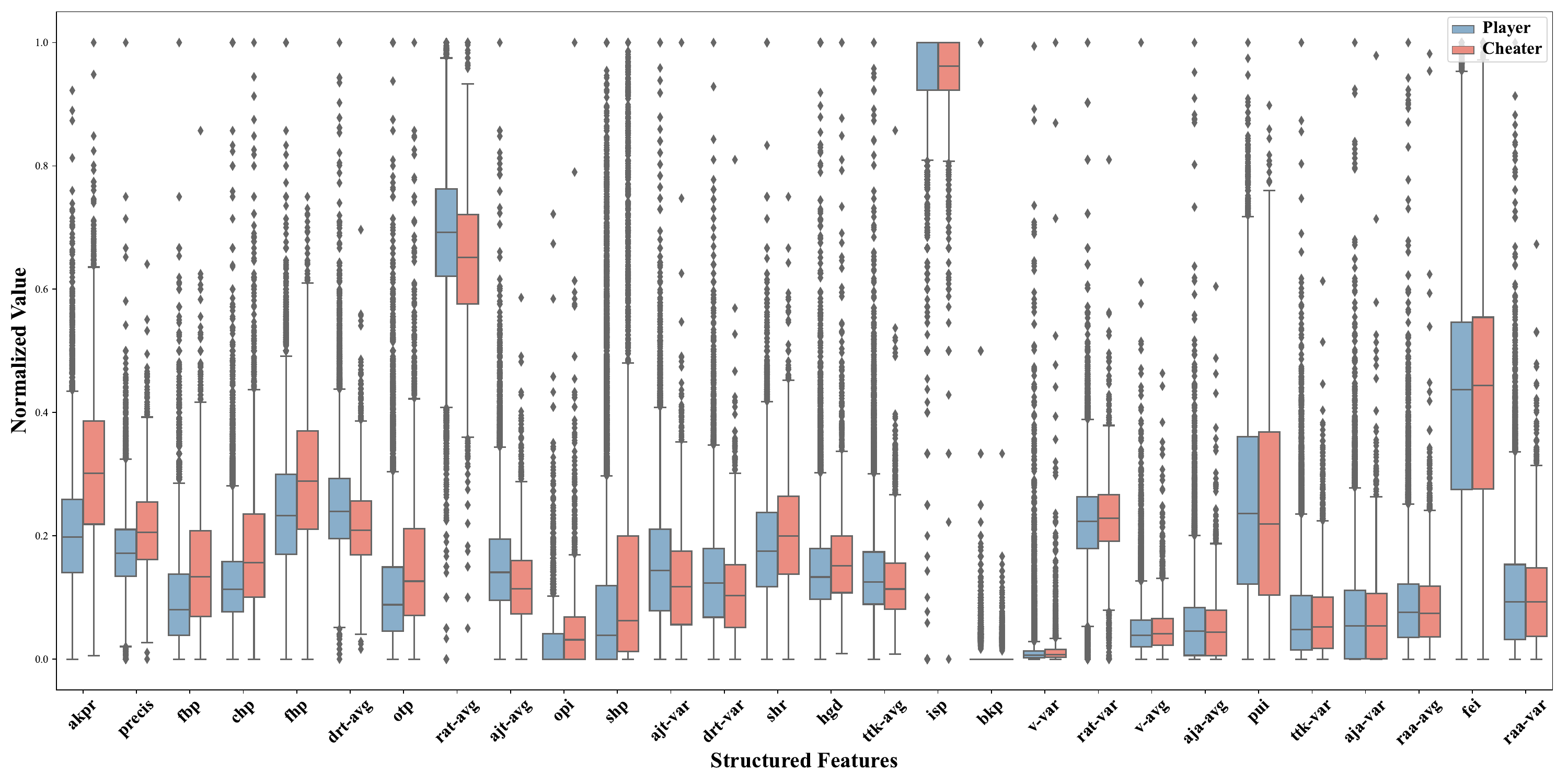}
\caption{Complete box plots of comparative visualization between honest players and cheaters on all structured features under Mann-Whitney U (Wilcoxon Rank-Sum) test with P-Value ascending (distinction descending) order.}
% \Description{Complete box plots of comparative visualization between honest players and cheaters on all structured features under Mann-Whitney U (Wilcoxon Rank-Sum) test with P-Value ascending (distinction descending) order.}
\label{fig:structured_boxplot_all}
\end{figure*}
\paragraph*{Velocity}
\textit{velocity average.} [C] Denotes the average mouse velocity of the current player among one match between $\textbf{\textit{t}}_{0}$ and $\textbf{\textit{t}}_{2}$, through which way two features of distance and time period can be highly compressed into one feature. The higher the feature is, the faster the player hits the targets.

{\normalsize
\begin{equation}V=\frac{dist(i, o)}{DRT_{i}}\end{equation}
}
\begin{align*}
\text{where } o \text{ denotes the victim in the specified } DRT\text{.}
\end{align*}
{\normalsize
\begin{equation}\mathbf{F_{\textbf{v-avg}}(i)}=\overline{V}=\frac{\sum_{j=1}^{N}V_{j}}{N}\end{equation}
}
\begin{align*}
\text{where } N \text{ denotes } \#(DRT) \text{ within a match.}
\end{align*}

\textit{velocity variance.} [C] Denotes the degree of dispersion of \textit{velocity} distributions of the current player among one match. 

{\normalsize\begin{equation}\mathbf{F_{\textbf{v-var}}(i)}=Var(V)=\frac{\sum_{j=1}^{N}\left(V_{j}-\overline{V}\right)^{2}}{N}\end{equation}
}
\begin{align*}
\text{where } N \text{ denotes } \#(DRT) \text{ within a match.}
\end{align*}

\paragraph*{Angles}
\autoref{fig:feature} illustrates the three different angles, in which $\theta$ denotes \textit{divergence of aiming} is extracted by Liu et al. \cite{liu2017detecting} and represents the initial angular divergence between the aiming position and the opponent position. However, the time complexity for obtaining such a feature is too expensive to be accepted for industrial implementation. Therefore we extracted the following features which not only preserve strong expressiveness but also significantly reduce the time cost, as well as innovatively segment each engagement process into three stages. $\delta$ represents the \textit{reaction angle (raa)}, which reflects the angle of two cross-hair directions at $\textbf{\textit{t}}_{0}$ and $\textbf{\textit{t}}_{1}$ respectively. $\gamma$ represents the \textit{adjustment angle (aja)}, which, similarly reflects the angle of two cross-hair directions at $\textbf{\textit{t}}_{1}$ and $\textbf{\textit{t}}_{2}$ respectively. The larger $\delta$ and $\gamma$ are the longer distance the cross-hair travels, which ultimately results in performance divergence. Similarly, the average and variance operation is applied accordingly.

\textit{raa average.} [C] Denotes the average performance regarding \textit{raa} of the current player among one match. The lower the feature is, the smaller the angle and the less effort the player aims for and puts into reaching their targets.

{\normalsize\begin{equation}\mathbf{F_{\textbf{raa-avg}}(i)}=\overline{RAA}=\frac{\sum_{j=1}^{N}RAA_{j}}{N}\end{equation}
}
\begin{align*}
\text{where } N \text{ denotes } \#(RAA) \text{ within a match.}
\end{align*}
\textit{raa variance.} [C] Denotes the degree of dispersion of \textit{raa} distributions of the current player among one match.

{\normalsize\begin{equation}\mathbf{F_{\textbf{raa-var}}(i)}=Var(RAA)=\frac{\sum_{j=1}^{N}\left(RAA_{j}-\overline{RAA}\right)^{2}}{N}\end{equation}
}
\begin{align*}
\text{where } N \text{ denotes } \#(RAA) \text{ within a match.}
\end{align*}

\textit{aja average.} [C] Denotes the average performance regarding \textit{aja} of the current player among one match. The lower the feature is, the smaller angle and the less effort the player fires for and puts into hitting their targets.

{\normalsize\begin{equation}\mathbf{F_{\textbf{aja-avg}}(i)}=\overline{AJA}=\frac{\sum_{j=1}^{N}AJA_{j}}{N}\end{equation}
}
\begin{align*}
\text{where } N \text{ denotes } \#(AJA) \text{ within a match.}
\end{align*}

\textit{aja variance.} [C] Denotes the degree of dispersion of \textit{aja} distributions of the current player among one match. 
{\normalsize\begin{equation}\mathbf{F_{\textbf{aja-var}}(i)}=Var(AJA)=\frac{\sum_{j=1}^{N}\left(AJA_{j}-\overline{AJA}\right)^{2}}{N}\end{equation}
}
\begin{align*}
\text{where } N \text{ denotes } \#(AJA) \text{ within a match.}
\end{align*}

\subsubsection{Firing features}
\paragraph*{Inertial shot \cite{liu2017detecting} percentage (isp)}

\textit{Inertial shot} [C] denotes that after an elimination is conducted by a player, who may apply several extra firings due to the inertia. \textit{isp} represents the ratio of the number of \textit{Inertial shot} times to the number of eliminations. $\alpha$ is a threshold and set as 0.15 seconds in our case.

{\normalsize\begin{equation}
\resizebox{0.48\textwidth}{!}{$\operatorname{\textbf{IS}}\left(\textbf{E}_{\textit{kill}}^{i \rightarrow o^{*}}[k]\right)=\left\{\begin{array}{ll}1, & \exists \textbf{E}_{\textit{fire}}^{i} \in \left(\textbf{\textit{t}}\left(\textbf{E}_{\textit{kill}}^{i \rightarrow o^{*}}[k]\right),\textbf{\textit{t}}\left(\textbf{E}_{\textit{kill}}^{i \rightarrow o^{*}}[k]\right)+\alpha\right) \\0,  & \text{otherwise}\end{array}\right.$}\end{equation}
}
{\normalsize\begin{equation}\mathbf{F_{\textbf{isp}}(i)}=\frac{\sum_{k} \operatorname{\textbf{IS}}\left(\textbf{E}_{\textit{kill}}^{i \rightarrow o^{*}}[k]\right)}{\sum_{n}\textbf{E}_{\textit{kill}}^{i \rightarrow o^{*}}[n]}\end{equation}
}
\paragraph*{First hit percentage (fhp)} [C] 
Denote how accurate a player is while spotting an opponent and reacting to fire. Event \textit{reload} represents the action of changing the magazine while the firing is performed and the bullet is consumed. \textit{\textbf{F}ire \textbf{R}ound} denotes the unit of the firing round. \textit{fhp} refers to the ratio of the number of the first firing event equivalent to a damage event while spotting at least one opponent within a time threshold $\varepsilon$ to the number of times that the aforementioned event occurs while disregarding the equivalency and within the same time threshold $\varepsilon$. In our case, $\varepsilon$ is set as 5 seconds (\textit{i.e}., 640 ticks).

{\normalsize\begin{equation}\textbf{E}_{\textit{reload}}^{i}=\Delta\left(\textit{mag}(i)\right)>0 \land \textbf{\textit{t}}(\textbf{E}_{\textit{fire}}^{i}[k])-\textbf{\textit{t}}(\textbf{E}_{\textit{fire}}^{i}[k+1])>0\end{equation}
}
{\normalsize
\begin{equation}
\textbf{FR}[i]={\{\textbf{E}_{\textit{fire}}^{i}[i]\mid\exists\textbf{E}_{\textit{reload}}^{i}\lor\textbf{\textit{t}}(\textbf{E}_{\textit{fire}}^{i}[i][k+1])-\textbf{\textit{t}}(\textbf{E}_{\textit{fire}}^{i}[i][k])>\varepsilon\}}
\end{equation}
}
{\normalsize\begin{equation}\mathbf{F_{\textbf{fhp}}(i)}=\frac{\#\left(\{\textbf{FR}[i][0]\mid\forall\textbf{FR}[i][0]=\textbf{E}_{\textit{damage}}^{i \rightarrow o^{*}}\}\right)}{\#\left(\textbf{FR}[i][0]\right)}\end{equation}
}
\paragraph*{Precision (precis)} [C] 
Denote the performance of firing accuracy. \textit{precis} refers to the ratio of the number of damage caused by weapons to the total number of fires applied.

{\normalsize\begin{equation}\mathbf{F_{\textbf{precis}}(i)}=\frac{\sum_{k}\textbf{E}_{\textit{damage}}^{i \rightarrow o^{*}}[k]}{\sum_{n}\textbf{E}_{\textit{fire}}^{i}[n]}\end{equation}
}
\paragraph*{Strafing hit percentage (shp)} [C] 
Denote fire accuracy while \textit{strafing}. \textit{Strafe} allows a player to keep the view focused on an opponent while moving in a different direction. \textit{shp} refers to the ratio of the number of damage caused while strafing to the number of damage caused.

{\normalsize\begin{equation}\mathbf{F_{\textbf{shp}}(i)}=\frac{\sum_{k}\textbf{E}_{\textit{damage}\land\textit{strafe}}^{i \rightarrow o^{*}}[k]}{\sum_{n}\textbf{E}_{\textit{damage}}^{i \rightarrow o^{*}}[n]}\end{equation}
}
\paragraph*{Hit-group distribution (hgd) variance} [C] 
\textit{Hit group} represents the body part that suffers damage, typically hit groups can be categorized as follows: \textit{chest}, \textit{generic}, \textit{head}, \textit{neck}, \textit{left arm}, \textit{right arm}, \textit{left leg}, \textit{right leg}, and \textit{stomach}. \textit{hgd} variance denotes the degree of dispersion of \textit{hit group} distributions for the current player among one match. 

{\normalsize\begin{equation} HG_{j}=\sum_{k}\textbf{E}_{\textit{damage}\land\textit{hg(j)}}^{i \rightarrow o^{*}}[k]\end{equation}
}
{\normalsize\begin{equation}\overline{HG}=\frac{\sum_{j=chest}^{stomach}HG_{j}}{\sum_{k}\textbf{E}_{\textit{damage}}^{i \rightarrow o^{*}}[k]}\end{equation}
}
{\normalsize\begin{equation}\mathbf{F_{\textbf{hgd}}(i)}=Var(HG)=\frac{\sum_{j=chest}^{stomach}\left(HG_{j}-\overline{HG}\right)^{2}}{\sum_{n}\textbf{E}_{\textit{damage}}^{i \rightarrow o^{*}}[n]}\end{equation}
}
\begin{align*}
\text{where } j \in \{\text{chest, generic, head, neck, left arm,} & \\
\text{right arm, left leg, right leg, stomach}\}
\end{align*}

\paragraph*{Special hit Ratio (shr)} [S] \textit{Special hit} denotes hit with sniper between 50 and 150 meters (\textit{SH50-150}), head shot with sniper between 40 and 170 meters (\textit{SHS40-170}), hit with normal weapon farther than 800 meters (\textit{NH800+}) and headshot with normal weapon farther than 700 meters (\textit{NHS700+}). In FPS games, hitting processes include using different weapons and shooting from different distances. In our game environment, the aforementioned four types of hits are considered to be potentially cheating combinations. \textit{shr} denotes the ratio of the sum of the number of the aforementioned four types of \textit{special hit} to the number of hits in one match.

{\normalsize\begin{equation}\resizebox{0.43\textwidth}{!}{$\mathbf{\textbf{SH}_{50-150}(i)}=\sum_{k}\textbf{E}_{\textit{damage}\land{sniper}\land{(50 <= dist(i, o)} <= 150)}^{i \rightarrow o^{*}}[k]$}\end{equation}
}
{\normalsize\begin{equation}\resizebox{0.43\textwidth}{!}{$\mathbf{\textbf{SHS}_{40-170}(i)}=\sum_{k}\textbf{E}_{\textit{headshot}\land{sniper}\land{(40 <= dist(i, o)} <= 170)}^{i \rightarrow o^{*}}[k]$}\end{equation}
}
{\normalsize\begin{equation}\mathbf{\textbf{NH}_{800+}(i)}=\sum_{k}\textbf{E}_{\textit{damage}\land{normal}\land{(dist(i, o)} >= 800)}^{i \rightarrow o^{*}}[k]\end{equation}
}
{\normalsize\begin{equation}\mathbf{\textbf{NHS}_{700+}(i)}=\sum_{k}\textbf{E}_{\textit{headshot}\land{normal}\land{(dist(i, o)} >= 700)}^{i \rightarrow o^{*}}[k]\end{equation}
}
{\normalsize\begin{equation}\resizebox{0.4\textwidth}{!}{$\mathbf{F_\textbf{shr}(i)}=\frac{
\mathbf{\textbf{SH}_{50-150}(i)} + \mathbf{\textbf{SHS}_{40-170}(i)} + \mathbf{\textbf{NH}_{800+}(i)} + \mathbf{\textbf{NHS}_{700+}(i)}}
{\sum_{k}\textbf{E}_{\textit{damage}}^{i \rightarrow o^{*}}[k]}$}\end{equation}
}
\subsubsection{Elimination features}
\paragraph*{Time to kill (ttk)}
Denote the duration from the moment of the first damage caused to an opponent to the moment of the elimination of the opponent. In order to control the \textit{ttk} to be each time one player attacks one enemy, the threshold of one statistical period of \textit{ttk} is set to be 10 seconds. For instance, if the attacking period time is over 10 seconds without killing the victims, this action time will not be recorded as \textit{ttk}.

{\normalsize\begin{equation}\mathbf{F_{\textbf{ttk}}(i)}=\operatorname{\textbf{ELAPSE}}\left(\textbf{E}_{\textit{damage}}^{i \rightarrow o}[0], \textbf{E}_{\textit{kill}}^{i \rightarrow o}\right)\end{equation}
}

\textit{ttk average.} [C] Denotes the average time regarding \textit{ttk} of the current player among one match. The lower this feature is, the shorter the time the player kills victims.

{\normalsize\begin{equation}\mathbf{F_{\textbf{ttk-avg}}(i)}=\overline{TTK}=\frac{\sum_{j=1}^{N}TTK_{j}}{N}\end{equation}
}
\begin{align*}
\text{where } N \text{ denotes } \#(TTK) \text{ within a match.}
\end{align*}

\textit{ttk variance.} [C] Denotes the degree of dispersion of \textit{ttk} distributions of the current player among one match. 
{\normalsize\begin{equation}\mathbf{F_{\textbf{ttk-var}}(i)}=Var(TTK)=\frac{\sum_{j=1}^{N}\left(TTK_{j}-\overline{TTK}\right)^{2}}{N}\end{equation}
}
\begin{align*}
\text{where } N \text{ denotes } \#(TTK) \text{ within a match.}
\end{align*}

\paragraph*{First-blood percentage (fbp)} [C] 
\textit{first-blood} denotes the very first elimination in one round. \textit{fbp} denotes the ratio of the number of \textit{first-blood} to the number of rounds in one match.

{\normalsize\begin{equation}\mathbf{F_{\textbf{fkp}}(i)}=\frac{\#\left(\textbf{FK}(i)\right)}{\#\left(round\right)}\end{equation}
}
\paragraph*{Onetap percentage (otp)} [C] 
% \textit{Onetap} denotes a head-shot fire event on an opponent that conducts an elimination without sniper rifles. \textit{otp} represents the ratio of the number of \textit{Onetap} times to the number of damage done times by the normal weapons.
\textit{Onetap} denotes a head-shot fire event on an opponent that results in elimination without the use of sniper rifles, and where the inflicted damage is equal to or greater than 100 (the full health of a player). The \textit{otp} represents the ratio of the number of \textit{Onetap} instances to the number of times damage is inflicted by normal weapons.

{\normalsize\begin{equation}\resizebox{0.43\textwidth}{!}{$
\operatorname{\textbf{ONETAP}}\left(\textbf{E}_{\textit{kill}}^{i \rightarrow o^*}[k]\right)=\left\{
\begin{array}{ll}
    1, & \left(\textbf{E}_{\textit{fire}}^{i}=\textbf{E}_{\textit{kill}}^{i \rightarrow o^{*}}[k]\right) \land \left(\textbf{E}_{\textit{damage}}^{i \rightarrow o^{*}} \geq 100\right) \\
    0, & \text{otherwise}
\end{array}
\right.
$}
\end{equation}
}
{\normalsize\begin{equation}\mathbf{F_{\textbf{otp}}(i)}=\frac{\sum_{k} \operatorname{\textbf{ONETAP}}\left(E_{\text{kill}}^{i \rightarrow o^{*}}[k]\right)}{\sum_{n}\textbf{E}_{\textit{damage}\land wp(i)\neq sniper\ rifle}^{i \rightarrow o^{*}}[n]}\end{equation}
}
\paragraph*{Occluder-penetration index (opi)} [S] 
\textit{Occluder-penetration} denotes the elimination event that happens when the victim is eliminated behind one layer or more layers of occluders by a penetrated bullet. The occluders are typically walls, players (namely wall bang), or smoke. \textit{opi} refers to a weighted index that reflects the extent of the average occluder-penetration eliminations conducted by the player. The use of the number of rounds within the division aims to eliminate bias brought by that. In our case, the weights in $\alpha_i$ are 0.5, 1, 2, 0.5 respectively.

{\normalsize
\begin{equation}
\mathbf{F_{\text{opi}}(i)}=\frac{\mathbf{W_{i}}\alpha_{i}}{\#\left(\textit{round}\right)}\
\end{equation}
}
$$\text{where}\ \alpha_{i}=\begin{bmatrix}w_1\\w_2\\w_3\\w_4\end{bmatrix},$$

$$
\resizebox{0.48\textwidth}{!}{${\mathbf{W_{i}}}=\begin{bmatrix}\sum_{k}\textbf{E}_{\textit{kill}\land p(1)}^{i \rightarrow o^{*}}[k] & \sum_{n}\textbf{E}_{\textit{kill}\land p(2)}^{i \rightarrow o^{*}}[n] & \sum_{m}\textbf{E}_{\textit{kill}\land p(>2)}^{i \rightarrow o^{*}}[m] & \sum_{j}\textbf{E}_{\textit{kill}\land ts}^{i \rightarrow o^{*}}[j]\end{bmatrix}$}
$$

\paragraph*{Blind kill percentage (bkp)} [S] 
Denote the percentage of a player performing elimination while blinded. \textit{bkp} refers to the ratio of the sum of elimination events where the attackers suffer a blinded effect to the number of eliminations.

{\normalsize\begin{equation}\mathbf{F_{\textbf{bkp}}(i)}=\frac{\sum_{k} \textbf{E}_{\textit{kill}\land\textit{blind}}^{i \rightarrow o^{*}}[k]}{\sum_{n}\textbf{E}_{\textit{kill}}^{i \rightarrow o^{*}}[n]}\end{equation}
}
\paragraph*{Critical hit percentage (chp)} [C] 
Denote the \textit{critical hit} rate. \textit{Critical hit} represents a successful damage event that applies to a particular hit group (typically, head) that deals more damage than a normal blow. \textit{chp} refers to the ratio of the number of critical hits to the number of damage caused.

{\normalsize\begin{equation}\mathbf{F_{\textbf{chp}}(i)}=\frac{\sum_{k}\textbf{E}_{\textit{headshot}}^{i \rightarrow o^{*}}[k]}{\sum_{n}\textbf{E}_{\textit{damage}}^{i}[n]}\end{equation}
}
\paragraph*{Average kill per round (akpr)}
Denote the average number of eliminations per round by a player in a match.
{\normalsize\begin{equation}\mathbf{F_{\textbf{akpr}}(i)}=\frac{\sum_{k} \textbf{E}_{\textit{kill}}^{i \rightarrow o^{*}}[k]}{\#\left(round\right)}\end{equation}
}
\subsubsection{Props utilization features}
\paragraph*{Flash efficiency index (fei)} [S] 
Denote how well a player utilizes the flash grenade. \textit{fei} refers to the ratio of the total affection duration on the opponents minus that on the friendlies to the total affection duration.

{\footnotesize\begin{equation}
\mathbf{F_{\textbf{fei}}(i)}=\frac{\sum_{k} \operatorname{\textbf{ELAPSE}}\left(\textbf{E}_{\textit{blind}}^{i \rightarrow o^{*}}[k]\right)-\sum_{m} \operatorname{\textbf{ELAPSE}}\left(\textbf{E}_{\textit{blind}}^{i \rightarrow a^{*}}[m]\right)}{\sum_{n}\operatorname{\textbf{ELAPSE}}\left(\textbf{E}_{\textit{blind}}^{i \rightarrow p^{*}}[n]\right)}\end{equation}}

\paragraph*{Props utilization index (pui)} [S] 
Denote how frequently a player utilizes props such as a high-explosive grenade, Molotov or incendiary, smoke grenade, or flash grenade. \textit{pui} refers to the ratio of the sum of a player and the number of times different grenades are used to a fixed constant which reflects the maximum number of grenades used within this match for all players. $\rho$ denotes the maximum number of props a player is allowed to carry per round. In our case, $\rho$ is 5.

{\footnotesize\begin{equation}
\mathbf{F_{\textbf{pui}}(i)}=\frac{\#\left(\textbf{HEG}(i)\right)+\#\left(\textbf{INC}(i)\right)+\#\left(\textbf{SMK}(i)\right)+\#\left(\textbf{FLS}(i)\right)}{\#\left(player\right)\times\#\left(round\right)\times\rho}
\end{equation}}

\subsection{Temporal Features} \label{appx:temporalfeatures}

\subsubsection{Engagement features}\label{appx:eng}
Engagement features are extracted at each tick of the event occurrence. The introduction of the Engagement features is as follows:
\begin{itemize}[topsep=6pt, itemsep=2pt, parsep=0pt, partopsep=0pt]
    \item \textbf{Damage features}:
    \begin{itemize}[topsep=6pt, itemsep=2pt, parsep=0pt, partopsep=0pt]
        \item \textbf{tick} [C]: the tick number;
        \item \textbf{seconds} [C]: the time elapsed within a round in second;
        \item \textbf{attackerSteamID, victimSteamID} [C] : the ID of the player;
        \item \textbf{attackerSide, victimSide} [C]: the team side of the player;
        \item \textbf{attackerX, attackerY, attackerZ, victimX, victimY, victimZ} [C]: the X, Y, and Z location coordination in $\mathbb{R}^{3}$ of the player;
        \item \textbf{attackerViewX, attackerViewY, victimViewX, victimViewY} [C]: the X and Y view direction coordination in $\mathbb{R}^{2}$ of the player;
        \item \textbf{attackerStrafe} [C]: the Boolean value indicates whether the attacker is moving sideways relative to their forward direction;
        \item \textbf{weapon} [C]: the ID of the damage-caused weapon;
        \item \textbf{weaponClass} [S]: the class ID of the damage-caused weapon;
        \item \textbf{hpDamage} [C]: the value of damage caused to the victim;
        \item \textbf{hpDamageTaken} [C]: the value of damage taken by the victim;
        \item \textbf{armorDamage} [S]: the value of the damage caused to the victim's armor;
        \item \textbf{armorDamageTaken} [S]: the value of damage taken by the victim's armor;
        \item \textbf{hitGroup} [C]: the ID of the damage-taken area on the body of the victim;
        \item \textbf{isFriendlyFire} [S]: the Boolean value indicates whether the damage is from the friendly;
        \item \textbf{distance} [C]: the distance between the attacker and the victim;
        \item \textbf{zoomLevel} [S]: the zoom level of the attacker's weapon;
        \item \textbf{roundNum} [S]: the round number of the current tick.
    \end{itemize}
    \item \textbf{Auxiliary props features}: The tick, second, IDs, sides, location, view direction, and round number are similar to above;
    \begin{itemize}[topsep=6pt, itemsep=2pt, parsep=0pt, partopsep=0pt]
        \item \textbf{flashDuration} [S]: the time elapsed for the blinding effect affects on the victim in seconds;
    \end{itemize}
    \item \textbf{Offensive props features}: The ID, side, location, view direction of the thrower and the prop, and round number are similar to above;
    \begin{itemize}[topsep=6pt, itemsep=2pt, parsep=0pt, partopsep=0pt]
        \item \textbf{throwTick, destroyTick} [S]: the tick of the prop thrown or destroyed;
        \item \textbf{throwSeconds, destroySeconds} [S]: the time elapsed within a round in seconds at the moment of the prop thrown or destroyed;
    \end{itemize}
    \item \textbf{Elimination features}: The tick, second, IDs, sides, location, view direction, distance, weapon, weapon class, and round number are similar as above;
    \begin{itemize}[topsep=6pt, itemsep=2pt, parsep=0pt, partopsep=0pt]
        \item \textbf{assisterSteamID} [C]: the ID of the assister;
        \item \textbf{assisterSide} [C]: the team side of the assister;
        \item \textbf{isSuicide} [C]: the Boolean value indicates whether the elimination is a suicide event;
        \item \textbf{isTeamkill} [S]: the Boolean value indicates whether the elimination is a team kill event;
        \item \textbf{isWallbang} [S]: the Boolean value indicates whether the elimination is conducted by occluder penetration;
        \item \textbf{penetratedObjects} [S]: the number of obstacles the bullet penetrated;
        \item \textbf{isFirstKill} [C]: the Boolean value indicates whether the elimination is the first kill within a round;
        \item \textbf{isHeadshot} [C]: the Boolean value indicates whether the elimination is a head-shot event;
        \item \textbf{victimBlinded} [S]: the Boolean value indicates whether the victim is blinded;
        \item \textbf{attackerBlinded} [S]: the Boolean value indicates whether the attacker is blinded;
        \item \textbf{flashThrowerSteamID} [S]: the ID of the flash-thrower;
        \item \textbf{flashThrowerSide} [S]: the team side of the flash-thrower;
        \item \textbf{noScope} [S]: the Boolean value indicates whether the elimination is conducted with no scoped;
        \item \textbf{thruSmoke} [S]: the Boolean value indicates whether the elimination is conducted through smoke;
        \item \textbf{isTrade} [S]: the Boolean value indicates whether the elimination is traded with the opponent;
    \end{itemize}
    \item \textbf{Weapon fire features}: The IDs, sides, location, view direction, weapon, weapon class, player strafe, zoom level of the attacker, and the tick, second, and round number are similar as above;
    \begin{itemize}[topsep=6pt, itemsep=2pt, parsep=0pt, partopsep=0pt]
        \item \textbf{ammoInMagazine} [C]: the number of ammunition in the magazine;
        \item \textbf{ammoInReserve} [C]: the number of ammunition in the reserve;
    \end{itemize}
\end{itemize}

\subsubsection{Movement features}\label{appx:mov}

Movement features are extracted per tick. The IDs, sides, location, view direction of the player, and the tick, second, and round numbers are similar to those above. The rest of the introduction of the Movement features is as follows:
\begin{itemize}[topsep=6pt, itemsep=2pt, parsep=0pt, partopsep=0pt]
    \item \textbf{velocityX, velocityY, velocityZ} [C]: the velocity of the player in 3D space;
    \item \textbf{isAlive} [C]: the Boolean value indicates whether the player is alive;
    \item \textbf{isBlinded} [S]: the Boolean value indicates whether the player is blinded;
    \item \textbf{isAirborne} [S]: the Boolean value indicates whether the player is feet dangling;
    \item \textbf{isDucking} [C]: the Boolean value indicates whether the player is ducking;
    \item \textbf{isDuckingInProgress} [C]: the Boolean value indicates whether the player is ducking in progress;
    \item \textbf{isUnDuckingInProgress} [C]: the Boolean value indicates whether the player is standing up from ducking in progress;
    \item \textbf{isDefusing} [S]: the Boolean value indicates whether the player is defusing the bomb;
    \item \textbf{isPlanting} [S]: the Boolean value indicates whether the player is planting the bomb;
    \item \textbf{isReloading} [C]: the Boolean value indicates whether the player is reloading;
    \item \textbf{isInBombZone} [S]: the Boolean value indicates whether the player is in the bomb planting area;
    \item \textbf{isStanding} [C]: the Boolean value indicates whether the player is standing;
    \item \textbf{isScoped} [S]: the Boolean value indicates whether the player is scoped with the weapon;
    \item \textbf{isWalking} [S]: the Boolean value indicates whether the player is walking;
    \item \textbf{IsolationDegree} [C]: the distance to the calculation center of each teammate.
\end{itemize}

\subsubsection{Economy features}\label{appx:eco}

Economy features are extracted per round. The introduction of the Economy features are as follows:
\begin{itemize}[topsep=6pt, itemsep=2pt, parsep=0pt, partopsep=0pt]
    \item \textbf{round} [S]: the number of round;
    \item \textbf{equipmentValueFreezetimeEnd} [S]: the value of the player's equipment after the end of the freeze time (the interval from the commencement of a round to the moment players are granted the ability to move and fire. This period typically allows players to purchase equipment);
    \item \textbf{equipmentValueRoundStart} [S]: the value of the player's equipment at the start of each round; 
    \item \textbf{cash} [S]: the amount of money the player obtains at the end of each round;
    \item \textbf{cashSpendTotal} [S]: the amount of money the player spends within each round.
\end{itemize}
\subsection{Sense and Performance Features Classification}
\label{appx:spcfeatures}

All features presented in this section are recombinations of features from the first two subsystems, categorized into two main types, i.e., sense features and performance features. For a detailed description of each feature, please refer to \autoref{appx:temporalfeatures} and \autoref{appx:structuredfeatures}. The following merely enumerates the specific categorizations of the features and some brief descriptions.

\subsubsection{Sense Features}

Sense features indicate a player's in-game cognition, emphasizing their strategic acumen and tactical foresight.

\begin{itemize}[topsep=6pt, itemsep=2pt, parsep=0pt, partopsep=0pt]
    \item \textbf{Temporal Sense Features}:
    \begin{itemize}[topsep=6pt, itemsep=2pt, parsep=0pt, partopsep=0pt]
        \item \textbf{Economy features}: Financial transactions and asset management, mirroring strategic resource allocation.
        \item \textbf{Movement features}: Spatial patterns, positioning, and navigation dynamics, indicating tactical awareness.
        \item \textbf{Flash features}: Strategic deployment of flash grenades during in-game engagements.
        \item \textbf{Grenade features}: Utilization of assorted grenades, highlighting understanding of map control.
    \end{itemize}
    
    \item \textbf{Structured Sense Features}:
    \begin{itemize}[topsep=6pt, itemsep=2pt, parsep=0pt, partopsep=0pt]
        \item \textbf{Flash efficiency index (fei)}: Metrics gauging flash grenade efficiency.
        \item \textbf{Occluder-penetration index (opi)}: Achieve eliminations at specific locations, demonstrating proficiency with the map.
        \item \textbf{Inertial shot percentage (isp)}: As a normal human being's habitual firing after destroying an opponent.
        \item \textbf{Blind kill percentage (bkp)}: Eliminations executed under visual impairment.
        \item \textbf{Props utilization index (pui)}: Metrics showcasing prop usage across the whole match.
    \end{itemize}
\end{itemize}

\subsubsection{Performance Features}

Performance features encapsulate a player's in-game efficacy, concentrating on engagement aptitude, aiming precision, and mechanism mastery.

\begin{itemize}[topsep=6pt, itemsep=2pt, parsep=0pt, partopsep=0pt]
    \item \textbf{Temporal Performance Features}:
    \begin{itemize}[topsep=6pt, itemsep=2pt, parsep=0pt, partopsep=0pt]
        \item \textbf{Elimination Features}: Efficiency of eliminations and some special, difficult eliminations.
        \item \textbf{Weapon Fire Features}: Proficiency metrics across weapon categories, illuminating combat adeptness.
        \item \textbf{Damage Features}: The player's performance with respect to inflicting effective damage in-game.
    \end{itemize}

    \item \textbf{Structured Performance Features}: The remaining structured features in \autoref{appx:structuredfeatures}. 
    % \begin{itemize}[topsep=6pt, itemsep=2pt, parsep=0pt, partopsep=0pt]
    %     \item \textbf{Time to kill (ttk)}: Time taken from start of damage to elimination.
    %     \item \textbf{Angle dynamics}: Aiming adaptability and responsiveness metrics.
    %     \item \textbf{Critical hit percentage (chp)}: Precision metrics targeting critical hits (head shot).
    %     \item \textbf{First-blood percentage (fbp)}: Ratio metrics for the first elimination in all rounds of a game match.
    %     \item \textbf{Onetap percentage (otp)}: Single-shot elimination metrics.
    %     \item \textbf{Strafing hit percentage (shp)}: Metrics firing accuracy during lateral movements.        
    %     \item \textbf{Average Kill per round (akpr)}: Metrics averaging round-wise eliminations.
    % \end{itemize}
\end{itemize}
\begin{figure*}[htbp]
\centering
\includegraphics[width=0.8\textwidth]{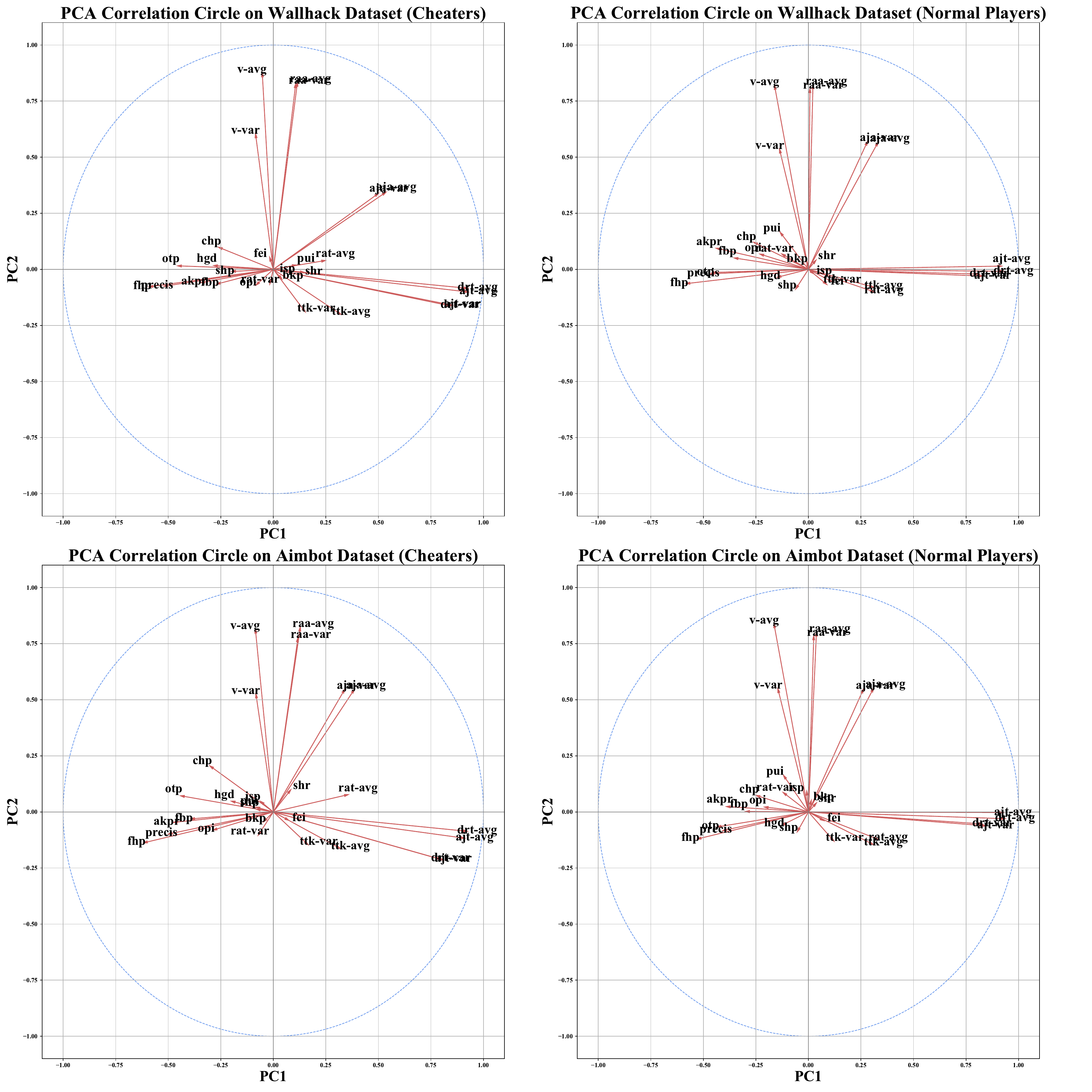}
\caption{Structured Features PCA results on both datasets.}
% \Description{Structured Features PCA results on both datasets.}
\label{fig:pca_vis}
\end{figure*}
\subsection{PCA on Structured Features}
\label{appx:SF_PCA}
In Section IV-A1, we have demonstrated the distribution distinction between cheaters and normal players.
In the following section, we will demonstrate the PCA visualization on the structured features to display their orthogonality on both \textit{Aimbot} and \textit{Wallhack} datasets.
\autoref{fig:pca_vis} illustrates the structured features PCA results for both datasets regarding the cheaters and normal players respectively.
Horizontal and vertical axes represent the first principal component (PC1) and the second principal component (PC2), respectively. 
These principal components capture the directions of maximum variance in the data, showing where the data varies most.
The unit circle in a correlation circle plot is a circle with a radius of 1, serving as a reference. 
Within this circle, the vectors (represented as arrows) demonstrate the magnitude and direction of the original variables' associations with the principal components.
The arrow represents feature vectors, where the direction and length of each arrow indicate the relationship of the original data features with the principal components.
The direction of the arrow shows the correlation direction of the feature with the principal components. 
If two arrows are pointing in similar directions, it means their corresponding features exhibit similar patterns of variation in the dataset.
Length indicates the impact or correlation strength of that feature on the corresponding principal component. 
Longer arrows suggest that the feature contributes more significantly to the variance captured by that principal component, and vice versa.
It is noticeable that within the same plot, most of the features are distinct from others by obtaining different directions and lengths of the arrow.
We can also observe that the direction and length of the arrows across \textit{Aimbot} and \textit{Wallhack} datasets are distinctive, indicating that the data distributions of the corresponding features are statistically different.
Additionally, there are differences in the direction and length of the arrows between cheaters and normal players within the same or across different datasets.
Above all, the original features can not be replaced by PCA. 

\subsection{Evaluation Metrics}
\label{appx:eval_metrics}
The following are the detailed evaluation matrices that are used in Section VI.
% Details about the evaluation metrics' descriptions in the table can be found in Appx.\autoref{app:evaluation}. 

% To set the stage for our detailed evaluation, here are some basic concepts pivotal in detecting cheating behaviors in FPS games. The forthcoming metrics, rooted in these elementary notions, will provide a lens through which we scrutinize the models' adeptness at identifying cheating instances amid the complex gaming environment.
\textbf{True Positives (TP)}, shown in \autoref{eq:TP}, represents instances where the model accurately predicts the occurrence of cheating behavior.

{\normalsize\begin{equation}\label{eq:TP}
TP = \sum_{i=1}^{n} I(y_i = 1 \land \hat{y}_i = 1)
\end{equation}
}

\textbf{True Negatives (TN)}, represented in \autoref{eq:TN}, denotes instances where the model correctly identifies the absence of cheating behavior.

{\normalsize\begin{equation}\label{eq:TN}
TN = \sum_{i=1}^{n} I(y_i = 0 \land \hat{y}_i = 0)
\end{equation}
}

\textbf{False Positives (FP)}, represented in \autoref{eq:FP}, encompass instances where the model erroneously predicts the presence of cheating behavior.

{\normalsize\begin{equation}\label{eq:FP}
FP = \sum_{i=1}^{n} I(y_i = 0 \land \hat{y}_i = 1)
\end{equation}
}

\textbf{False Negatives (FN)}, represented in \autoref{eq:FN}, are instances where the model inaccurately predicts the absence of cheating behavior.

{\normalsize\begin{equation}\label{eq:FN}
FN = \sum_{i=1}^{n} I(y_i = 1 \land \hat{y}_i = 0)
\end{equation}
}
In these equations, $y_i$ denotes the true label, $\hat{y}_i$ signifies the predicted label, and $I(\cdot)$ represents the indicator function.
% With these basic elements, here are some metrics that are used in the evaluation.

\textbf{Recall}, represented in \autoref{eq:Recall}, measures the proportion of actual positive instances that were correctly identified by the model.
{\normalsize\begin{equation}\label{eq:Recall}
Recall = \frac{TP}{TP + FN}
\end{equation}
}

\textbf{Accuracy}, as shown in \autoref{eq:Accuracy}, gives a holistic measure of the model's performance across both positive and negative classes.

{\normalsize\begin{equation}\label{eq:Accuracy}
Accuracy = \frac{TP + TN}{TP + TN + FP + FN}
\end{equation}
}
% \textbf{Specificity}, illustrated in \autoref{eq:Specificity}, computes the proportion of actual negative instances that were correctly identified by the model.
% {\normalsize\begin{equation}\label{eq:Specificity}
% Specificity = \frac{TN}{TN + FP}
% \end{equation}
% \textbf{Balanced Accuracy}, depicted in \autoref{eq:BalancedAccuracy}, is the arithmetic mean of the True Positive Rate (TPR) and the True Negative Rate (TNR), offering a balanced perspective of the model's performance across both classes.
% {\normalsize\begin{equation}\label{eq:BalancedAccuracy}
% BA = \frac{TPR + TNR}{2}
% \end{equation}

\textbf{AUC-ROC}, denoted in \autoref{eq:AUC-ROC}, is the area under the receiver operating characteristic curve, indicating the model's capability to differentiate between positive and negative classes. It is computed by integrating the ROC curve, as shown below, where \( TPR \) (True Positive Rate) is synonymous with Recall, and \( FPR \) (False Positive Rate) is defined in \autoref{eq:FPR}.

{\normalsize\begin{equation}\label{eq:AUC-ROC}
AUC-ROC = \int_{0}^{1} TPR(FPR) \, dFPR
\end{equation}
}

\textbf{NPV}, shown in \autoref{eq:NPV}, is the proportion of actual negative instances among all instances predicted as negative.

{\normalsize\begin{equation}\label{eq:NPV}
NPV = \frac{TN}{TN + FN}
\end{equation}
}

\textbf{FPR}, indicated in \autoref{eq:FPR}, is the proportion of actual negative instances that are incorrectly predicted as positive.

{\normalsize\begin{equation}\label{eq:FPR}
FPR = \frac{FP}{FP + TN}
\end{equation}
}
% Lastly, \textbf{Informedness}, represented in \autoref{eq:Informedness}, encapsulates the excess of true positive rate over false positive rate, offering a single-value summary of the model's diagnostic ability.

% \begin{equation}\label{eq:Informedness}
% Informedness = Recall + Specificity - 1
% \end{equation}
\bibliographystyle{IEEEtran}
\bibliography{mybibliography}